\documentclass[12pt]{article}
\usepackage{graphicx}
\usepackage{amsmath}
\usepackage{amssymb}
\usepackage{longtable}
\usepackage{rotating}
\usepackage{rotfloat}
\usepackage[small]{caption2}

\newcommand{\beq}{\begin{equation}}
\newcommand{\eeq}{\end{equation}}
\newcommand{\beqn}{\begin{eqnarray}}
\newcommand{\eeqn}{\end{eqnarray}}

\begin{document}
\begin{center}
{\bf \Large A model of \boldmath \(\bar{B}^0\to D^{*+}\omega\pi^-\) decay}

\vspace{0.5cm}

D.V.~Matvienko$^a$, A.S.~Kuzmin$^b$ and S.I.~Eidelman$^c$

\vspace{0.5cm}

Budker Institute of Nuclear Physics, SB RAS,\\
11, Lavrentieva prospect, Novosibirsk, Russia\\
\vspace{3mm}
Novosibirsk State University, \\
2, Pirogova street, Novosibirsk, Russia

\vspace{0.5cm}

$^a$d.v.matvienko@inp.nsk.su\\
$^b$a.s.kuzmin@inp.nsk.su\\
$^c$s.i.eidelman@inp.nsk.su

\end{center}

\vspace{1cm}

\begin{abstract}
We suggest a parameterization of the matrix element
for \(\bar{B}^0\to D^{*+}\omega\pi^-\) decay using kinematic variables
convenient for experimental analysis.
The contributions of intermediate \(\omega\pi\)- and \(D^{**}\)-states
up to spin 3 have been taken into account. The angular distributions
for each discussed hypothesis have been obtained and analysed
using Monte Carlo simulation.
\end{abstract}

\section{Introduction}

The discovery of excited $D$-states (referred to as $D^{**}$-states)
stimulates interest in their spectroscopy and $D^{**}\to D^{(*)}\pi$
decay properties. There are four $P$-wave states, which are usually
labeled $D^*_0$ ($J^P_{j_q}=0^+_{1/2}$), $D'_1$ ($J^P_{j_q}=1^+_{1/2}$),
$D_1$ ($J^P_{j_q}=1^+_{3/2}$), $D^*_2$ ($J^P_{j_q}=2^+_{3/2}$),
where $J$ is the spin of the meson and $j_q$ is the total angular
momentum of a light quark $q=(u,d)$, which is the sum of the
orbital momentum $l$ and the light quark spin $s_q$. In the heavy
quark limit, the angular
momentum $j_q$ is a good quantum number. Conservation of parity and
angular momentum imposes constraints on the strong decays of the
$D^{**}$ to $D^{(*)}\pi$. Two states with $j_q=1/2$ decay to the
$D^{(*)}\pi$-state in $S$-wave while two other with $j_q=3/2$
decay in $D$-wave. Since the decay width $\Gamma \sim {\mathbf{Q}}^{2L+1}$,
where ${\mathbf{Q}}$ is the magnitude of the daughter particle momentum,
$L$ is the orbital momentum between decay products,
and $\mathbf{Q}$ is small,
$D_1$ and $D^*_2$ have small decay width of about $20$ MeV, but
$D^*_0$ and $D'_1$ are expected to be quite broad with decay width
of about $300$ MeV~\cite{peskin,pdg}.

 A further study of these states will allow a more in-depth comparison
to be made with theoretical predictions such as Heavy Quark Effective
Theory (HQET)~\cite{hqet,hqet2} and QCD sum rules~\cite{qcd}. The last
experimental studies of $D^{**}$ mesons were performed in
$B^-\to D^{(*)+}\pi^-\pi^-$~\cite{belle2,babar1} and
$\bar{B}^0\to D^{0}\pi^+\pi^-$~\cite{belle1} decays. These states have
also been studied in semileptonic $B$-decays~\cite{semb}. Thus,
understanding of their properties is significant for reducing
uncertainties in the measurement of semileptonic decays and
determination of the Cabibbo-Kobayashi-Maskawa (CKM)~\cite{ckm}
matrix elements $|V_{cb}|$ and $|V_{ub}|$.

 $D^{**}$-states can be also produced in other hadronic $B$ decays,
e.g., $B\to D^*\omega\pi$. Here, \(D^{**}\) production is described
by the \(W\) vertex instead of the transition Isgur-Wise
functions~\cite{hqet}, which describe these states in the  \(D^{(*)}\pi\pi\)
modes. This channel was first observed by the CLEO~\cite{cleo} and
BaBar~\cite{babar} collaborations, the latter finding an enhancement
in $D^*\pi$ mass due to the broad $D_1(2430)^0$-state, representing
a $P$-wave of a $D$ meson.

Let us note that light mesons decaying to the $\omega\pi$ final
state (e.g., $\rho(1450)$, $b_1(1235)$ and their excitations)
appear in the color-favored mode of this process. Thus, a possible
contribution of these resonant structures to the total branching fraction
can be measured. The $\rho(1450)$-resonance, dominant in this mode,
was observed by both collaborations~\cite{cleo,babar},
but the $b_1(1235)$-state was not observed in this channel.

An amplitude of three-body decay can be written as a sum of the
contributions corresponding to the quasi-two-body
resonances~\cite{belle2,babar1,belle1}.
Analysing experimental data one has to determine relative amplitudes
and phases of different intermediate states. To do this, one needs the
amplitudes expressed via kinematic variables convenient for Dalitz plot
analysis\footnote{In the case of decays with more than three particles
in the final state, the term Dalitz plot is used in a general sense to
refer to the distribution of the chosen degrees of freedom used to
describe the decay.}. These expressions can be used for optimization
of selection criteria and creation of efficient Monte-Carlo generators.

\section{The general method}

A weak $B(0^-)\to R(J^P)1^-$ decay amplitude (for $J>0$) includes
three independent terms while a strong $R(J^P)\to 1^-0^-$ amplitude can have
one or two independent terms.
We can parameterize a decay matrix element using a set of different
independent bases. In general, we can use the basis of covariant
amplitudes or helicity basis etc.
Since the real particles $D_1$ and $D'_1$ are expected to be close to the
pure $j_q=3/2$- and $j_q=1/2$-states and their decays have particular orbital
momenta, it is convenient to use the basis of amplitudes describing
decay with fixed angular orbital momenta in the \(B\) and resonance rest frames.

\begin{figure}[h]
\center
\begin{tabular}{c c}
\includegraphics[scale=1.3]{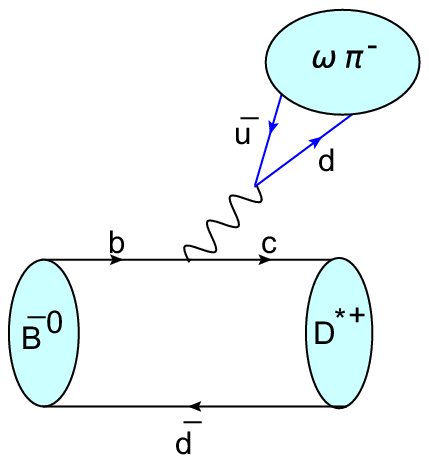} &
\includegraphics[scale=1.3]{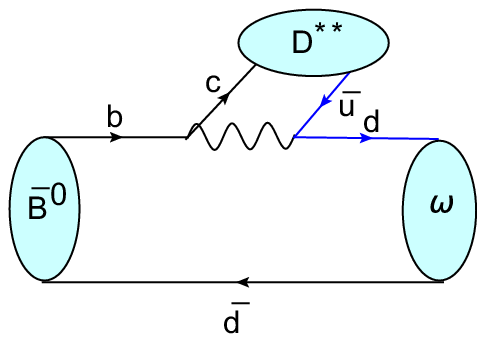}  \\
a) & b) \\
\end{tabular}
\caption{a) Color-favored and b) color-suppressed channel.}\label{fig1}
\end{figure}

In this paper we use an isobar model formulation in which our decay
is described by a coherent sum of a number of quasi-two-body amplitudes.
The amplitudes can be subdivided into two channels.
The effective Hamiltonian for Cabibbo-favored decays can be reduced to
the color-favored and color-suppressed forms \cite{wilson,kamal}:
\begin{eqnarray}
H_{CF}\,&=&\,\frac{G_F}{\sqrt{2}}V_{cb}V^*_{ud}(a_1(\bar{c}\Gamma_{\mu}b)(\bar{u}\Gamma^{\mu}d)+C_2 H^8_w){,}\nonumber \\
H_{CS}\,&=&\,\frac{G_F}{\sqrt{2}}V_{cb}V^*_{ud}(a_2(\bar{c}\Gamma_{\mu}d)(\bar{u}\Gamma^{\mu}b)+C_1 \tilde{H}^8_w){,}
\end{eqnarray}
where $G_F$ is the Fermi constant, $C_1$ and $C_2$ are the Wilson
coefficients and $\Gamma_{\mu}=\gamma_{\mu}(1-\gamma_5)$. The coefficients
$a_1=C_1+C_2/N$ and $a_2=C_2+C_1/N$, where $N$ is an effective number
of colors. The terms $H^8_w=\frac{1}{2}\sum^8_{a=1}(\bar{c}\lambda^a \Gamma_{\mu} b)(\bar{u}\lambda^a \Gamma^{\mu} d)$ and $\tilde{H}^8_w=\frac{1}{2}\sum^8_{a=1}(\bar{c}\lambda^a \Gamma_{\mu} d)(\bar{u}\lambda^a \Gamma^{\mu} b)$ ($\lambda^a$
are the Gell-Mann matrices), involving color-octet currents,
generate non-factorized contributions. The other non-factorization source
is the non-factorized matrix element of the product of the color-singlet
currents. It includes loop current-current terms.
The color-favored and color-suppressed channels are shown in Fig.~\ref{fig1}.
We show tree diagrams only, however, not all the intermediate states
are described by them.
In this paper we do not apply the factorization method but consider all
intermediate resonant contributions up to spin $3$ allowed
by the momentum-parity conservation.
The color-favored term receives a contribution from the
$\omega\pi$-resonances, e.g., $\rho(1450)$ and $b_1(1235)$. Since
these resonances are broad, this channel allows factorization to
be precisely tested~\cite{ligeti}.
The color-suppressed term receives a contribution from the $D^{**}$-states,
which are $P$- and $D$-wave excitations of the $c\bar{u}$ states.

Let us consider briefly the spectroscopy of the
$D$-wave $c\bar{u}$ excitations.
We have $J^P_{j_u}=1^-_{3/2}$-, $J^P_{j_u}=2^-_{3/2}$-, $J^P_{j_u}=2^-_{5/2}$-,
and  $J^P_{j_u}=3^-_{5/2}$ states.
Again, as discussed above, two states with $j_u=3/2$ decay to the
$D^{(*)}\pi$-state in $P$-wave and two other with $j_u=5/2$ decay in $F$-wave.

Observable $c\bar{u}$-states with the same $J^P=1^+$ ($J^P=2^-$)
quantum numbers are two linear combinations of pure $j_u=1/2$ ($j_u=3/2$)-
and $j_u=3/2$ ($j_u=5/2$)-states.
Thus, the physical $D_1$ and $D'_1$-states are as follows:
\begin{eqnarray}
|D_1>\,&=&\,\sin\vartheta_1\,|j_u=1/2>\,+\,\cos\vartheta_1\,e^{-i\vartheta_2}\,|j_u=3/2>{,}\nonumber\\
|D'_1>\,&=&\,\cos\vartheta_1\,|j_u=1/2>\,-\,\sin\vartheta_1\,e^{i\vartheta_2}\,|j_u=3/2>{,}\nonumber
\end{eqnarray}
where $\vartheta_1$ and $\vartheta_2$ are mixing angles.

Let us discuss kinematic properties of the considered process.
In the final state we have six particles, namely, $D^0$ and $\pi^+$
from the $D^{*+}$ decay, $\pi^+$, $\pi^-$ and $\pi^0$ from the $\omega$
decay and $\pi^-$ from the $\bar{B}^0$ decay. The $\bar{B}^0$ decay is
described by two invariant masses squared of the $D^*\pi$ ($m^2_{D^*\pi}$)
and $\omega\pi$ ($m^2_{\omega\pi}$) systems, the one corresponding
to resonance mass labeled as $q^2$.

The $\omega$ decay is described by five variables. We use invariant masses
squared $M^2_0=(P_++P_-)^2$ and $M^2_+=(P_++P_0)^2$ (here $P_i$ is a
$4$-momentum of the pion $\pi^i$ from the $\omega$ decay,
$i=\pm,0$)\footnote{The $\omega$ invariant mass squared is
$p^2=(P_++P_-+P_0)^2$.},
the azimuthal angle of the $\pi^0$ in the $\omega$ decay plane, and two
angles (polar $\theta$ and azimuthal $\phi$) for a vector $\vec{n}$
normal to the $\omega$ decay plane. Let us note that the
$\omega \to \pi\pi\pi$ decay proceeds through two mechanisms. The first one
involves an intermediate $\rho$-meson. Experimental studies of the
$e^+e^- \to 3\pi$ reaction have confirmed the Gell-Mann-Sharp-Wagner
suggestion~\cite{gsw} that the $\omega \to 3\pi$ transition is dominated by this
contribution. The second mechanism represents the non-resonant contribution.
This contact contribution can not be excluded because interference between
these mechanisms leads to a sizeable effect in the decay rate.

The $D^{*+}$ decay is described by two variables. We use polar
$\beta$ and azimuthal $\psi$ angles for the $D^0$ momentum in the $D^{*+}$
rest frame. For further applications we assume the width of the
$D^*$-meson to be negligible
($\Gamma_{D^{*+}} \approx 0.1\,{\rm MeV}\ll m_{D^{*+}}-(m_{D^0}+m_{\pi^+}) \approx 10\,{\rm MeV} $).
To describe the intermediate resonance decay, we use polar $\xi$ and
azimuthal $\zeta$ angles for the daughter particle momentum in the
resonance rest frame.
The polar angle $\xi$ is expressed via the invariant mass squared
$m^2_{D^*\pi}$ for the $\omega\pi$-states and $m^2_{\omega\pi}$ for
the $D^{*}\pi$-states. Moreover, the matrix element does not depend on the
azimuthal angle $\zeta$ for the $\omega\pi$- as well as for
the $D^*\pi$-states.

\begin{figure}[h]
\center
\includegraphics[scale=1.7]{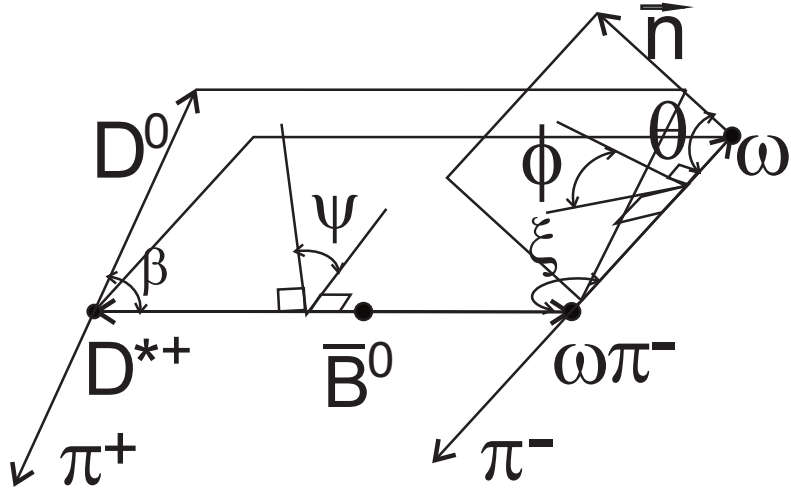}
\caption{Complete visual definition of the angles for
the $\omega\pi$-resonances. The angles $\theta$ and $\phi$ are defined in
the $\omega$ rest frame, the angles $\beta$ and $\psi$ are defined in the
$D^*$ rest frame and the angle $\xi$ is defined in the $\omega\pi$ rest
frame.}
\label{fig2}
\end{figure}

A further definition of angles depends on the decay channel.
Figure~\ref{fig2} shows the decay scheme and definition of the angles for
the $\omega\pi$-resonances.

\begin{figure}[h]
\center
\begin{tabular}{c c c}
\includegraphics[scale=1.5]{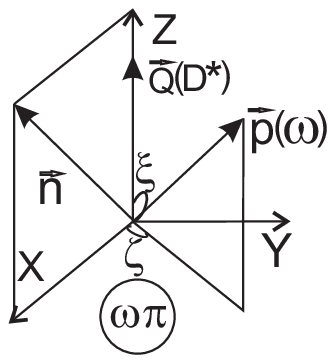} &
\includegraphics[scale=1.5]{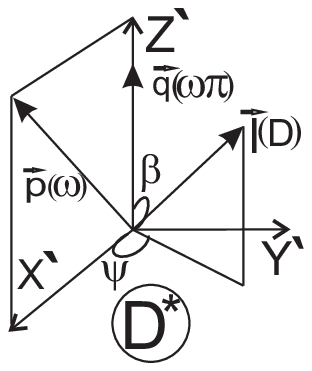} &
\includegraphics[scale=1.5]{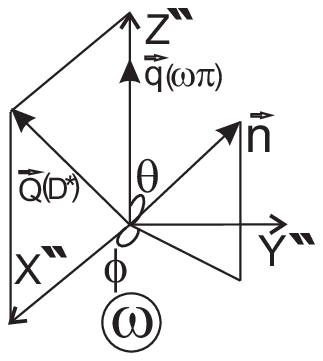}\\
a) & b) & c) \\
\end{tabular}
\caption{Definition of the angles for the
$\omega\pi$-resonances. Color-favored channel. a) The $\omega\pi$ rest frame,
b) the $D^*$ rest frame and c) the $\omega$ rest frame.}
\label{fig3}
\end{figure}

\begin{figure}[h]
\center
\begin{tabular}{c c c}
\includegraphics[scale=1.5]{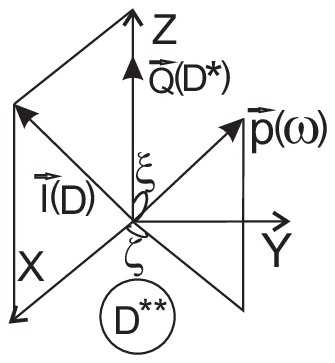} &
\includegraphics[scale=1.5]{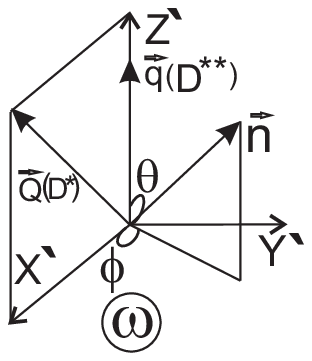} &
\includegraphics[scale=1.5]{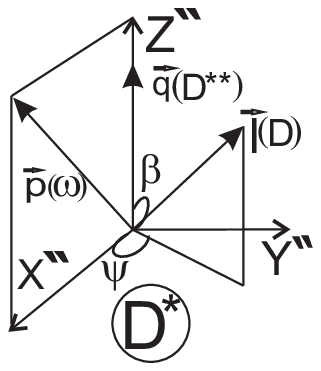}\\
a) & b) & c) \\
\end{tabular}
\caption{Definition of the angles for the
$D^{**}$-resonances. Color-suppressed channel. a) The $D^{**}$ rest frame,
b) the $\omega$ rest frame and c) the $D^*$ rest frame.}
\label{fig4}
\end{figure}

Figures~\ref{fig3} and~\ref{fig4} define these angles using momentum variables for the
$\omega\pi$- and $D^{*}\pi$-resonances, respectively. The notations are
as follows:
the variables $p$, $Q$, $l$, $q$ are the four-momenta of the
$\omega$-, $D^*$-, $D$-meson and an intermediate resonance,
respectively, while $\mathbf{p}$, $\mathbf{Q}$, $\mathbf{l}$,
$\mathbf{q}$ are the magnitudes of their three-momenta in the mother
particle rest frames. In Figs.~\ref{fig3},~\ref{fig4} the directions of these momenta
define angular variables $\theta$ and $\phi$ in the $\omega$ rest frame,
$\beta$ and $\psi$ in the $D^*$ rest frame and
$\xi$ in the resonance rest frame.

In this paper each compound particle is described by a relativistic
Breit-Wigner (BW) with a $q^2$-dependent width. Such an approach is not
exact since it does not take into account final state interactions and
is neither analytic nor unitary. Nevertheless, it describes the main features
of the amplitude behaviour and allows one to find and distinguish the
contributions of different quasi-two-body intermediate states. Thus,
the denominator of the BW propagator is:
\begin{equation}
D_R(q^2)\,=\,q^2-m^2_R+i m_R\Gamma_R(q^2){.}
\end{equation}
It corresponds to the intermediate resonance $R$ with mass $m_R$ and
$q^2$-dependent width $\Gamma_R$. The numerator of the propagator is
to be the sum over polarizations of the resonance and depends on its spin.

\section{\boldmath{$\omega\pi$}-resonances}

We consider such $\omega\pi$-states, which can be combined to
$J^P=0^-$-, $J^P=1^+$ (\(b_1(1235)\))-, $J^P=1^-$ (\(\rho(1450)\))-,
$J^P=2^-$-, $J^P=2^+$- and $J^P=3^-$(\(\rho_3(1690)\))-states.
Let us note that $J^P=0^-$, $J^P=2^-$ and $J^P=2^+$ charged states,
which decay to the $\omega\pi$-final system, have not yet been
observed at the present time~\cite{pdg}. Such states have the isotopic
quantum numbers $I^G=1^+$.
It is natural to assume that these states are members of the
\(b\) and \(\rho\)-families.

The matrix element for production of the $J^P=0^-$ intermediate state
(labeled as $\rho_0$) is given by:
\begin{equation}
M_{\bar{B}\to D^* \rho_0}\,=\,\frac{G_F}{\sqrt{2}}V_{cb}V^*_{ud}<D^* \rho_0|(\bar{u}\Gamma_{\mu}d)(\bar{c}\Gamma^{\mu}b)|\bar{B}>{,}
\end{equation}
Parameterizing this amplitude in the covariant form,
we have:\footnote{Here the term $(\varepsilon^* Q)$ is neglected
because the longitudinal currents arise far from the resonance,
where they should be suppressed by transition form factor behavior.
However, they also modify the angular dependence of the amplitude.
Throughout this paper the longitudinal currents are neglected.}
\begin{equation}
M_{\bar{B}\to D^* \rho_0}\,=\,\frac{G_F}{\sqrt{2}}V_{cb}V^*_{ud} g_{\rho_0}F_P(q^2)(\varepsilon^*q){,}
\end{equation}
where $\varepsilon_{\mu}$ is a polarization vector of
$D^*$, $g_{\rho_0}= a_1 f_{\rho_0}$\footnote{The coefficient $a_1$
is expressed via Wilson coefficients, as discussed in the previous section.},
$f_{\rho_0}$ is a weak decay constant of the $\rho_0$ and $F_P(q^2)$
is a transition  form factor.

The strong amplitude for the $\rho_0$-decay is presented as follows:
\begin{equation}
M_{\rho_0 \to \omega\pi}\,=\,\tilde{g}_{\rho_0\omega\pi}\tilde{F}_P(q^2, p^2)(v^*q){,}
\end{equation}
where $v_{\mu}$ is a polarization vector of $\omega$,
$\tilde{g}_{\rho_0\omega\pi}$ is a coupling constant and
$\tilde{F}_P(q^2,p^2)$ is a transition form factor.
The amplitude describing the $\omega$ decay comprises the contributions
from the intermediate $\rho$-meson and $3\pi$ phase space:
\begin{equation}
\label{om3pi}
M_{\omega\to 3\pi}\,=g_{\omega\rho\pi}(p^2)\left(a_{3\pi}\,+\,\sum_{i=\pm,0}\frac{g_{\rho\pi\pi}}{D_{\rho^i}(M^{2}_i)Z(M^{2}_i)}\right)\sqrt{\Delta(p,P_+,P_0)}(nv){,}
\end{equation}
where
\begin{equation}
n^{\mu}\,=\,\frac{\epsilon^{\mu\nu\rho\sigma}P_{+\nu}P_{0\rho}p_{\sigma}}{\sqrt{\Delta(p,P_+,P_0)}}
\end{equation}
is a unit 4-vector normal to the $\omega$ decay plane and
\(\Delta(p,P_+,P_0)\) is the Kibble determinant. Other notations are
described in the Appendix. Here and further \(\epsilon_{\mu\nu\rho\sigma}\)
is the Levi-Civita symbol and $\epsilon_{0123}=+1$. The amplitude
corresponding to the $D^*$ decay is
\begin{equation}
\label{dpi}
M_{D^*\to D\pi}\,=\,g_{D^*D\pi}(\varepsilon l){.}
\end{equation}

The factor
\begin{equation}
g_{D^*D\pi}(Q^2)g_{\omega\rho\pi}(p^2)\left(a_{3\pi}\,+\,\sum_{i=\pm,0}\frac{g_{\rho\pi\pi}}{D_{\rho^i}(M^{2}_i)Z(M^{2}_i)}\right)\frac{\sqrt{\Delta(p,P_+,P_0)}}{D_{D^*}(Q^2)D_{\omega}(p^2)}\mathbf{l}
\end{equation}
is common for all intermediate states, and $\omega$-decay part can be expressed
via the phase integral $W(p^2)$, presented in the Appendix.

The total rate for $B\to D^*\omega\pi$ decay expressed via the branching
fraction $\mathcal{B}_{D^{*+}\to D^0\pi^+}$ and the phase integral $W(p^2)$ can be presented as follows:
\begin{align}
\label{decompi}
d\Gamma\,&=\,\frac{6\mathcal{B}_{D^{*+}\to D^0\pi^+}}{(4\pi)^{10}m^2_B}\,\frac{|M|^2\mathbf{p}\mathbf{Q}}{\sqrt{q^2}}\,\frac{W(p^2)}{|D_{\omega}(p^2)|^2}\,dp^2\,\,(d\cos\theta\,d\phi)\,(d\cos\beta\,d\psi)\,(dq^2\,d\cos\xi){,}
\end{align}
where
$m_B$ is a $B$-meson mass, and the matrix element $M$ describes particular
dependencies for the different intermediate channels.

The matrix element for the \(\bar{B} \to D^*R_J\) transition, where \(R_J\)
is the intermediate resonance with the integer total spin
\(J\)\footnote{As emphasized above, in this paper we discuss
resonances with \(J=1,\,2,\,3\).}, can be parameterized in terms of
the amplitudes with the definite angular orbital momentum \(L\) as follows:
\begin{align}
\label{bdstr}
M_{\bar{B} \to D^* R_J}\,&=\,\frac{G_F}{\sqrt{2}}V_{cb}V^*_{ud}\,g_J\,\left[\vphantom{\frac{1}{f_{J,J+1}}} C_J\epsilon^{\mu\nu\rho\sigma} \varepsilon^{'*(J)}_{\mu} \varepsilon^*_{\nu}q_{\rho}Q_{\sigma} F_{L=J}(q^2)+ \nonumber  \right.\\& \left. {} +i m^2_B C_{J-1}((\varepsilon^{'*(J)}\varepsilon^*)-\frac{1}{f_{J,J-1}(q^2)} (\varepsilon^{'*(J)} Q) (\varepsilon^* q))F_{L=J-1}(q^2)+ \nonumber \right. \\& \left. {} + i C_{J+1} ((\varepsilon^{'*(J)} Q) (\varepsilon^* q)-f_{J,J+1}(q^2)(\varepsilon^{'*(J)}\varepsilon^*))F_{L=J+1}(q^2)\vphantom{\frac{1}{f_{J,J+1}}} \right]{.}
\end{align}
Here, \(C_J\) is the relative amplitude, which is in general complex;
\(F_L(q^2)\) is a transition form factor corresponding to the orbital
momentum \(L\); \(g_1=a_1 f_R\), \(f_R\) is a weak decay constant of the
vector resonance, when  \(q^2=m^2_R\), \(g_2=m_B g_{\bar{B}D^*R_2}\)
and \(g_3=m^2_B g_{\bar{B}D^*R_3}\) are appropriate coupling constants;
\(\varepsilon^{' (J)}\) is a convolution of the resonant polarization tensor
of rank \(J\) and momentum \(Q\)\footnote{The notation
\(\varepsilon^{'(J)}\) is not related to the resonance helicity state.}
\begin{align}
\varepsilon^{'(J=1)}_{\mu}&=\varepsilon^{'}_{\mu}, \quad \varepsilon^{'(J=2)}_{\mu}=\varepsilon^{'}_{\mu\alpha}Q^{\alpha}/m_B, \quad \varepsilon^{'(J=3)}_{\mu}=\varepsilon^{'}_{\mu\alpha\beta}Q^{\alpha}Q^{\beta}/m^2_B{;}
\end{align}
\begin{equation}
\label{fom}
f_{J,J\pm 1}(q^2)=\frac{2 m^2_B \mathbf{Q}^2}{m^2_B-m^2_{D^*}-q^2+2 a_{J,J\pm 1} m_{D^*} \sqrt{q^2}}
\end{equation}
and
\begin{align}
a_{1,0}=-1, \quad a_{1,2}=+2, \quad a_{2,1}=-1, \quad a_{2,3}=+3/2, \quad a_{3,2}=-1, \quad a_{3,4}=+4/3{.}
\end{align}

The parameterization of the matrix element describing the resonance decay
depends on its \(J^P\) quantum numbers. Thus, resonances with
\(J^P=1^-,2^+,3^-\) are described by the following matrix element:
\begin{align}
\label{rompi1}
M_{R_J \to \omega \pi}\,&=\,\tilde{g}_J\,\epsilon^{\mu\nu\rho\sigma}\tilde{\varepsilon}^{' (J)}_{\mu} v^*_{\nu} q_{\rho} p_{\sigma} \tilde{F}_{L=J}(q^2,p^2){,}
\end{align}
where \(\tilde{g}_1=g_{R_1\omega\pi}\), \(\tilde{g}_2=m_R g_{R_2\omega\pi}\)
and \(\tilde{g}_3=m^2_R g_{R_3\omega\pi}\) are appropriate coupling constants;
\(\tilde{F}_{L}(q^2,p^2)\) is a transition form factor and
\begin{align}
\tilde{\varepsilon}^{'(J=1)}_{\mu}=\varepsilon^{'}_{\mu}, \quad \tilde{\varepsilon}^{'(J=2)}_{\mu}=\varepsilon^{'}_{\mu\alpha}p^{\alpha}/m_R, \quad \tilde{\varepsilon}^{'(J=3)}_{\mu}=\varepsilon^{'}_{\mu\alpha\beta}p^{\alpha}p^{\beta}/m^2_R{.}
\end{align}
The discussed resonances with \(J^P=1^+,2^-\) are described by the
following matrix element:
\begin{align}
\label{rompi2}
M_{R_J \to \omega\pi}\,&=\,\tilde{g}_J\,\left[\vphantom{\frac{1}{f_{J,J+1}}} \tilde{C}_{J-1} m^2_R ((\tilde{\varepsilon}^{'(J)} v^*)-\frac{1}{\tilde{f}_{J,J-1}(q^2)}(\tilde{\varepsilon}^{'(J)} p)(v^* q)) \tilde{F}_{L=J-1}(q^2,p^2)+ \nonumber \right. \\& \left. {} + \tilde{C}_{J+1} ((\tilde{\varepsilon}^{'(J)} p)(v^*q) - \tilde{f}_{J,J+1}(q^2) (\tilde{\varepsilon}^{' (J)} v^*)) \tilde{F}_{L=J+1}(q^2,p^2) \vphantom{\frac{1}{f_{J,J+1}}} \right]{,}
\end{align}
where \(\tilde{C}_J\) is the relative amplitude,
which is in general complex, and
\begin{equation}
\label{tfom}
\tilde{f}_{J,J\pm 1}(q^2)\,=\,\frac{2 q^2 \mathbf{p}^2}{q^2+p^2-m^2+2 a_{J,J\pm 1}\sqrt{p^2 q^2}}{,}
\end{equation}
where $m$ is the charged pion mass.

Then we move from the covariant amplitudes to the expressions depending on
the selected angles, which are defined in the intermediate particle rest frames.

\section{$D^{**}$-resonances}

The decay rate for the channel with  $D^{**}$-resonance production
has a form similar to (\ref{decompi}).
As already mentioned, in this case the angles
$(\theta,\phi,\xi,\beta,\psi)$ differ from their analogues for
the \(\omega\pi\) states and are described in Fig.~\ref{fig4}. Here we
discuss two $J^P=1^+$-states and a $J^P=2^+$-state, which correspond
to $P$-wave in the spectroscopy of the $c\bar{u}$ excitations as well as a
$J^P=1^-$-state, two $J^P=2^-$-states and a $J^P=3^-$-state
corresponding to the $D$-wave excited $c\bar{u}$-states.
Pure \(J^P_{j_u}=1^+_{1/2}\) (\(J^P_{j_u}=2^-_{3/2}\))- and
\(J^P_{j_u}=1^+_{3/2}\) (\(J^P_{j_u}=2^-_{5/2}\))-states decay to the
\(D^*\pi\) in \(S\)- (\(P\)-) wave and \(D\)- (\(F\)-) wave, respectively.
As discussed above, observable \(J^P=1^+\) (\(J^P=2^-\)) states can be
a mixture of  pure \(j_u=1/2\) (\(j_u=3/2\)) and \(j_u=3/2\) (\(j_u=5/2\))
states.
This fact has to be taken into account for the total amplitude construction.
The parameterization of the matrix elements for all \(D^{**}\)-states is
similar to the case of the \(\omega\pi\)-states. However,
mutual substitutions of the four-momenta \(p\) and \(Q\) and polarizations
\(\varepsilon_{\mu}\) and \(v_{\mu}\) have to be made.
The functions \(f_{J,J\pm 1}(q^2)\) and \(\tilde{f}_{J,J\pm 1}(q^2)\)
for the \(D^{**}\)-states are as follows:
\begin{align}
\label{fddst}
f_{J,J\pm 1}(q^2)\,&=\,\frac{2 m^2_B \mathbf{p}^2}{m^2_B-p^2-q^2+2 a_{J,J\pm 1} \sqrt{p^2 q^2}}{,}\\
\label{tfddst}
\tilde{f}_{J,J\pm 1}(q^2)\,&=\,\frac{2 q^2 \mathbf{Q}^2}{q^2+m^2_{D^*}-m^2+2 a_{J,J\pm 1} m_{D^*}\sqrt{q^2}}{.}
\end{align}

\section{Results}
Using the technique described in the previous sections,
we present the final expressions for matrix elements with
different intermediate resonances. The total matrix element squared
is as follows:
\begin{align}
|M|^2\,&=\, |M_6+M_{\rho_0}+M_{\rho(1450)}+M_{b_1(1235)}+ M_{b_2}+M_{\rho_2}+M_{\rho_3}
+M_{D_1}+M_{D'_1}+M_{D^*_2}+\nonumber \\&+M_{1^-_{3/2}}+ M_{D_2}+M_{D'_2}+M_{3^-_{5/2}}|^2{.}
\end{align}
Here, \(M_6\) presents the non-resonant contributions to the matrix element.
The amplitudes $M_{D_1}$ and $M_{D'_1}$ are as follows:
\begin{eqnarray}
M_{D_1}\,&=&\,\sin\vartheta_1\,|j_u=1/2>\,+\,\cos\vartheta_1\,e^{-i\vartheta_2}\,\,|j_u=3/2>{,}\nonumber\\
M_{D'_1}\,&=&\,\cos\vartheta_1\,|j_u=1/2>\,-\,\sin\vartheta_1\,e^{i\vartheta_2}\,|j_u=3/2>{,}
\end{eqnarray}
where $\vartheta_1$ and $\vartheta_2$ are mixing angles
and similar expressions can be used for $2^-$-states.

The resonant matrix element can be presented as follows:
\begin{align}
M_{R_J}\,&=\,\frac{G_F}{\sqrt{2}}V_{cb}V^*_{ud}\frac{g_{\bar{B}D^*(\omega)R_J} \tilde{g}_{R_J\omega(D^*)\pi}}{D_R(q^2)}\sum_{L_1 L_2}C_{L_1} \tilde{C}_{L_2} F_{L_1}(q^2) \tilde{F}_{L_2}(q^2)\mathcal{P}_{L_1 L_2} A_{L_1 L_2}{.}
\end{align}
Here, $L_1(L_2)$ is the angular orbital momentum in the $\bar{B}^0
(R)$ rest frame; $C_{L_1}, \tilde{C}_{L_2}$ are relative amplitudes
defined above, $\mathcal{P}_{L_1 L_2}$ is the expression for the
momentum dependence; $A_{L_1 L_2}$ is the expression for the angular
dependence. The expressions $\mathcal{P}_{L_1 L_2}$ and $A_{L_1 L_2}$
are combined in Table~\ref{t:LongTable} for different intermediate states.
The notations \(c_{\alpha}=\cos\alpha\) and \(s_{\alpha}=\sin\alpha\)
are used. The functions \(f_{J,J\pm 1}(q^2)\) and \(\tilde{f}_{J,J\pm 1}(q^2)\)
used in Table~\ref{t:LongTable} are defined by (\ref{fom}) and (\ref{tfom})
for the \(\omega\pi\)-resonances and by (\ref{fddst}) and (\ref{tfddst})
for the \(D^{**}\)-resonances.
\newpage
\begin{sidewaystable}[H]
\begin{longtable}[c]{c c c c c}
\hline
& & & & \\
Resonance & $L_1$ & $L_2$ & $\mathcal{P}_{L_1 L_2}$ & $A_{L_1 L_2}$ \\
$\omega\pi$& & & & \\
\hline \hline
& & & & \\
$\rho_0$ & $P$ & $P$ & $\frac{m_B \sqrt{q^2}}{m_{D^*}\sqrt{p^2}} \mathbf{p} \mathbf{Q}$ & $c_{\theta}c_{\beta}$ \\
& & & & \\
& & & & \\
$\rho(1450)$ & $S$ & $P$ & $-i m^2_B \sqrt{q^2} \mathbf{p}$ & $-s_{\theta}s_{\phi}c_{\beta}s_{\xi}+s_{\theta}c_{\phi}s_{\beta}s_{\psi}-s_{\theta}s_{\phi}s_{\beta}c_{\psi}c_{\xi}$ \\
& & & & \\
& $P$ & $P$ & $m_B \sqrt{q^2} \mathbf{p} \mathbf{Q}$ & $s_{\theta}s_{\phi}s_{\beta}s_{\psi}c_{\xi}+s_{\theta}c_{\phi}s_{\beta}c_{\psi}$ \\
& & & &   \\
& $D$ & $P$ & $i \sqrt{q^2} f_{1,2}(q^2) \mathbf{p}$ & $2s_{\theta}s_{\phi}c_{\beta}s_{\xi}+s_{\theta}c_{\phi}s_{\beta}s_{\psi}- s_{\theta}s_{\phi}s_{\beta}c_{\psi}c_{\xi}$ \\
& & & & \\
& & & & \\
$b_1(1235)$ & $S$ & $S$ & $-i m^2_B m^2_R$ & $-c_{\theta}c_{\beta}c_{\xi}+s_{\theta}c_{\phi}c_{\beta}s_{\xi}-s_{\theta}s_{\phi}s_{\beta}s_{\psi}+$ \\
& & & & $+s_{\theta}c_{\phi}s_{\beta}c_{\psi}c_{\xi}+c_{\theta}s_{\beta}c_{\psi}s_{\xi}$ \\
& & & & \\
& $S$ & $D$ & $i m^2_B \tilde{f}_{1,2}(q^2)$ & $2 c_{\theta}c_{\beta}c_{\xi}-2 s_{\theta}c_{\phi}c_{\beta}s_{\xi}-s_{\theta}s_{\phi}s_{\beta}s_{\psi}+$ \\
& & & & $+s_{\theta}c_{\phi}s_{\beta}c_{\psi}c_{\xi}+c_{\theta}s_{\beta}c_{\psi}s_{\xi}$ \\
& & & & \\
& $P$ & $S$ &  $m^2_R m_B \mathbf{Q}$ & $-c_{\theta}s_{\beta}s_{\psi}s_{\xi}+s_{\theta}s_{\phi}s_{\beta}c_{\psi}-s_{\theta}c_{\phi}s_{\beta}s_{\psi}c_{\xi}$ \\
& & & & \\
& $P$ & $D$ &  $-m_B \tilde{f}_{1,2}(q^2) \mathbf{Q}$ & $2 c_{\theta}s_{\beta}s_{\psi}s_{\xi}+s_{\theta}s_{\phi}s_{\beta}c_{\psi}-s_{\theta}c_{\phi}s_{\beta}s_{\psi}c_{\xi}$ \\
& & & & \\
& $D$ & $S$ &  $i m^2_R f_{1,2}(q^2)$ & $2 c_{\theta}c_{\beta}c_{\xi}+s_{\theta}c_{\phi}c_{\beta}s_{\xi}-s_{\theta}s_{\phi}s_{\beta}s_{\psi}+$ \\
& & & & $+s_{\theta}c_{\phi}s_{\beta}c_{\psi}c_{\xi}-2 c_{\theta}s_{\beta}c_{\psi}s_{\xi}$ \\
& & & & \\
& $D$ & $D$ & $-i f_{1,2}(q^2) \tilde{f}_{1,2}(q^2)$ & $-4 c_{\theta}c_{\beta}c_{\xi}-2 s_{\theta}c_{\phi}c_{\beta}s_{\xi}-s_{\theta}s_{\phi}s_{\beta}s_{\psi}+$ \\
& & & & $+s_{\theta}c_{\phi}s_{\beta}c_{\psi}c_{\xi}-2 c_{\theta}s_{\beta}c_{\psi}s_{\xi}$ \\
& & & &  \\
\end{longtable}
\end{sidewaystable}

\newpage
\begin{sidewaystable}[H]
\begin{longtable}[c]{c c c c c}
\hline
& & & & \\
Resonance & $L_1$ & $L_2$  & $\mathcal{P}_{L_1 L_2}$ & $A_{L_1 L_2}$ \\
$\omega\pi$& & & &  \\
\hline \hline
& & & & \\
$b_2$ & $P$ & $D$ &  $-\frac{i}{2} m^3_B \mathbf{p}^2 \mathbf{Q}$ & $s_{\theta}s_{\phi}c_{\beta}s_{2\xi}+s_{\theta}s_{\phi}s_{\beta}c_{\psi}c_{2\xi}-s_{\theta}c_{\phi}s_{\beta}s_{\psi}c_{\xi}$ \\
& & & & \\
& $D$ & $D$ &   $\frac{1}{2}m^2_B \mathbf{p}^2 \mathbf{Q}^2$ & $s_{\theta}s_{\phi}s_{\beta}s_{\psi}+s_{\theta}c_{\phi}s_{\beta}c_{\psi}c_{\xi}$ \\
& & & &  \\
& $F$ & $D$ & $ \frac{i}{2}m_B f_{2,3}(q^2) \mathbf{p}^2 \mathbf{Q}$ &$-3/2 s_{\theta}s_{\phi}c_{\beta}s_{2\xi}+s_{\theta}s_{\phi}s_{\beta}c_{\psi}c_{2\xi}-s_{\theta}c_{\phi}s_{\beta}s_{\psi}c_{\xi}$ \\
& & & & \\
& & & & \\
$\rho_2$ & $P$ & $P$ & $\frac{i}{\sqrt{q^2}} m^3_B m^2_R \mathbf{p} \mathbf{Q}$ & $c_{\theta}c_{\beta}(c^2_{\xi}-1/3)-1/2c_{\theta}s_{\beta}c_{\psi}s_{2\xi}-1/2s_{\theta}c_{\phi}c_{\beta}s_{2\xi}-$ \\
& & & & $-1/2s_{\theta}s_{\phi}s_{\beta}s_{\psi}c_{\xi}-s_{\theta}c_{\phi}s_{\beta}c_{\psi}(c^2_{\xi}-1/2)$ \\
& & & & \\
& $P$ & $F$ & $-\frac{i}{\sqrt{q^2}} m^3_B \tilde{f}_{2,3}(q^2) \mathbf{p} \mathbf{Q}$ & $-3/2 c_{\theta}c_{\beta}(c^2_{\xi}-1/3)+3/4c_{\theta}s_{\beta}c_{\psi}s_{2\xi}-1/2s_{\theta}c_{\phi}c_{\beta}s_{2\xi}-$ \\
& & & & $-1/2s_{\theta}s_{\phi}s_{\beta}s_{\psi}c_{\xi}-s_{\theta}c_{\phi}s_{\beta}c_{\psi}(c^2_{\xi}-1/2)$ \\
& & & & \\
& $D$ & $P$ & $-\frac{m^2_R m^2_B}{2\sqrt{q^2}} \mathbf{p} \mathbf{Q}^2$ & $c_{\theta}s_{\beta}s_{\psi}s_{2\xi}+s_{\theta}c_{\phi}s_{\beta}s_{\psi}c_{2\xi}-s_{\theta}s_{\phi}s_{\beta}c_{\psi}c_{\xi}$ \\
& & & & \\
& $D$ & $F$ & $\frac{m^2_B}{2\sqrt{q^2}} \tilde{f}_{2,3}(q^2) \mathbf{p} \mathbf{Q}^2$ & $-3/2 c_{\theta}s_{\beta}s_{\psi}s_{2\xi}+s_{\theta}c_{\phi}s_{\beta}s_{\psi}c_{2\xi}-s_{\theta}s_{\phi}s_{\beta}c_{\psi}c_{\xi}$ \\
& & & & \\
& $F$ & $P$ & $-\frac{i}{\sqrt{q^2}} m^2_R m_B f_{2,3}(q^2) \mathbf{p} \mathbf{Q}$ & $-3/2 c_{\theta}c_{\beta}(c^2_{\xi}-1/3)-1/2c_{\theta}s_{\beta}c_{\psi}s_{2\xi}+3/4 s_{\theta}c_{\phi}c_{\beta}s_{2\xi}-$ \\
& & & & $-1/2s_{\theta}s_{\phi}s_{\beta}s_{\psi}c_{\xi}-s_{\theta}c_{\phi}s_{\beta}c_{\psi}(c^2_{\xi}-1/2)$ \\
& & & & \\
& $F$ & $F$ & $\frac{i}{\sqrt{q^2}} m_B f_{2,3}(q^2) \tilde{f}_{2,3}(q^2) \mathbf{p} \mathbf{Q}$ & $9/4 c_{\theta}c_{\beta}(c^2_{\xi}-1/3)+3/4 c_{\theta}s_{\beta}c_{\psi}s_{2\xi}+3/4 s_{\theta}c_{\phi}c_{\beta}s_{2\xi}-$ \\
& & & & $-1/2s_{\theta}s_{\phi}s_{\beta}s_{\psi}c_{\xi}-s_{\theta}c_{\phi}s_{\beta}c_{\psi}(c^2_{\xi}-1/2)$ \\
& & & & \\
& & & & \\
$\rho_3$ & $D$ & $F$ & $\frac{i}{\sqrt{q^2}} m^4_B \mathbf{p}^3 \mathbf{Q}^2$ & $1/3(s_{\theta}c_{\phi}s_{\beta}s_{\psi}-s_{\theta}s_{\phi}s_{\beta}c_{\psi}c_{\xi})(c^2_{\xi}-1/5)-$\\
& & & & $-s_{\theta}s_{\phi}c_{\beta}s_{\xi}(c^2_{\xi}-1/5)+2/3 s_{\theta}s_{\phi}s_{\beta}c_{\psi}c_{\xi}s^2_{\xi}$ \\
& & & &  \\
& $F$ & $F$ & $\frac{m^3_B}{3\sqrt{q^2}} \mathbf{p}^3 \mathbf{Q}^3$ & $(s_{\theta}s_{\phi}s_{\beta}s_{\psi}c_{\xi}+s_{\theta}c_{\phi}s_{\beta}c_{\psi})(c^2_{\xi}-1/5)-2 s_{\theta}s_{\phi}s_{\beta}s_{\psi}c_{\xi}s^2_{\xi}$ \\
& & & &  \\
& $G$ & $F$ & $-\frac{i}{\sqrt{q^2}} m^2_B f_{3,4}(q^2) \mathbf{p}^3 \mathbf{Q}^2$ & $1/3(s_{\theta}c_{\phi}s_{\beta}s_{\psi}-s_{\theta}s_{\phi}s_{\beta}c_{\psi}c_{\xi})(c^2_{\xi}-1/5)+$\\
& & & & $+4/3s_{\theta}s_{\phi}c_{\beta}s_{\xi}(c^2_{\xi}-1/5)+2/3 s_{\theta}s_{\phi}s_{\beta}c_{\psi}c_{\xi}s^2_{\xi}$ \\
& & & & \\
\end{longtable}
\end{sidewaystable}

\newpage
\begin{sidewaystable}[H]
\begin{longtable}[c]{c c c c c}
\hline
& & & & \\
Resonance & $L_1$ & $L_2$ & $\mathcal{P}_{L_1 L_2}$ & $A_{L_1 L_2}$ \\
$D^{**}$& & & & \\
\hline \hline
& & & & \\
$1^+_{1/2}$ & $S$ & $S$ &-$i m^2_B m^2_R$ & $-c_{\theta}c_{\beta}c_{\xi}+s_{\theta}c_{\phi}c_{\beta}s_{\xi}-s_{\theta}s_{\phi}s_{\beta}s_{\psi}+$  \\
& & & & $+s_{\theta}c_{\phi}s_{\beta}c_{\psi}c_{\xi}+c_{\theta}s_{\beta}c_{\psi}s_{\xi}$  \\
& & & & \\
& $P$ & $S$ & $m^2_R m_B \mathbf{p}$ & $-s_{\theta}s_{\phi}c_{\beta}s_{\xi}-s_{\theta}s_{\phi}s_{\beta}c_{\psi}c_{\xi}+s_{\theta}c_{\phi}s_{\beta}s_{\psi}$  \\
& & & & \\
& $D$ & $S$ & $i m^2_R f_{1,2}(q^2)$ & $2 c_{\theta}c_{\beta}c_{\xi}+s_{\theta}c_{\phi}c_{\beta}s_{\xi}-s_{\theta}s_{\phi}s_{\beta}s_{\psi}+$  \\
& & & & $+s_{\theta}c_{\phi}s_{\beta}c_{\psi}c_{\xi}-2 c_{\theta}s_{\beta}c_{\psi}s_{\xi}$  \\
& & & & \\
$1^+_{3/2}$ & $S$ & $D$ & $i m^2_B \tilde{f}_{1,2}(q^2)$ & $2 c_{\theta}c_{\beta}c_{\xi}-2 s_{\theta}c_{\phi}c_{\beta}s_{\xi}-s_{\theta}s_{\phi}s_{\beta}s_{\psi}+$ \\
& & & & $+s_{\theta}c_{\phi}s_{\beta}c_{\psi}c_{\xi}+c_{\theta}s_{\beta}c_{\psi}s_{\xi}$ \\
& & & & \\
& $P$ & $D$ & $-m_B \tilde{f}_{1,2}(q^2) \mathbf{p}$ & $2 s_{\theta}s_{\phi}c_{\beta}s_{\xi}-s_{\theta}s_{\phi}s_{\beta}c_{\psi}c_{\xi}+s_{\theta}c_{\phi}s_{\beta}s_{\psi}$  \\
& & & & \\
& $D$ & $D$ & $-i f_{1,2}(q^2) \tilde{f}_{1,2}(q^2)$ & $-4 c_{\theta}c_{\beta}c_{\xi}-2 s_{\theta}c_{\phi}c_{\beta}s_{\xi}-s_{\theta}s_{\phi}s_{\beta}s_{\psi}+$ \\
& & & & $+s_{\theta}c_{\phi}s_{\beta}c_{\psi}c_{\xi}-2 c_{\theta}s_{\beta}c_{\psi}s_{\xi}$ \\
& & & & \\
& & & & \\
$2^+_{3/2}$ & $P$ & $D$ & $-\frac{i}{2} m^3_B \mathbf{Q}^2 \mathbf{p}$ & $c_{\theta}s_{\beta}s_{\psi}s_{2\xi}+s_{\theta}c_{\phi}s_{\beta}s_{\psi}c_{2\xi}-s_{\theta}s_{\phi}s_{\beta}c_{\psi}c_{\xi}$ \\
& & & & \\
& $D$ & $D$ & $\frac{1}{2} m^2_B \mathbf{Q}^2 \mathbf{p}^2$ & $s_{\theta}s_{\phi}s_{\beta}s_{\psi}+s_{\theta}c_{\phi}s_{\beta}c_{\psi}c_{\xi}$ \\
& & & & \\
& $F$ & $D$ & $\frac{i}{2} m_B f_{2,3}(q^2) \mathbf{Q}^2 \mathbf{p} $ & $-3/2 c_{\theta}s_{\beta}s_{\psi}s_{2\xi}+s_{\theta}c_{\phi}s_{\beta}s_{\psi}c_{2\xi}-s_{\theta}s_{\phi}s_{\beta}c_{\psi}c_{\xi}$ \\
& & & & \\
& & & & \\
$1^-_{3/2}$ & $S$ & $P$ & $-i m^2_B \sqrt{q^2} \mathbf{Q}$ & $-c_{\theta}s_{\beta}s_{\psi}s_{\xi}+s_{\theta}s_{\phi}s_{\beta}c_{\psi}-s_{\theta}c_{\phi}s_{\beta}s_{\psi}c_{\xi}$ \\
& & & &  \\
& $P$ & $P$ & $m_B \sqrt{q^2} \mathbf{Q} \mathbf{p}$ & $s_{\theta}s_{\phi}s_{\beta}s_{\psi}c_{\xi}+s_{\theta}c_{\phi}s_{\beta}c_{\psi}$ \\
& & & & \\
& $D$ & $P$ & $i \sqrt{q^2} f_{1,2}(q^2) \mathbf{Q}$ & $2 c_{\theta}s_{\beta}s_{\psi}s_{\xi}+s_{\theta}s_{\phi}s_{\beta}c_{\psi}-s_{\theta}c_{\phi}s_{\beta}s_{\psi}c_{\xi}$ \\
& & & & \\
& & & & \\
\end{longtable}
\end{sidewaystable}

\setcounter{table}{0}
\newpage
\begin{sidewaystable}[H]
\begin{longtable}[c]{c c c c c}
\hline
& & & & \\
Resonance & $L_1$ & $L_2$ & $\mathcal{P}_{L_1 L_2}$ & $A_{L_1 L_2}$ \\
$D^{**}$& & & & \\
\hline \hline
& & & & \\
$2^-_{3/2}$ & $P$ & $P$ & $\frac{i}{\sqrt{q^2}} m^3_B m^2_R \mathbf{Q} \mathbf{p}$ & $c_{\theta}c_{\beta}(c^2_{\xi}-1/3)-1/2c_{\theta}s_{\beta}c_{\psi}s_{2\xi}-1/2s_{\theta}c_{\phi}c_{\beta}s_{2\xi}-$ \\
& & & & $-1/2s_{\theta}s_{\phi}s_{\beta}s_{\psi}c_{\xi}-s_{\theta}c_{\phi}s_{\beta}c_{\psi}(c^2_{\xi}-1/2)$ \\
& & & & \\
& $D$ & $P$ & $-\frac{m^2_R m^2_B}{2\sqrt{q^2}} \mathbf{Q} \mathbf{p}^2$ & $s_{\theta}s_{\phi}c_{\beta}s_{2\xi}+s_{\theta}s_{\phi}s_{\beta}c_{\psi}c_{2\xi}-s_{\theta}c_{\phi}s_{\beta}s_{\psi}c_{\xi}$ \\
& & & & \\
& $F$ & $P$ &  $-\frac{i}{\sqrt{q^2}} m^2_R m_B f_{2,3}(q^2) \mathbf{Q} \mathbf{p}$ & $-3/2 c_{\theta}c_{\beta}(c^2_{\xi}-1/3)-1/2c_{\theta}s_{\beta}c_{\psi}s_{2\xi}+3/4 s_{\theta}c_{\phi}c_{\beta}s_{2\xi}-$ \\
& & & & $-1/2s_{\theta}s_{\phi}s_{\beta}s_{\psi}c_{\xi}-s_{\theta}c_{\phi}s_{\beta}c_{\psi}(c^2_{\xi}-1/2)$  \\
& & & & \\
$2^-_{5/2}$ & $P$ & $F$ & $-\frac{i}{\sqrt{q^2}} m^3_B \tilde{f}_{2,3}(q^2) \mathbf{Q} \mathbf{p}$ & $-3/2 c_{\theta}c_{\beta}(c^2_{\xi}-1/3)+3/4c_{\theta}s_{\beta}c_{\psi}s_{2\xi}-1/2s_{\theta}c_{\phi}c_{\beta}s_{2\xi}-$ \\
& & & & $-1/2s_{\theta}s_{\phi}s_{\beta}s_{\psi}c_{\xi}-s_{\theta}c_{\phi}s_{\beta}c_{\psi}(c^2_{\xi}-1/2)$ \\
& & & & \\
& $D$ & $F$ & $\frac{m^2_B}{2\sqrt{q^2}} \tilde{f}_{2,3}(q^2) \mathbf{Q} \mathbf{p}^2$ & $-3/2 s_{\theta}s_{\phi}c_{\beta}s_{2\xi}+s_{\theta}s_{\phi}s_{\beta}c_{\psi}c_{2\xi}-s_{\theta}c_{\phi}s_{\beta}s_{\psi}c_{\xi}$ \\
& & & & \\
& $F$ & $F$ & $\frac{i}{\sqrt{q^2}} m_B f_{2,3}(q^2) \tilde{f}_{2,3}(q^2) \mathbf{Q} \mathbf{p}$ & $9/4 c_{\theta}c_{\beta}(c^2_{\xi}-1/3)+3/4 c_{\theta}s_{\beta}c_{\psi}s_{2\xi}+3/4 s_{\theta}c_{\phi}c_{\beta}s_{2\xi}-$ \\
& & & & $-1/2s_{\theta}s_{\phi}s_{\beta}s_{\psi}c_{\xi}-s_{\theta}c_{\phi}s_{\beta}c_{\psi}(c^2_{\xi}-1/2)$ \\
& & & & \\
& & & & \\
$3^-_{5/2}$ & $D$ & $F$ &  $\frac{i}{\sqrt{q^2}} m^4_B \mathbf{Q}^3  \mathbf{p}^2$ & $s_{\theta}c_{\phi}s_{\beta}s_{\psi}c_{\xi}s^2_{\xi}-4/15 s_{\theta}c_{\phi}s_{\beta}s_{\psi}c_{\xi}+$ \\
& & & & $+1/3 s_{\theta}s_{\phi}s_{\beta}c_{\psi}(c^2_{\xi}-1/5)-c_{\theta}s_{\beta}s_{\psi}s_{\xi}(c^2_{\xi}-1/5)$ \\
& & & &  \\
& $F$ & $F$ & $\frac{m^3_B}{3\sqrt{q^2}} \mathbf{Q}^3 \mathbf{p}^3$ & $(s_{\theta}s_{\phi}s_{\beta}s_{\psi}c_{\xi}+s_{\theta}c_{\phi}s_{\beta}c_{\psi})(c^2_{\xi}-1/5)-2 s_{\theta}s_{\phi}s_{\beta}s_{\psi}c_{\xi}s^2_{\xi}$ \\
& & & & \\
& $G$ & $F$ & $-\frac{i}{\sqrt{q^2}} m^2_B f_{3,4}(q^2) \mathbf{Q}^3 \mathbf{p}^2 $ & $s_{\theta}c_{\phi}s_{\beta}s_{\psi}c_{\xi}s^2_{\xi}-4/15 s_{\theta}c_{\phi}s_{\beta}s_{\psi}c_{\xi}+$ \\
& & & & $+1/3 s_{\theta}s_{\phi}s_{\beta}c_{\psi}(c^2_{\xi}-1/5)+4/3 c_{\theta}s_{\beta}s_{\psi}s_{\xi}(c^2_{\xi}-1/5)$ \\
& & & & \\
\caption{Summary of momentum and angular distributions
for different intermediate states, which are described in this
paper.}
\label{t:LongTable}\\
\end{longtable}
\end{sidewaystable}

\section{Decay chain simulation}

To demonstrate the angular distributions for each intermediate
resonance in the \(D^*\omega\pi\) final state, we generate
$2\times10^6$ $\bar{B}^0 \to D^{*+}\omega\pi^-$ events according to
the phase space distribution using the qq98 program package~\cite{qq98}.
For a further study we fill profile angular spectra with the appropriate
weight density functions for each resonant hypothesis, which have been
obtained above.

A description of each vertex includes transition form factors. Since it is
not yet possible to obtain these form factors from rigorous theoretical
calculations, we rely on the simple phenomenological Blatt-Weisskopf
model~\cite{blwe,hippel}.
For $L>0$ this simple form factor suppresses growth of the matrix element
with final particle momentum.
The Blatt-Weisskopf functions $B_L(x)$
are chosen as follows:
\begin{equation}
B_L(x)\,=\,\frac{x_0^{L+1} |h_L(x_0)|}{x^{L+1} |h_L(x)|}{,}
\end{equation}
where
\begin{equation}
h_L(x)\,=\,\frac{-i}{x}e^{i\left(x-\frac{\pi L}{2}\right)}\sum^L_{n=0}(-1)^n\frac{(L+n)!}{n!(L-n)!}(2ix)^{-n}
\end{equation}
is a spherical Hankel function, $x=\mathbf{k}r$, $x_0=\mathbf{k}_0 r$,
$\mathbf{k},\,\mathbf{k}_0$ are the magnitudes of the daughter particle
three-momentum in the mother particle rest frame for the case when
the resonance four-momentum squared is equal to $q^2$ and $m^2_R$,
respectively, and $r=1.6\,\mathrm{GeV}^{-1}$ is a hadron scale.
According to our normalization, these functions are equal to one, when
$\sqrt{q^2}=m_R$. Another common normalization gives
\(B_L(x)=1\) for \(x=1\). The Blatt-Weisskopf functions corresponding
to $L$ discussed here are given below for convenience:
\begin{align}
B_0(x)\,&=\,1{,} \nonumber  \\
B_1(x)\,&=\,\sqrt{\frac{1+x^2_0}{1+x^2}}{,} \nonumber \\
B_2(x)\,&=\,\sqrt{\frac{(x^2_0-3)^2+9 x^2_0}{(x^2-3)^2+9 x^2}}{,} \nonumber \\
B_3(x)\,&=\,\sqrt{\frac{x^2_0(x^2_0-15)^2+9(2 x^2_0-5)^2}{x^2 (x^2-15)^2+9(2 x^2-5)^2}}{,} \nonumber \\
B_4(x)\,&=\,\sqrt{\frac{(x^4_0-45 x^2_0+105)^2+25 x^2_0 (2 x^2_0 -21)^2}{(x^4-45 x^2+105)^2+25 x^2 (2 x^2 -21)^2}} {.}
\end{align}

\begin{longtable}{c c c}
\includegraphics[scale=0.23]{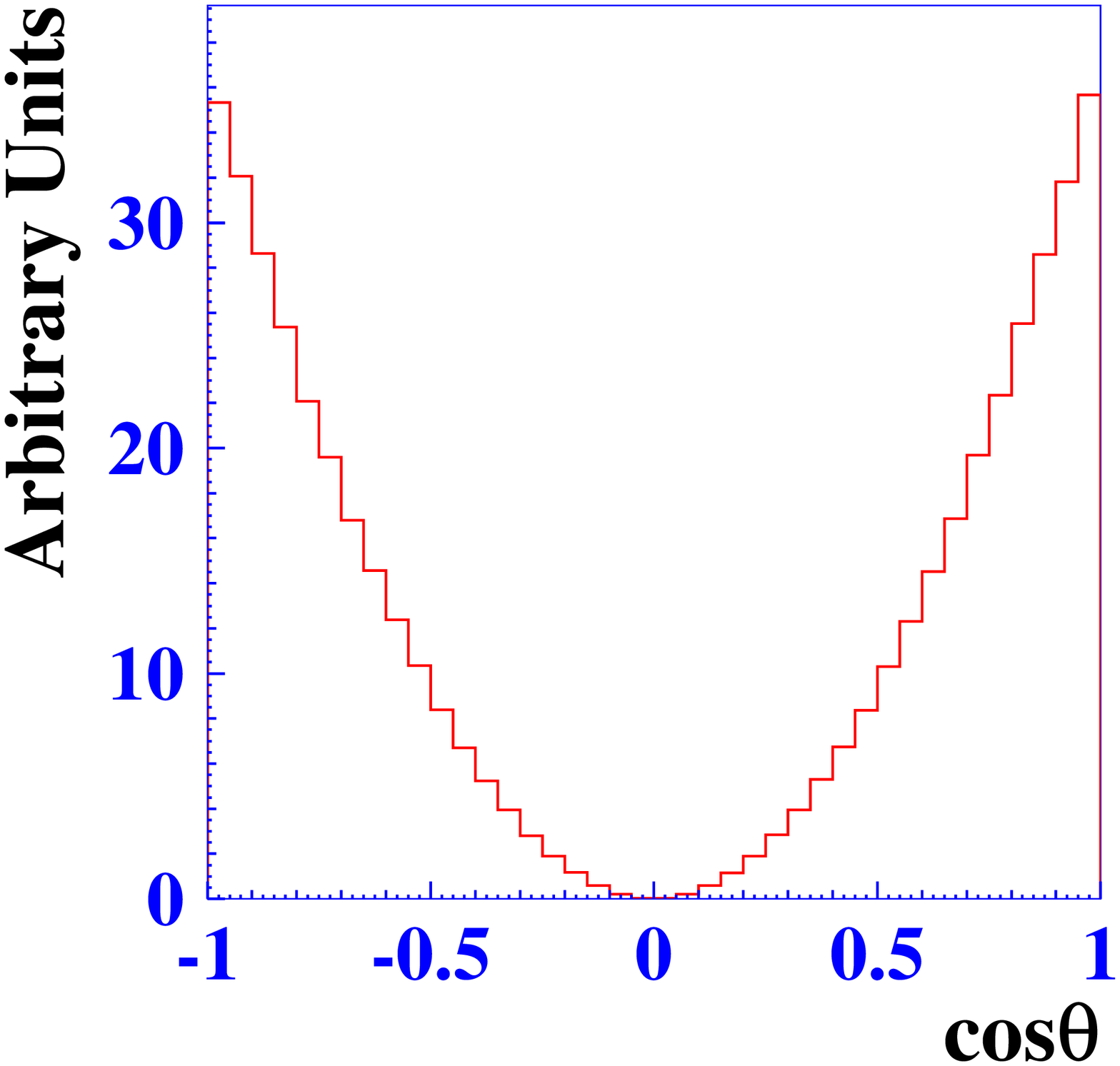} &
\includegraphics[scale=0.23]{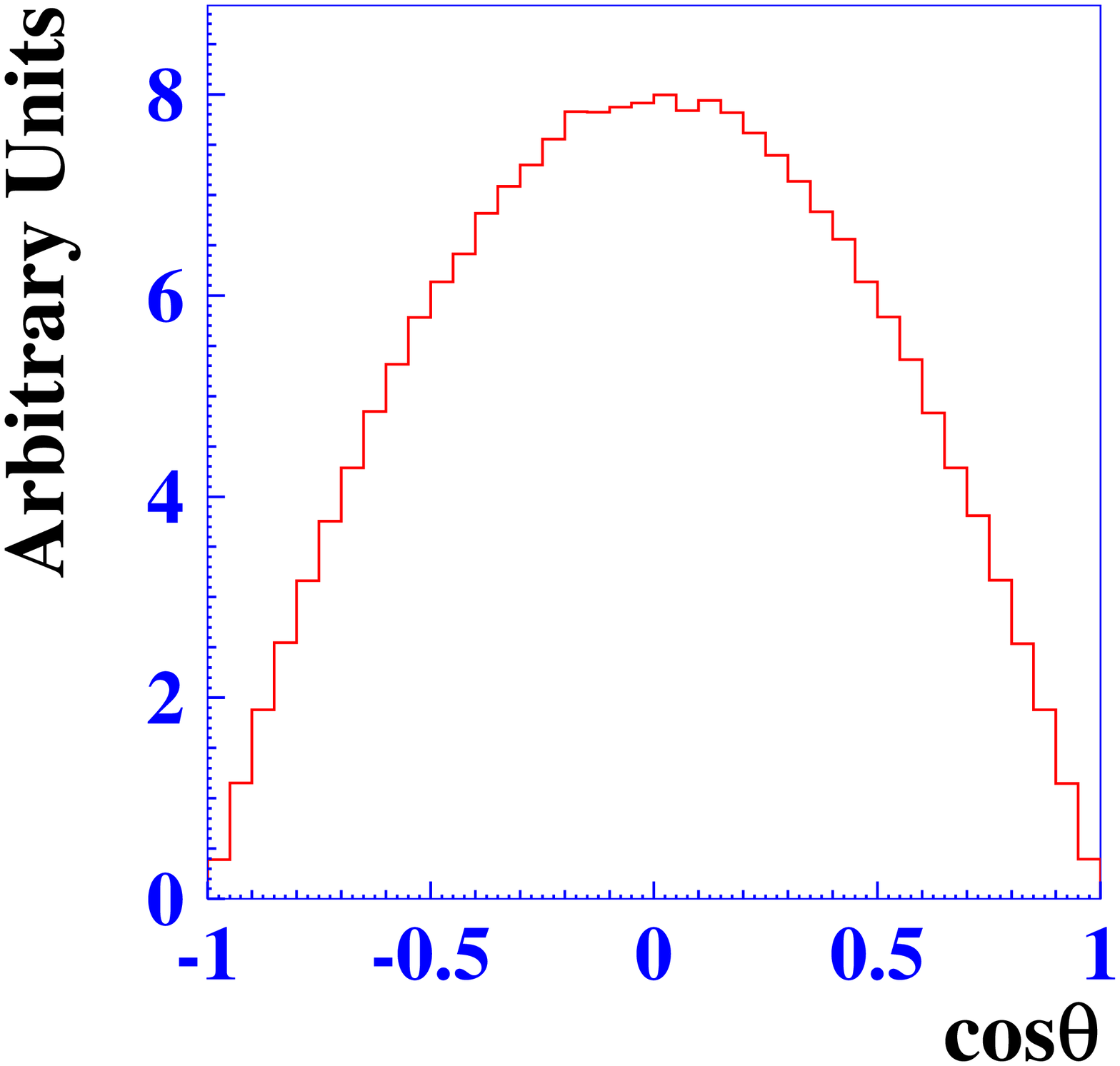}  &
\includegraphics[scale=0.23]{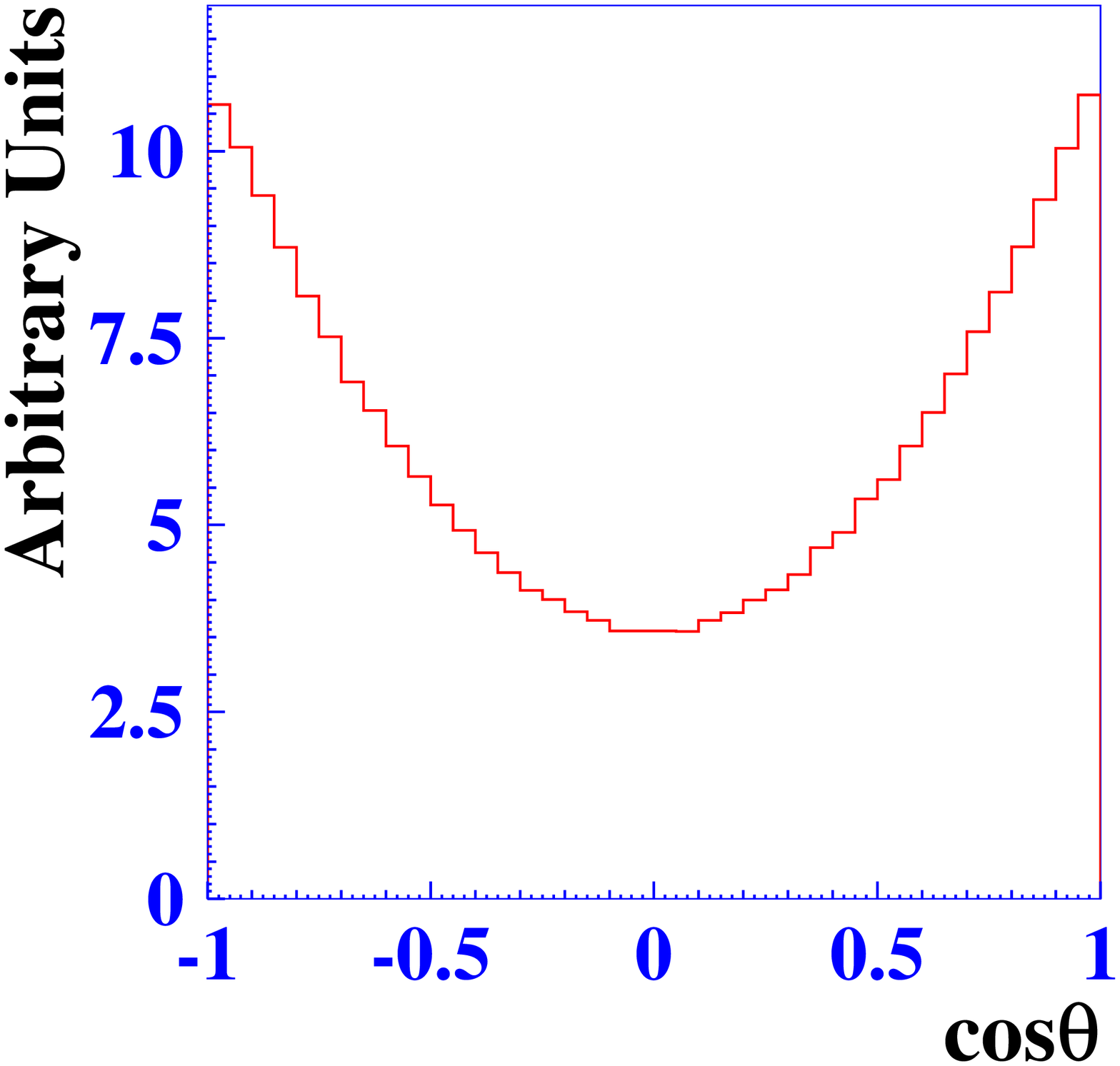} \\
a1) & a2)  & a3)  \\
\includegraphics[scale=0.23]{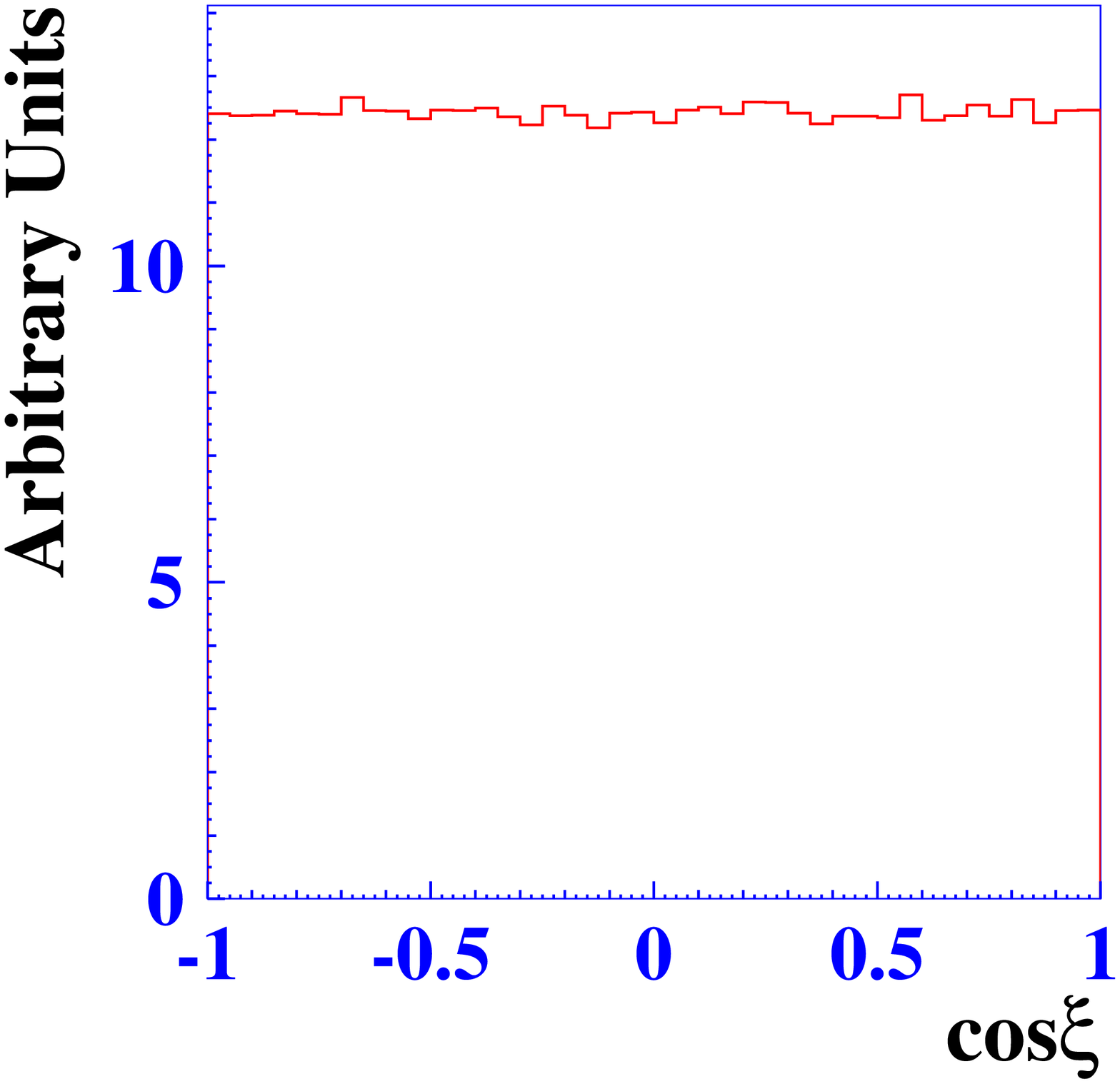} &
\includegraphics[scale=0.23]{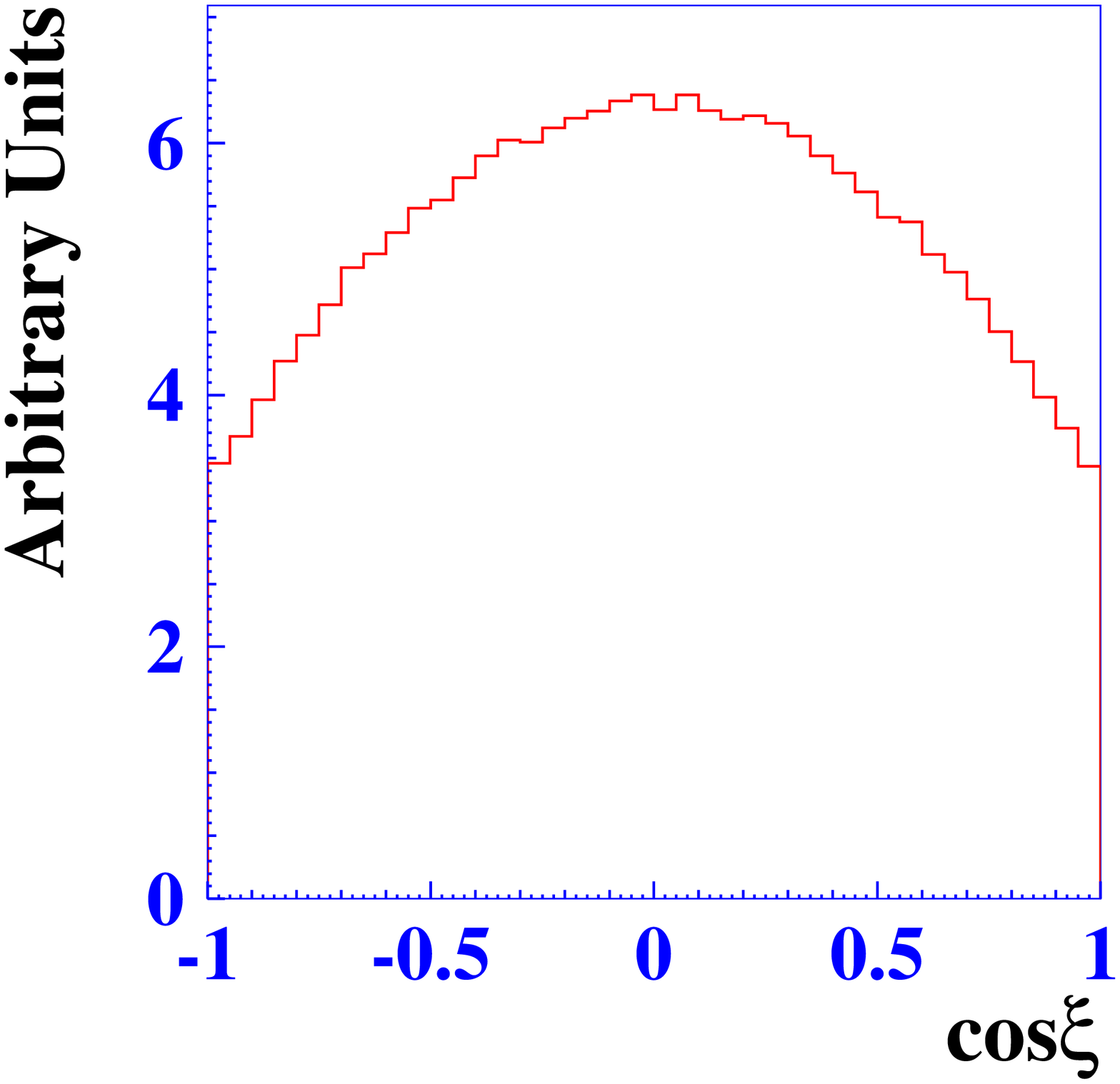}  &
\includegraphics[scale=0.23]{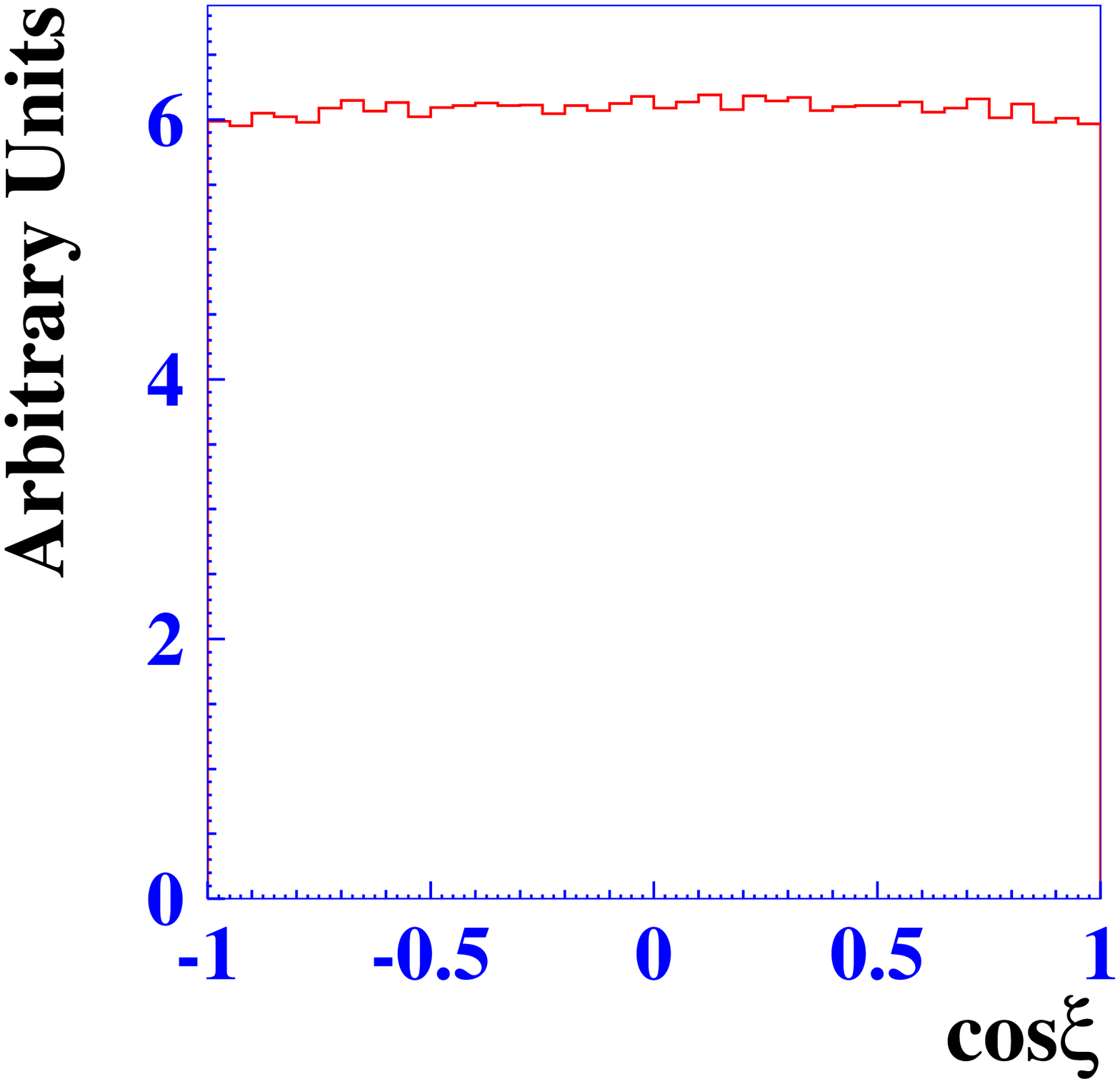}  \\
b1) & b2)  & b3)  \\
\includegraphics[scale=0.23]{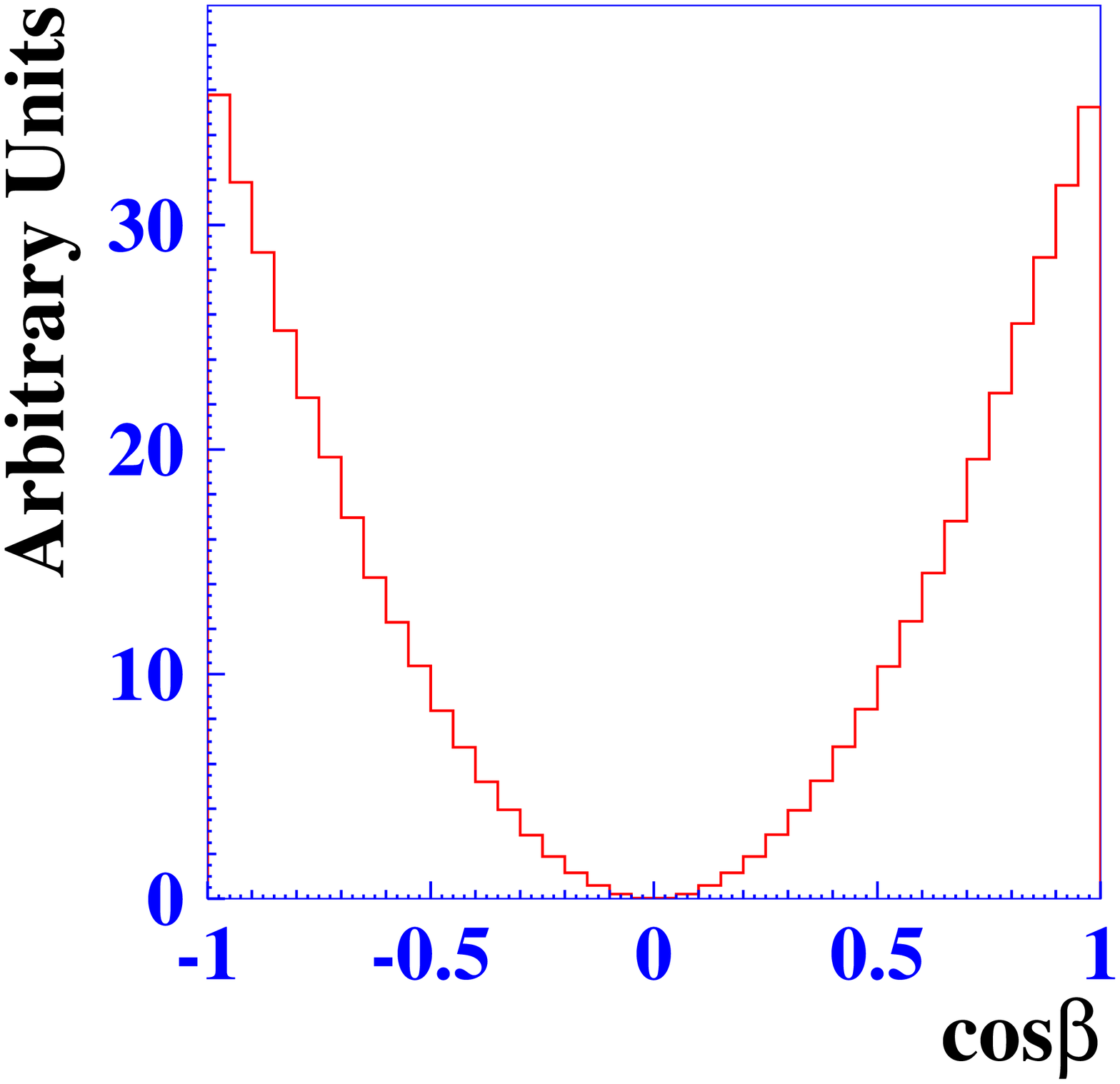} &
\includegraphics[scale=0.23]{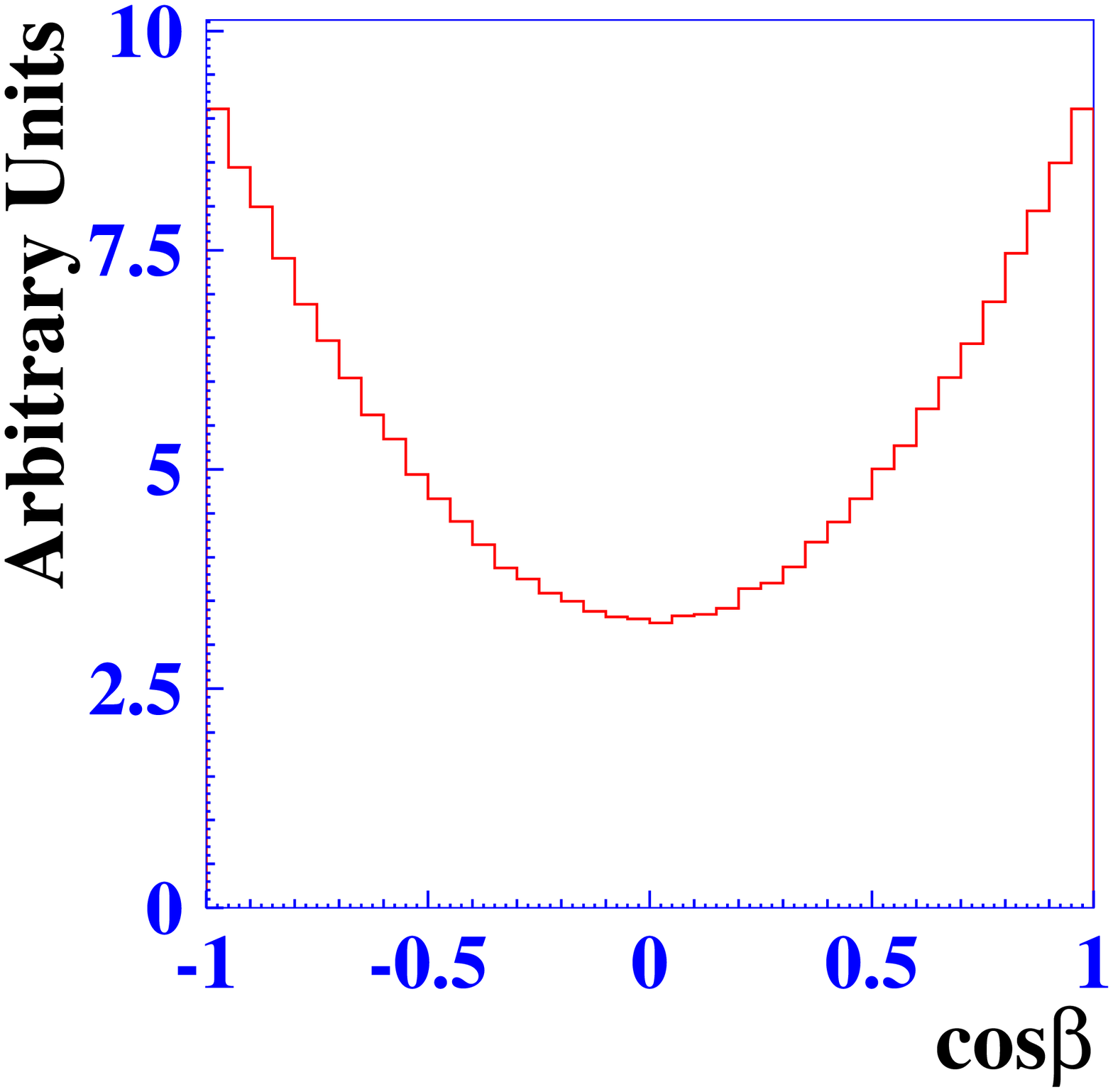}  &
\includegraphics[scale=0.23]{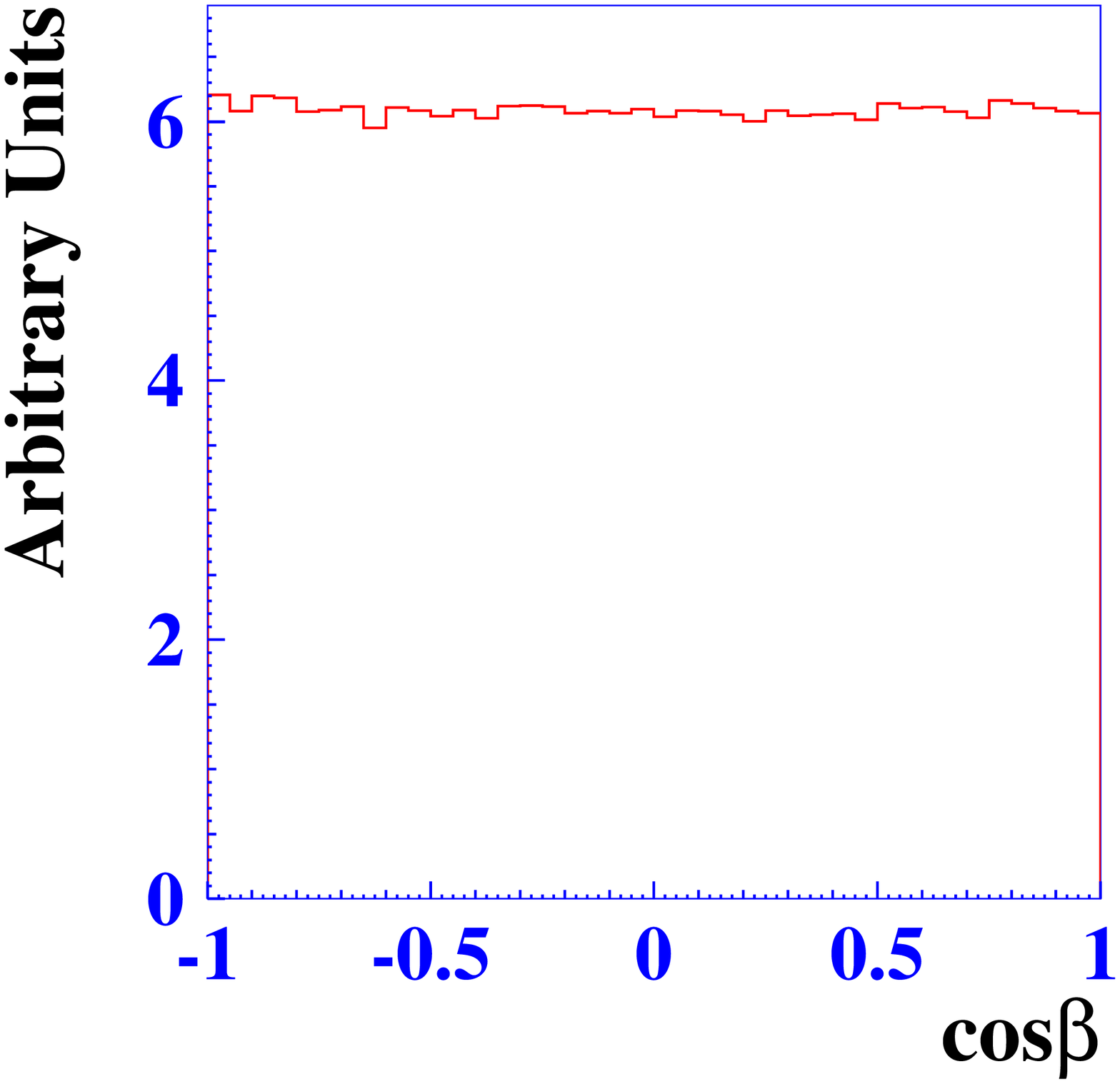}  \\\
c1) & c2) & c3)  \\
\includegraphics[scale=0.23]{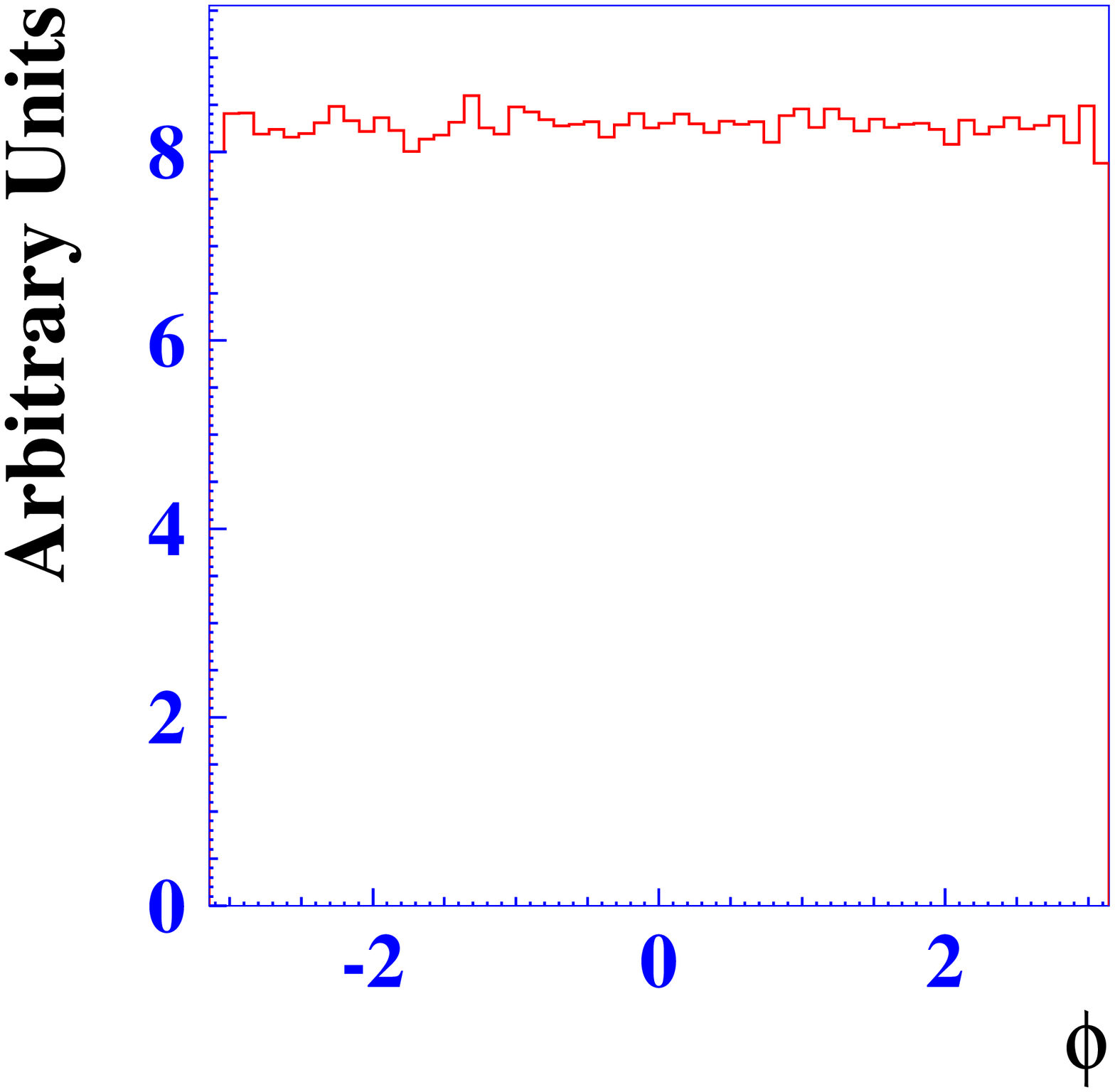} &
\includegraphics[scale=0.23]{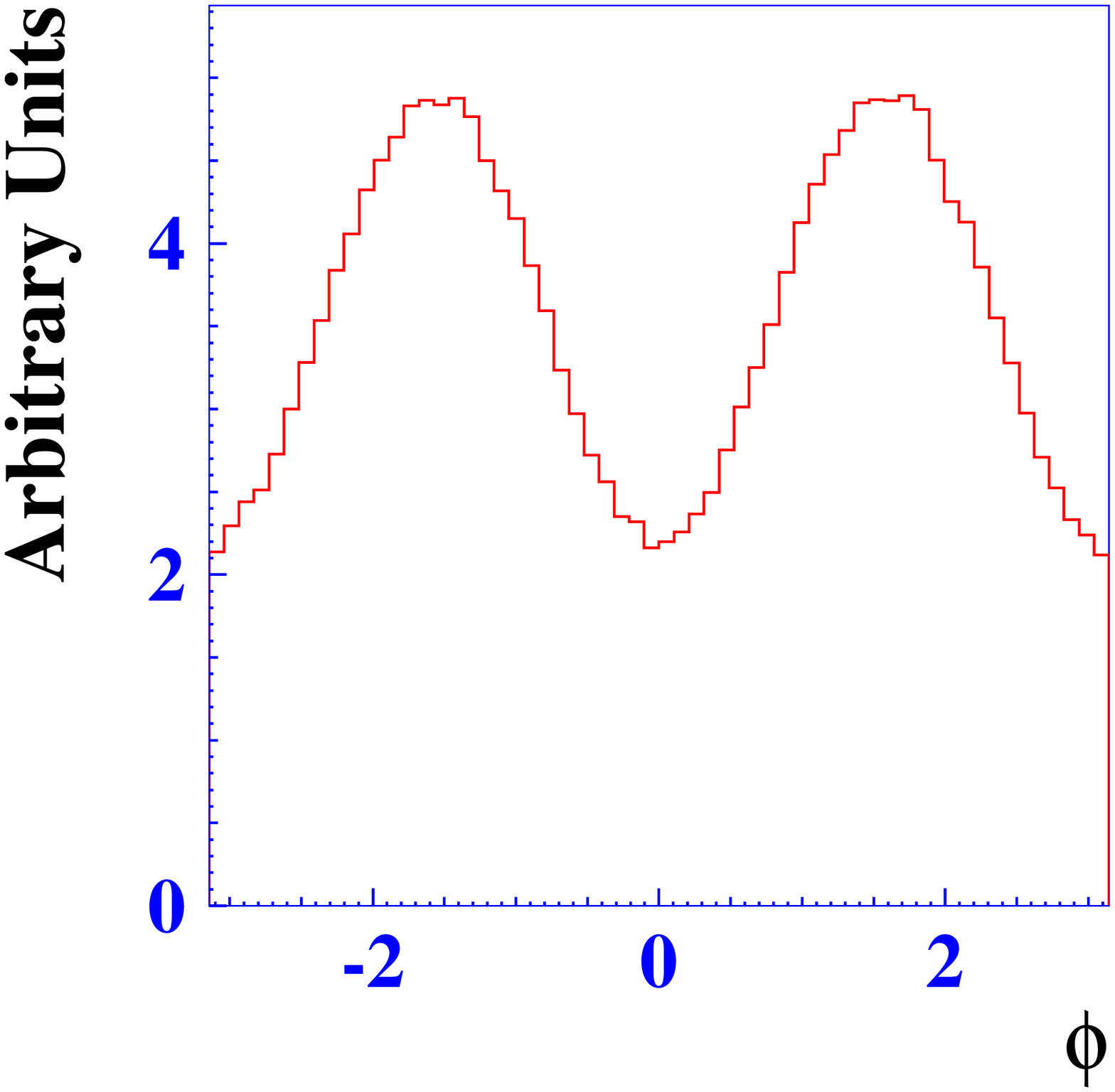}  &
\includegraphics[scale=0.23]{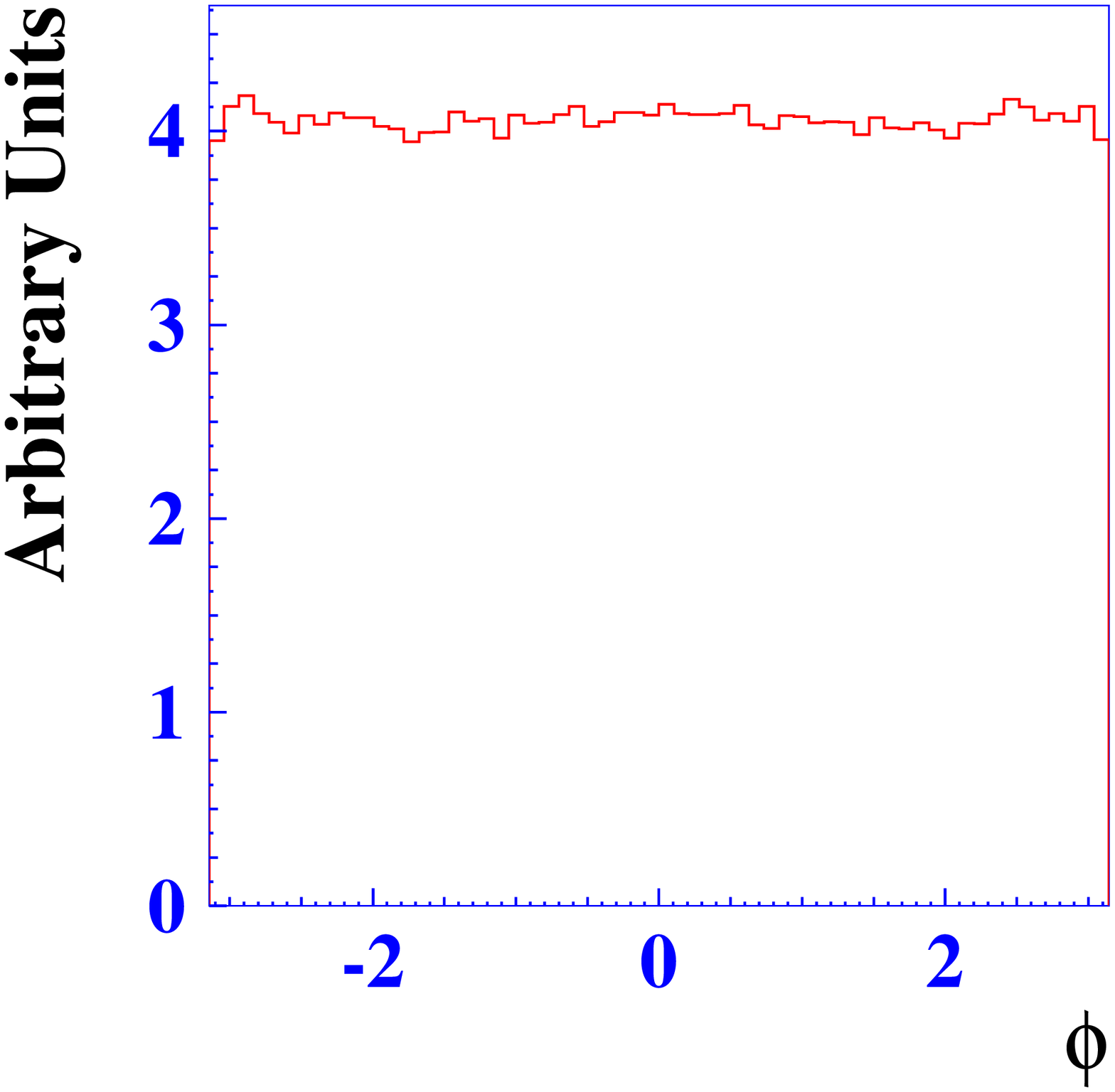} \\
d1) & d2) & d3) \\
\end{longtable}
\begin{figure}[h]
\begin{longtable}{c c c}
\includegraphics[scale=0.23]{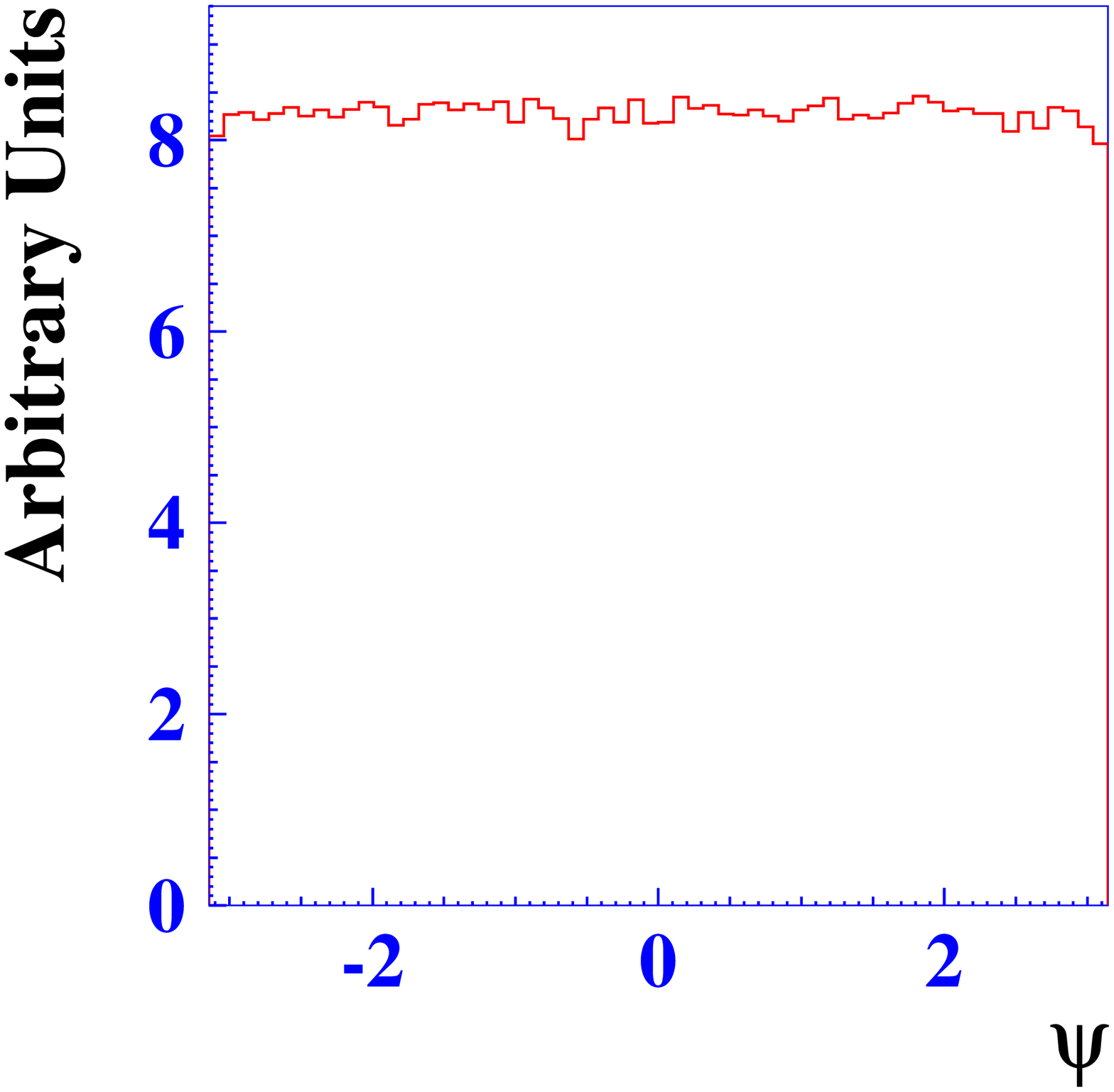} &
\includegraphics[scale=0.23]{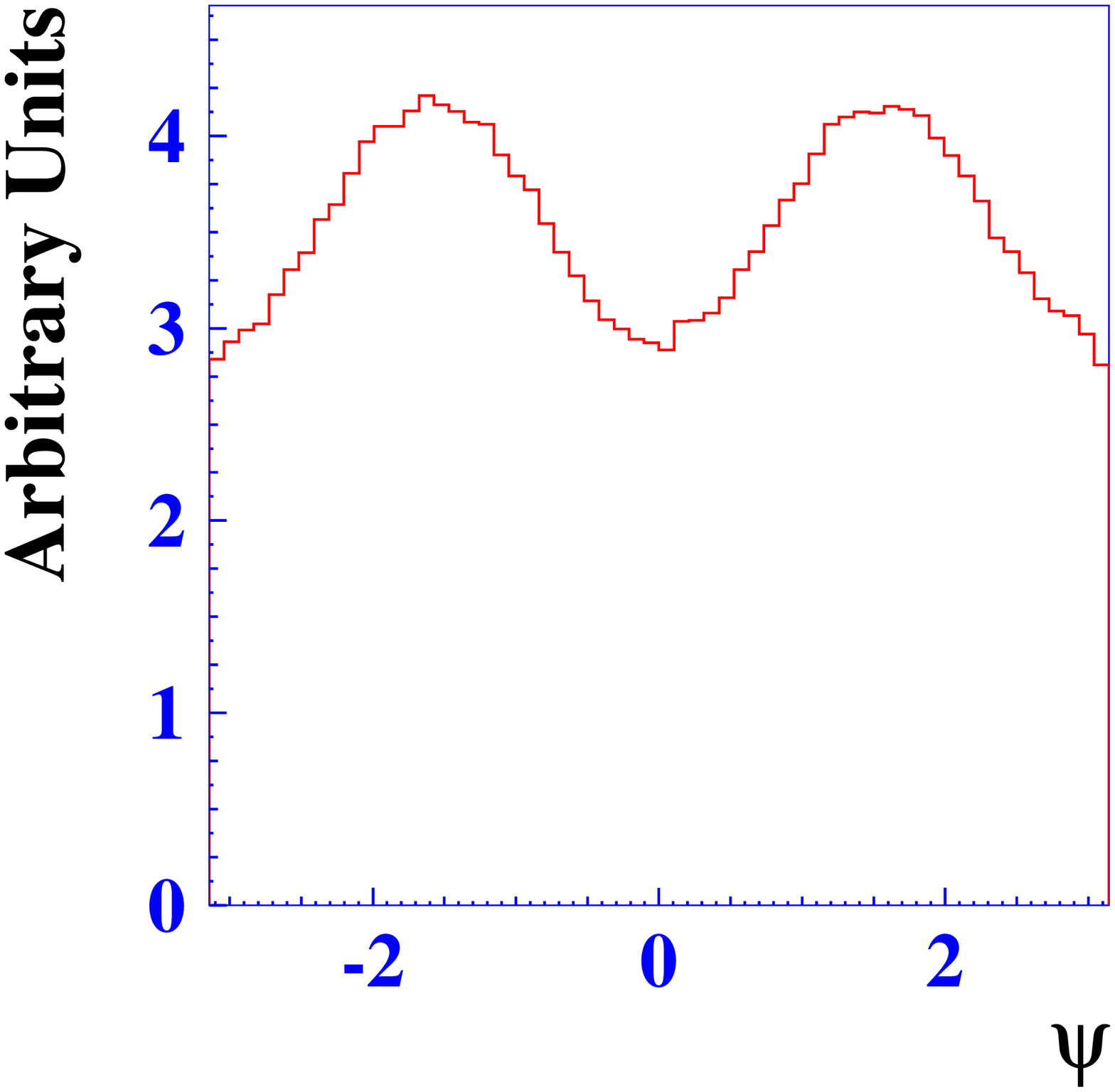}  &
\includegraphics[scale=0.23]{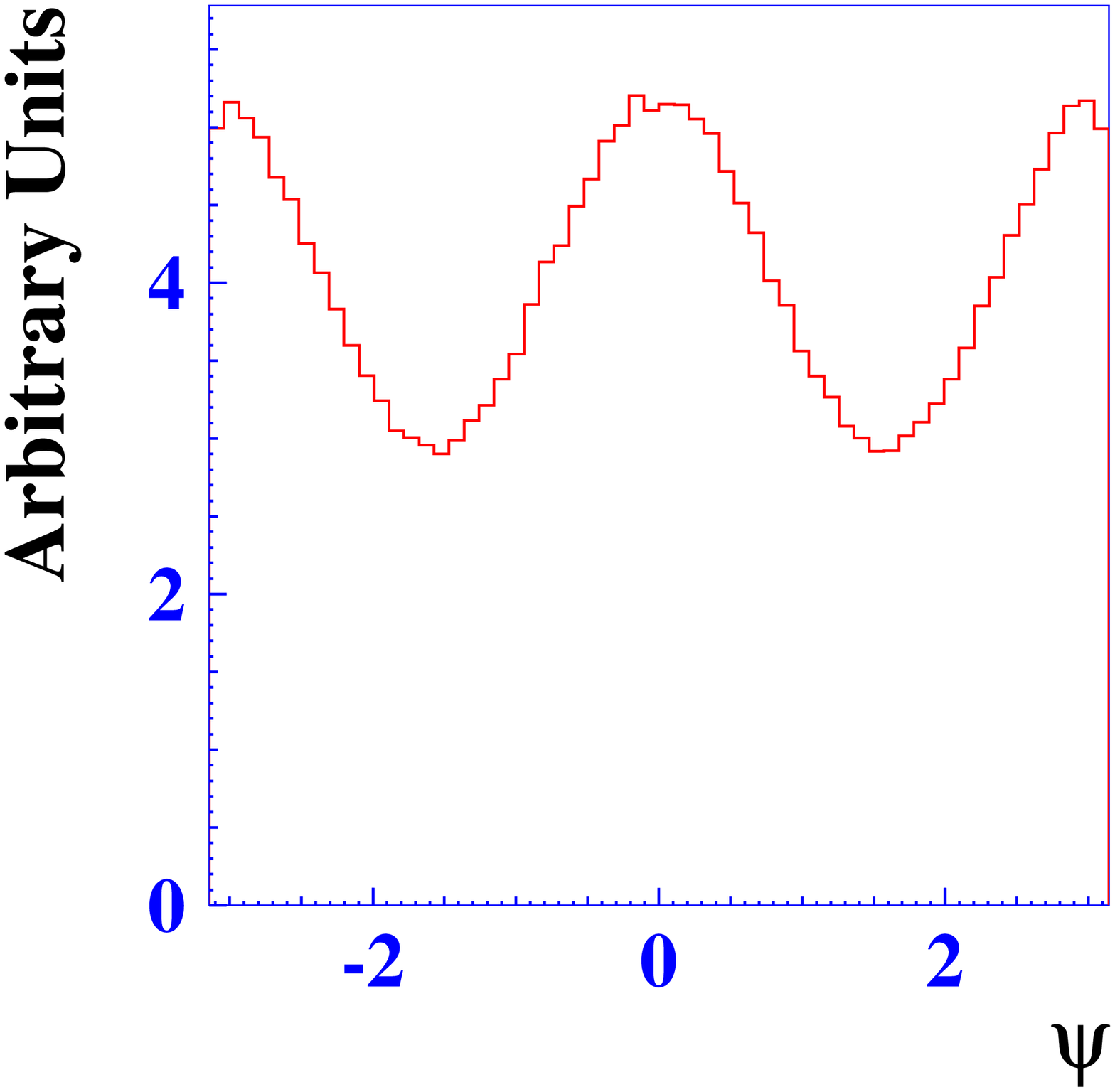} \\
e1) & e2) & e3)\\
\end{longtable}
\caption{Simulated angular distributions
for the $\omega \pi$-resonances. The figures a1), b1), c1), d1), e1)
correspond to the $J^P=0^-$ ($\rho_0^-$) intermediate state;
a2), b2), c2), d2), e2) ---  $J^P=1^-$ ($\rho(1450)^-$)-state;
a3), b3), c3), d3), e3) --- $J^P=1^+$ ($b_1(1235)^-$)-state.}
\label{fig5}
\end{figure}

\begin{longtable}{c c c}
\includegraphics[scale=0.23]{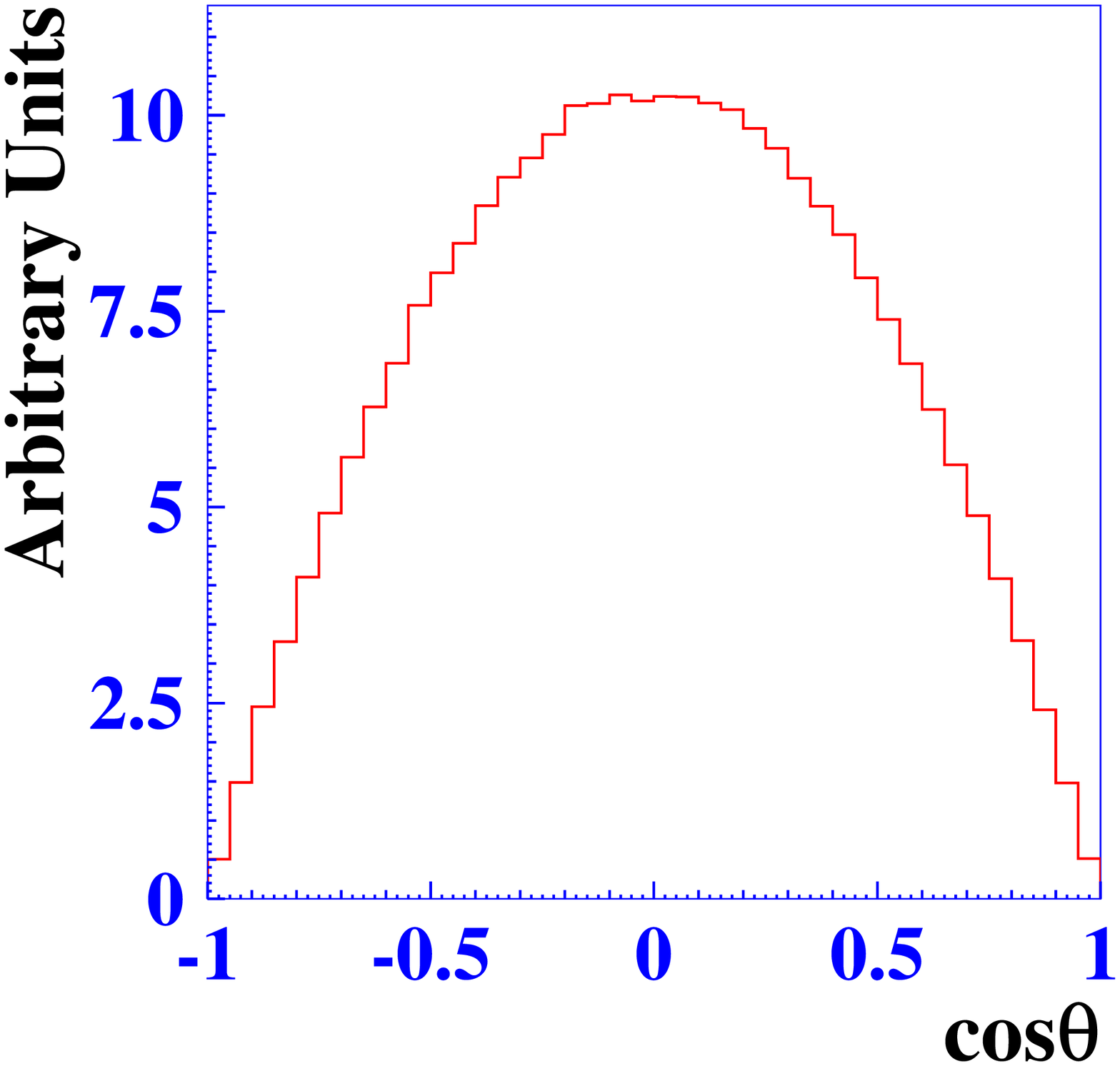}  &
\includegraphics[scale=0.23]{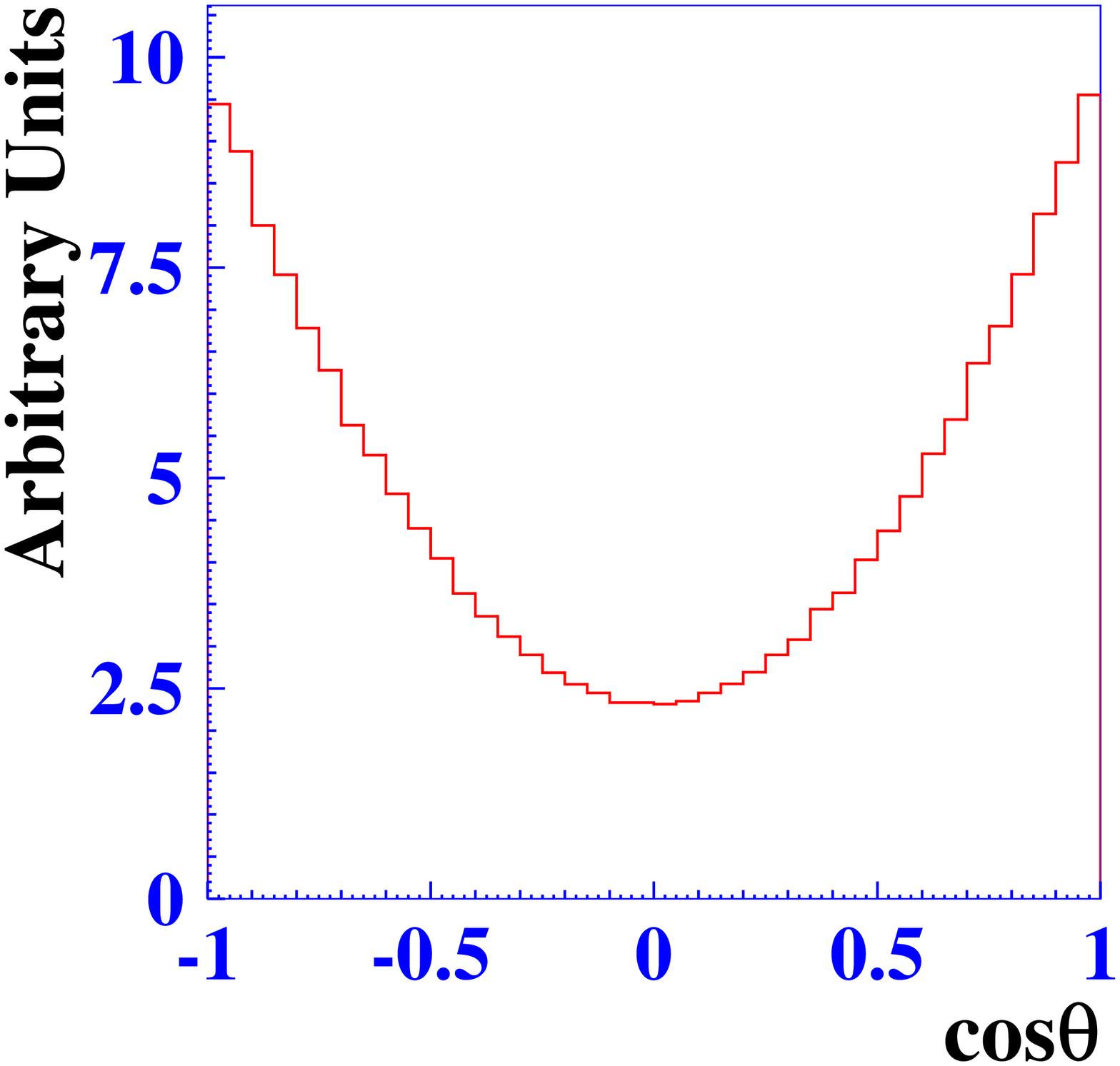}  &
\includegraphics[scale=0.23]{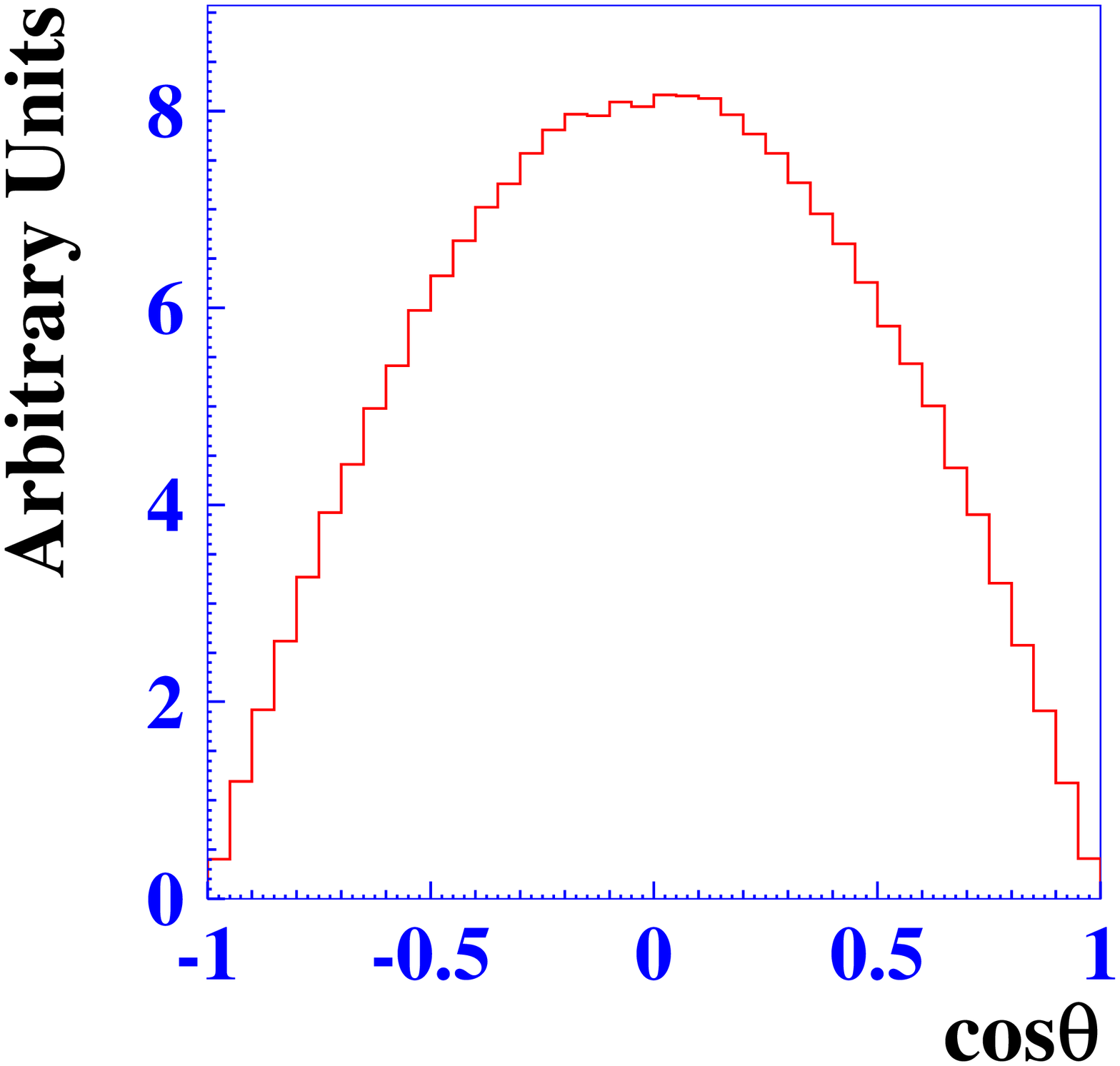} \\
a1) & a2)  & a3)  \\
\includegraphics[scale=0.23]{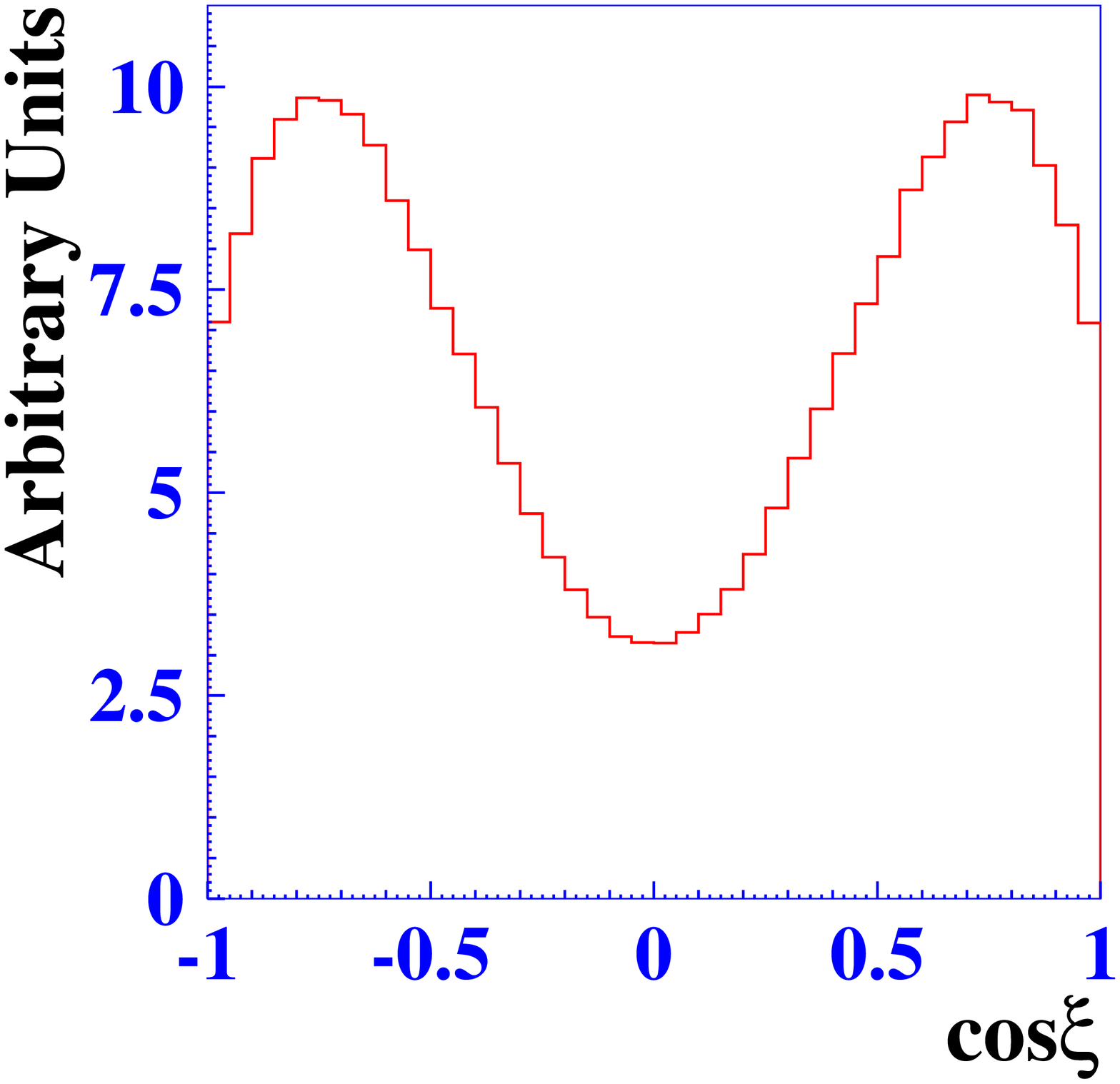}  &
\includegraphics[scale=0.23]{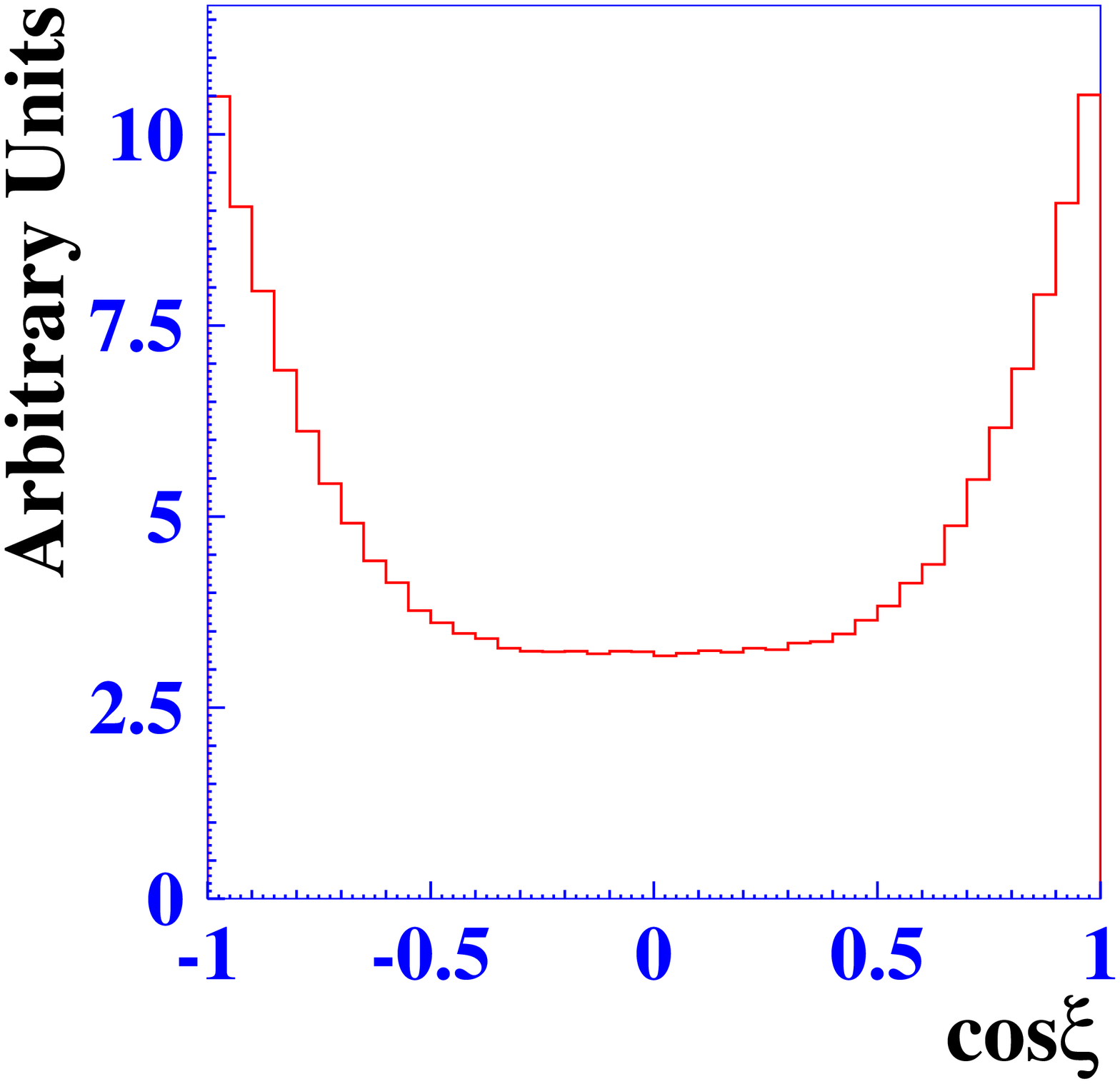}  &
\includegraphics[scale=0.23]{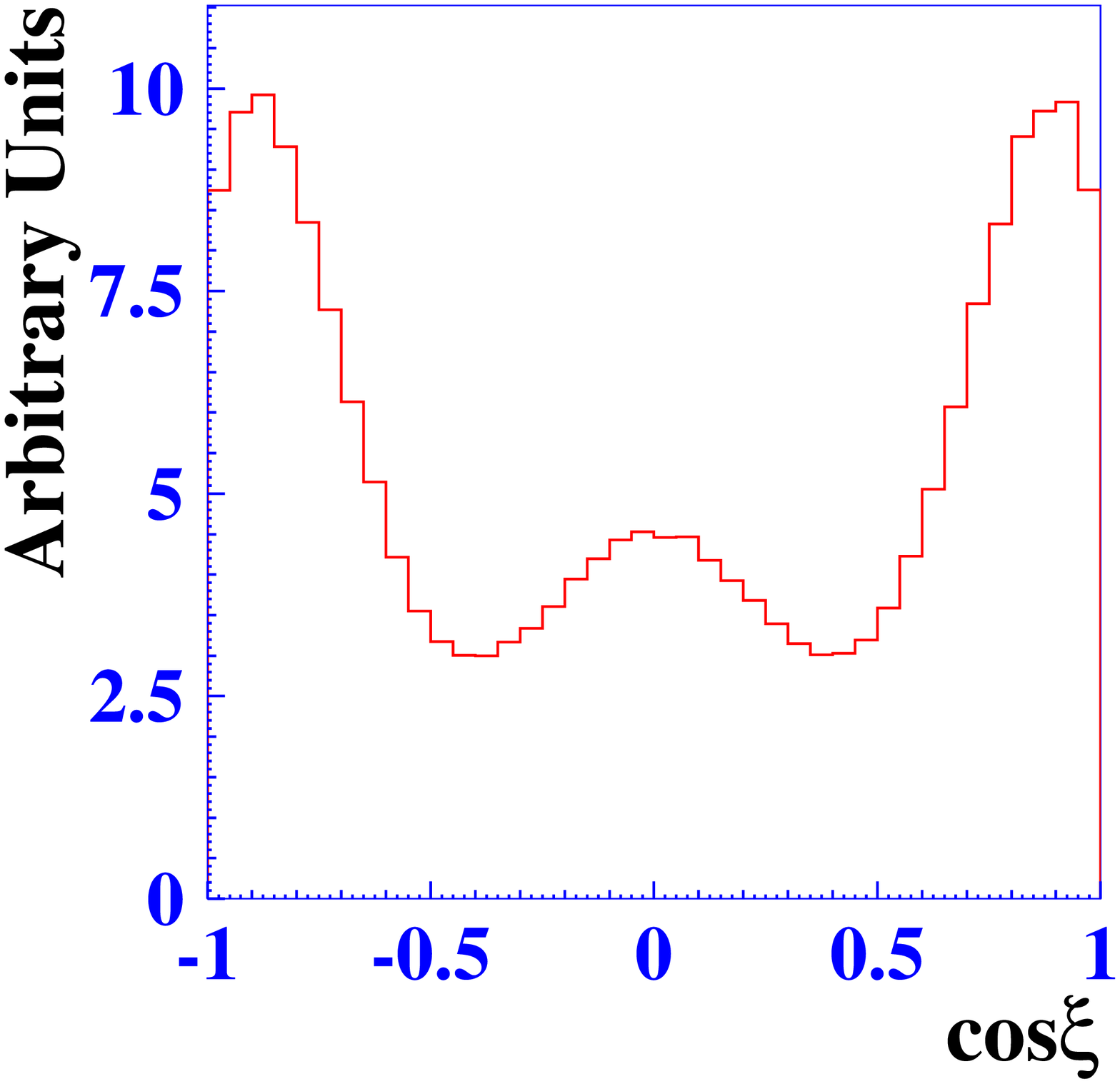} \\
b1) & b2)  & b3)  \\
\end{longtable}
\newpage
\begin{figure}[h]
\begin{longtable}{c c c}
\includegraphics[scale=0.23]{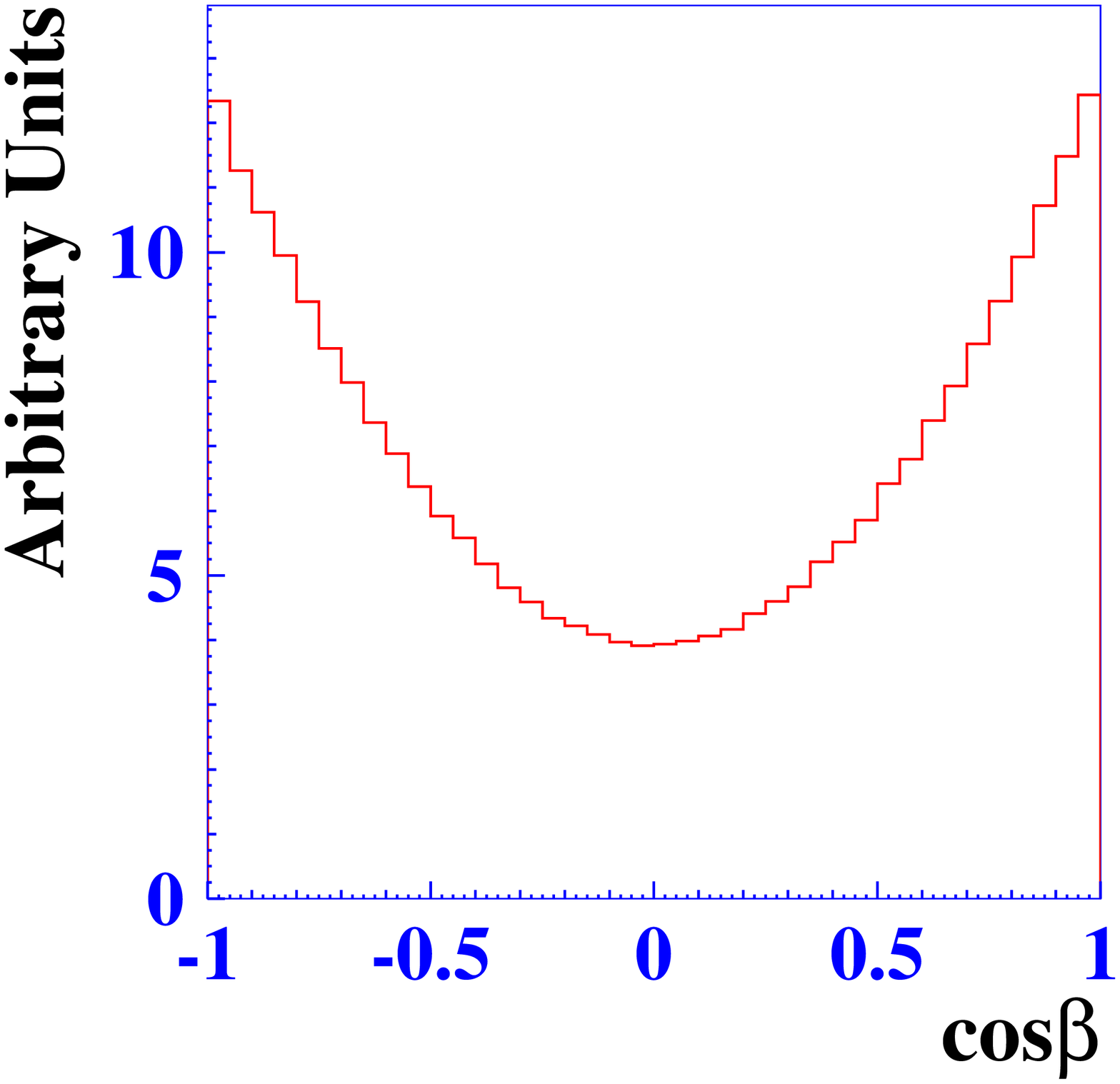}  &
\includegraphics[scale=0.23]{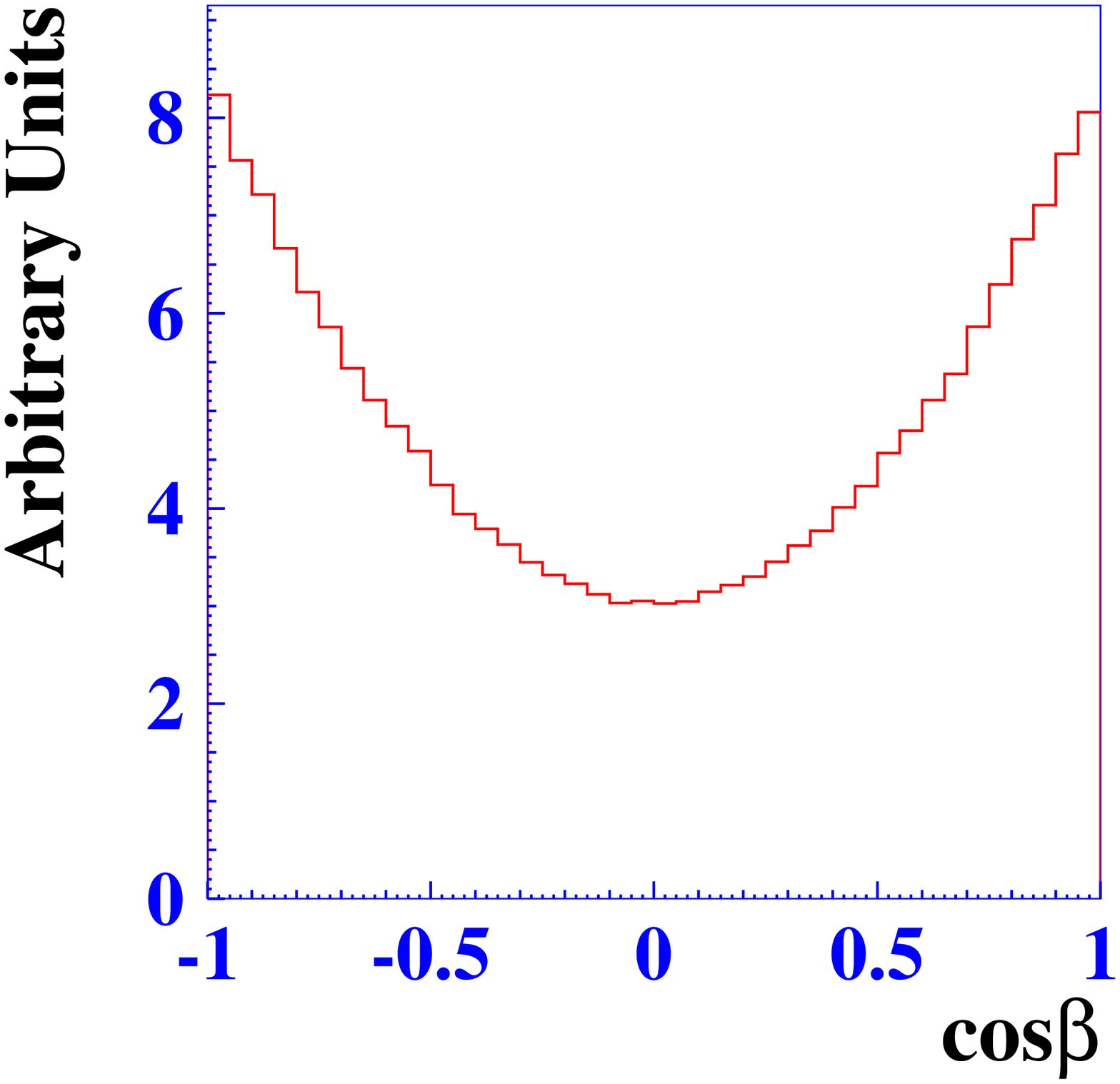} &
\includegraphics[scale=0.23]{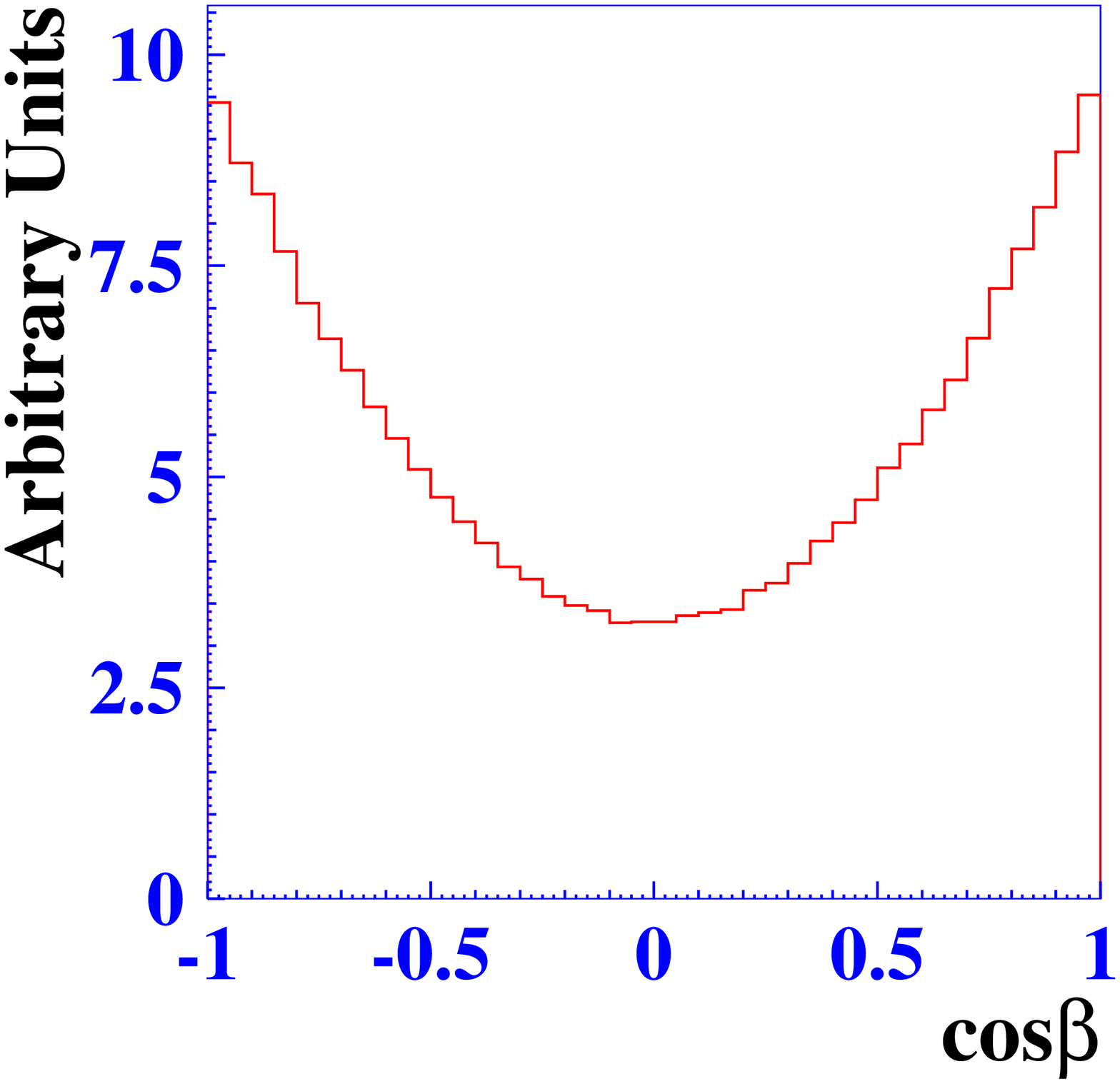} \\
c1) & c2) & c3)  \\
\includegraphics[scale=0.23]{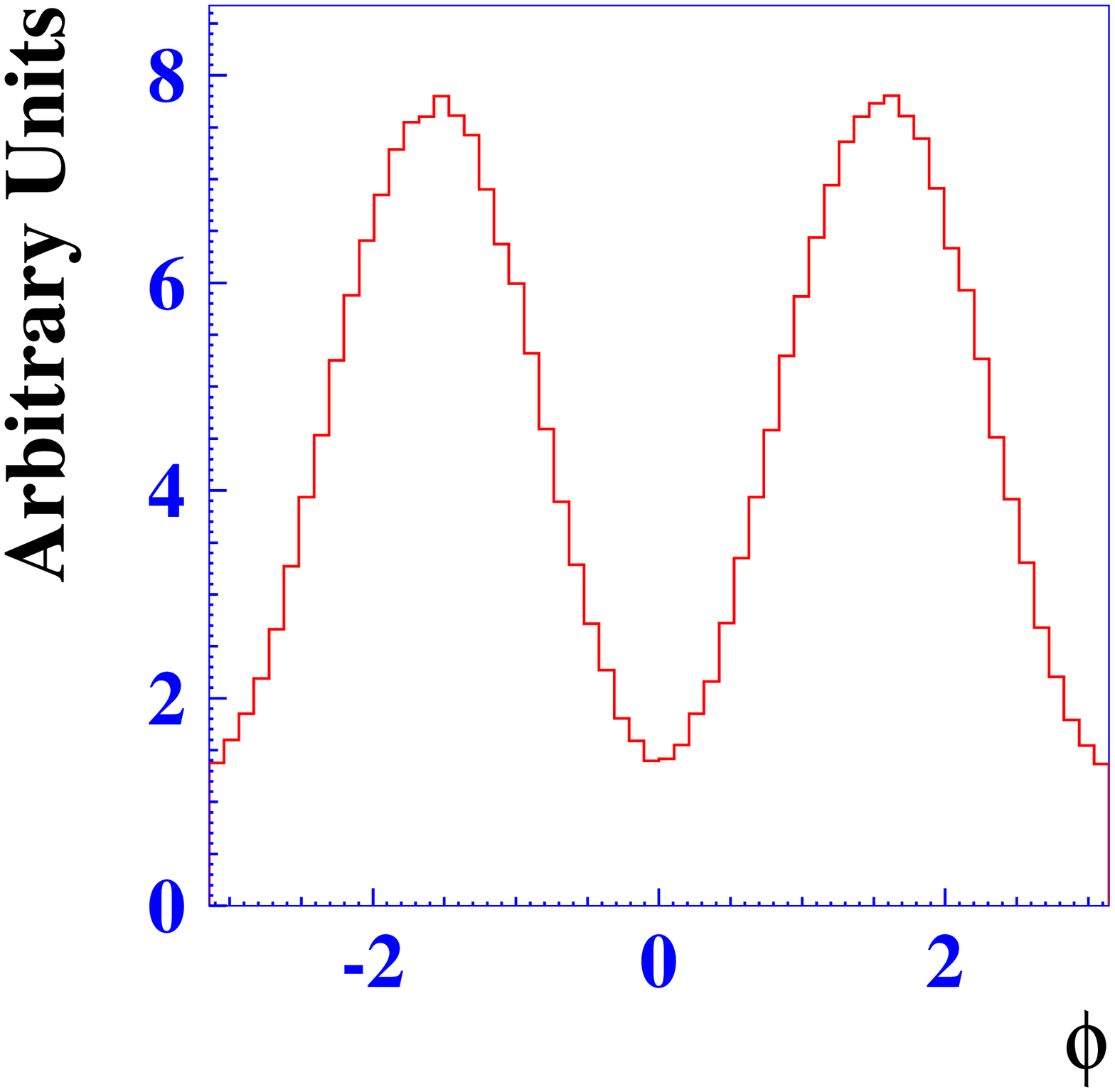}  &
\includegraphics[scale=0.23]{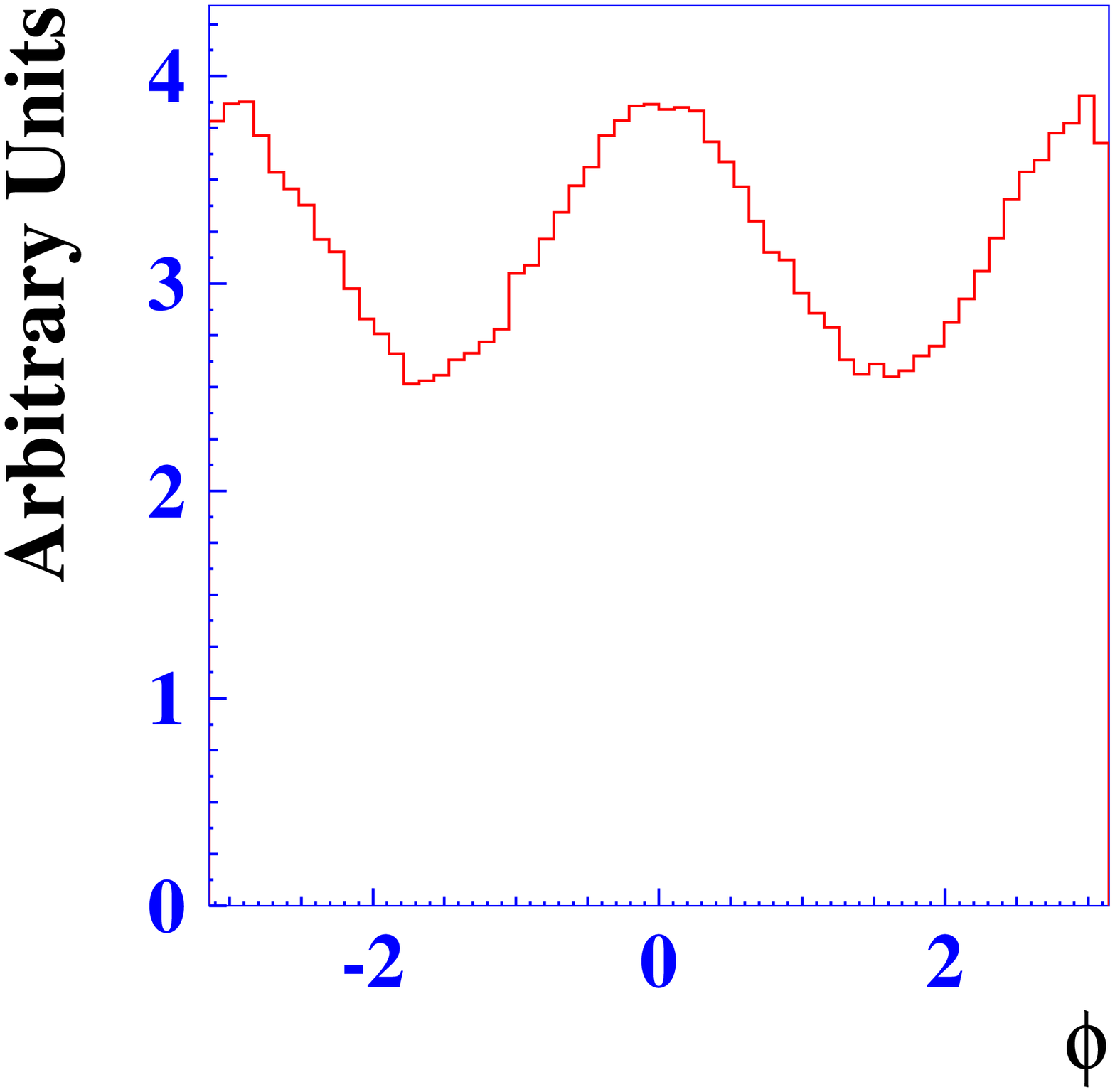} &
\includegraphics[scale=0.23]{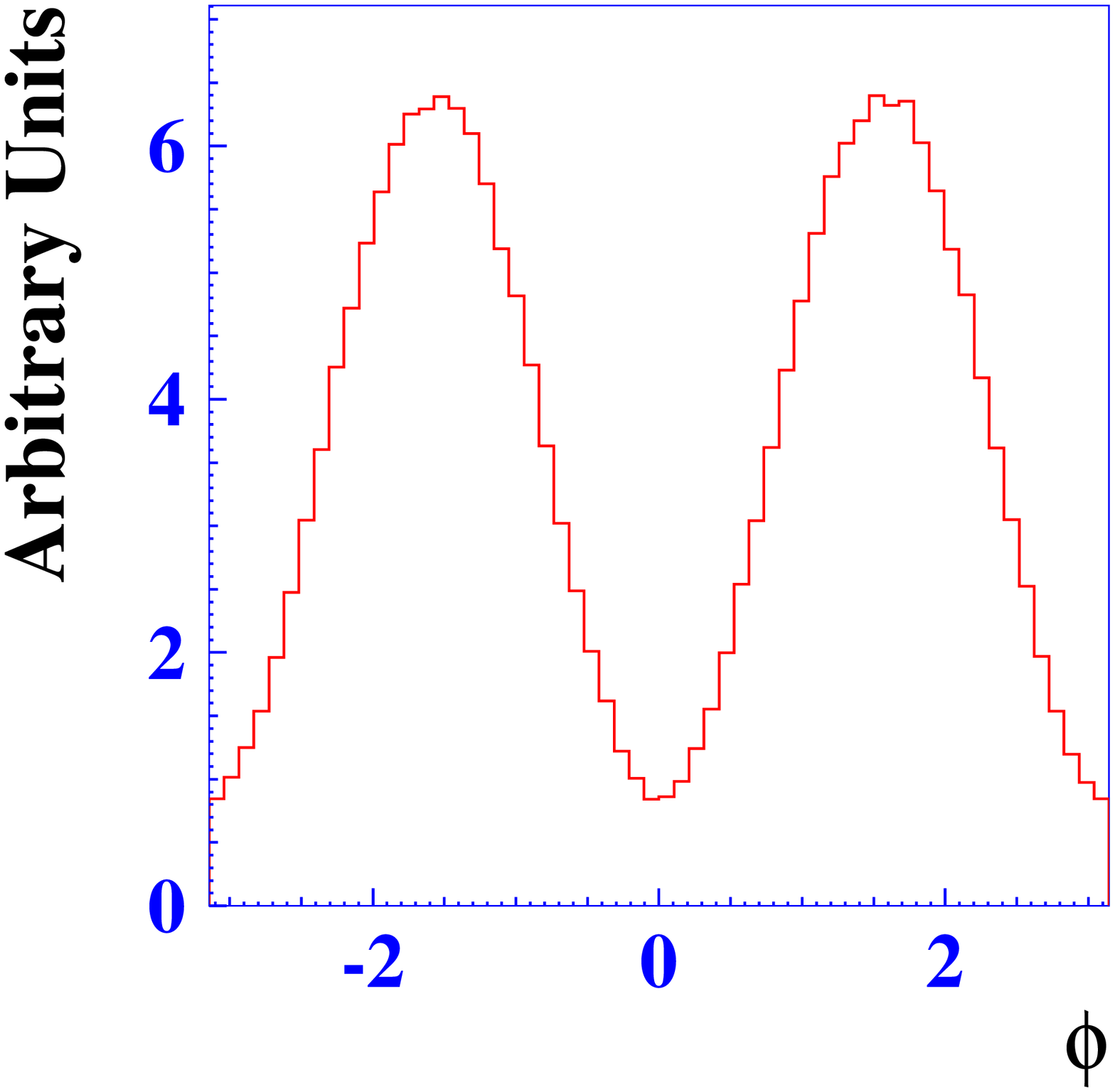} \\
d1) & d2) & d3) \\
\includegraphics[scale=0.23]{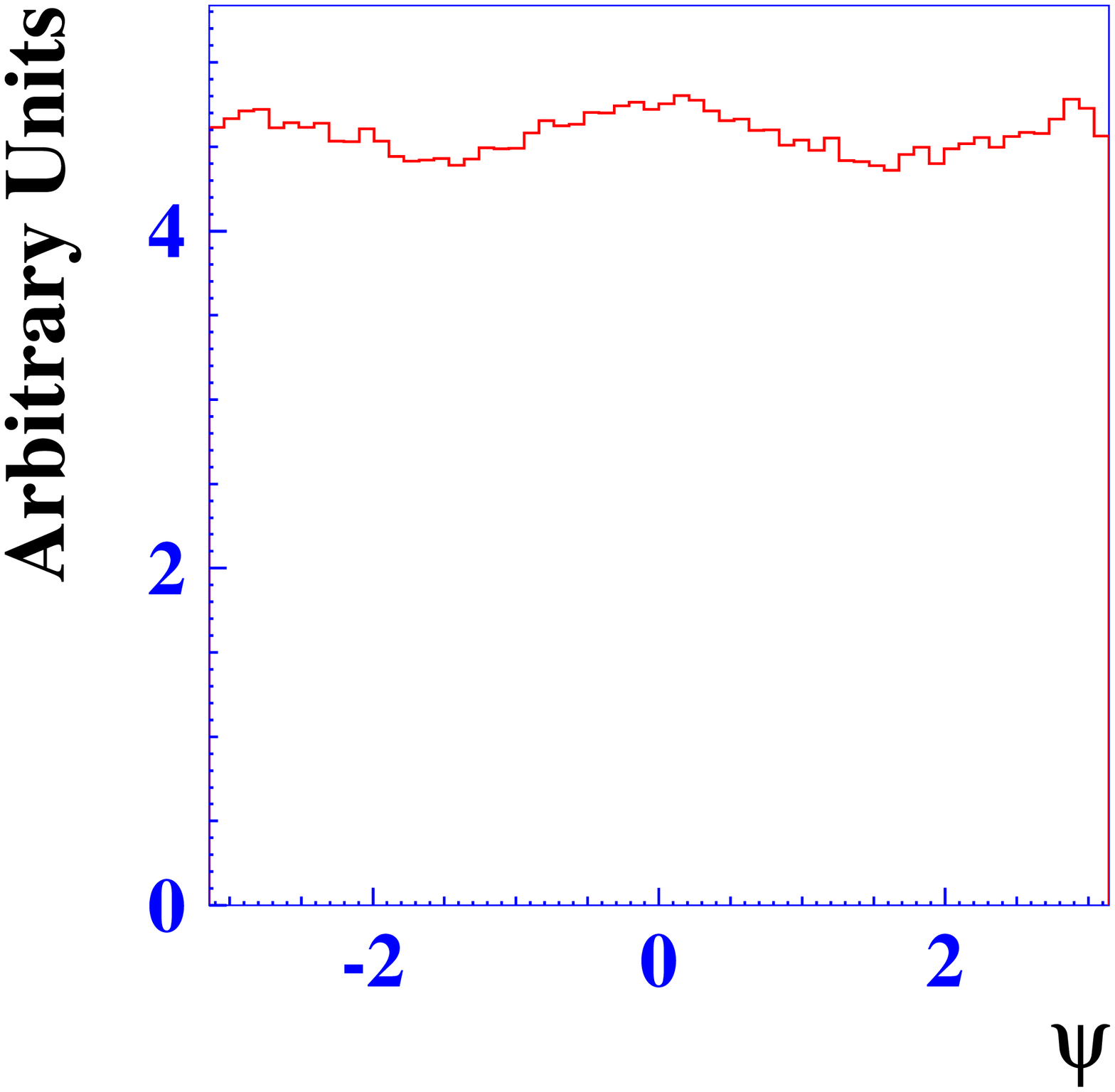}  &
\includegraphics[scale=0.23]{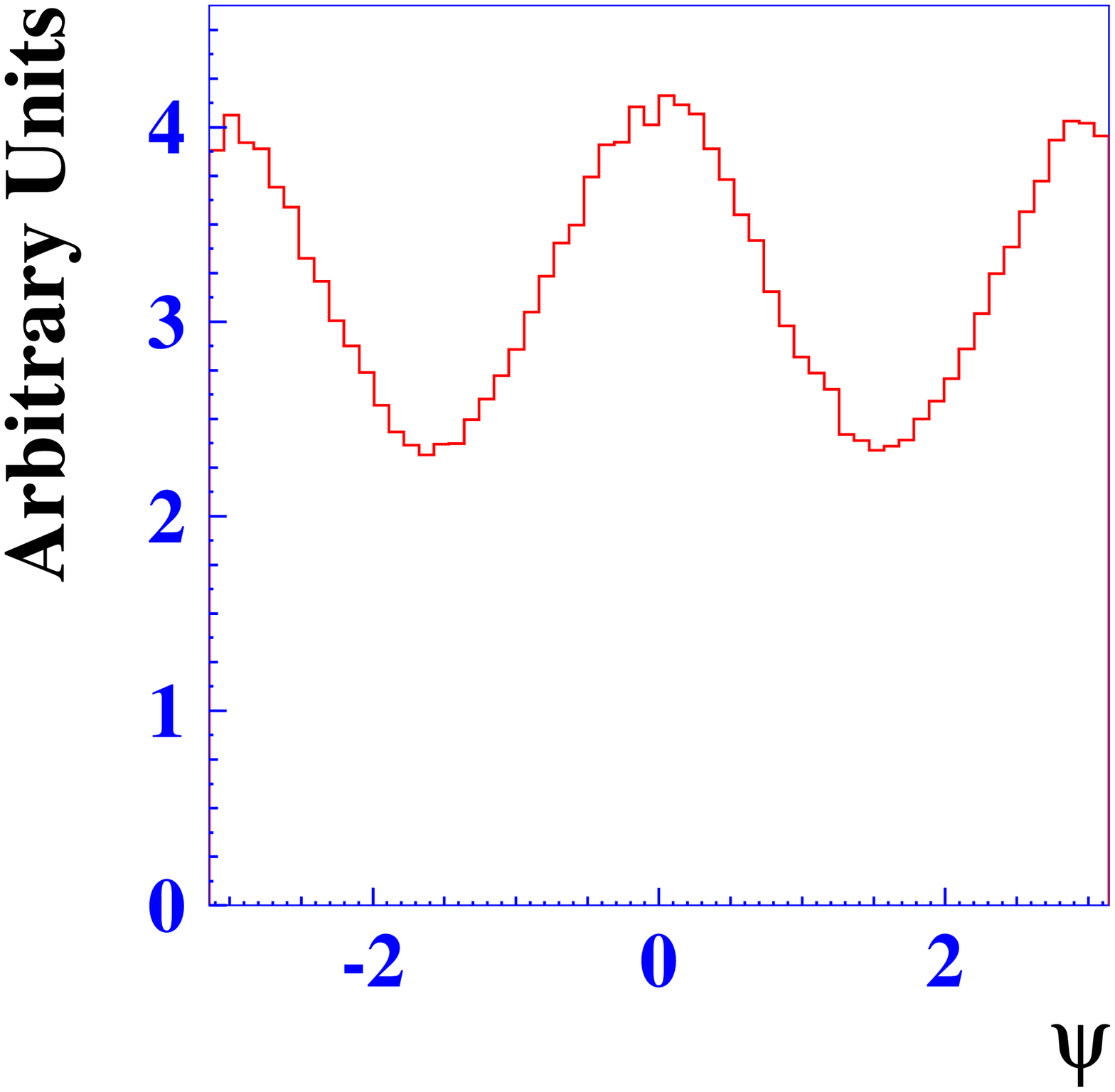} &
\includegraphics[scale=0.23]{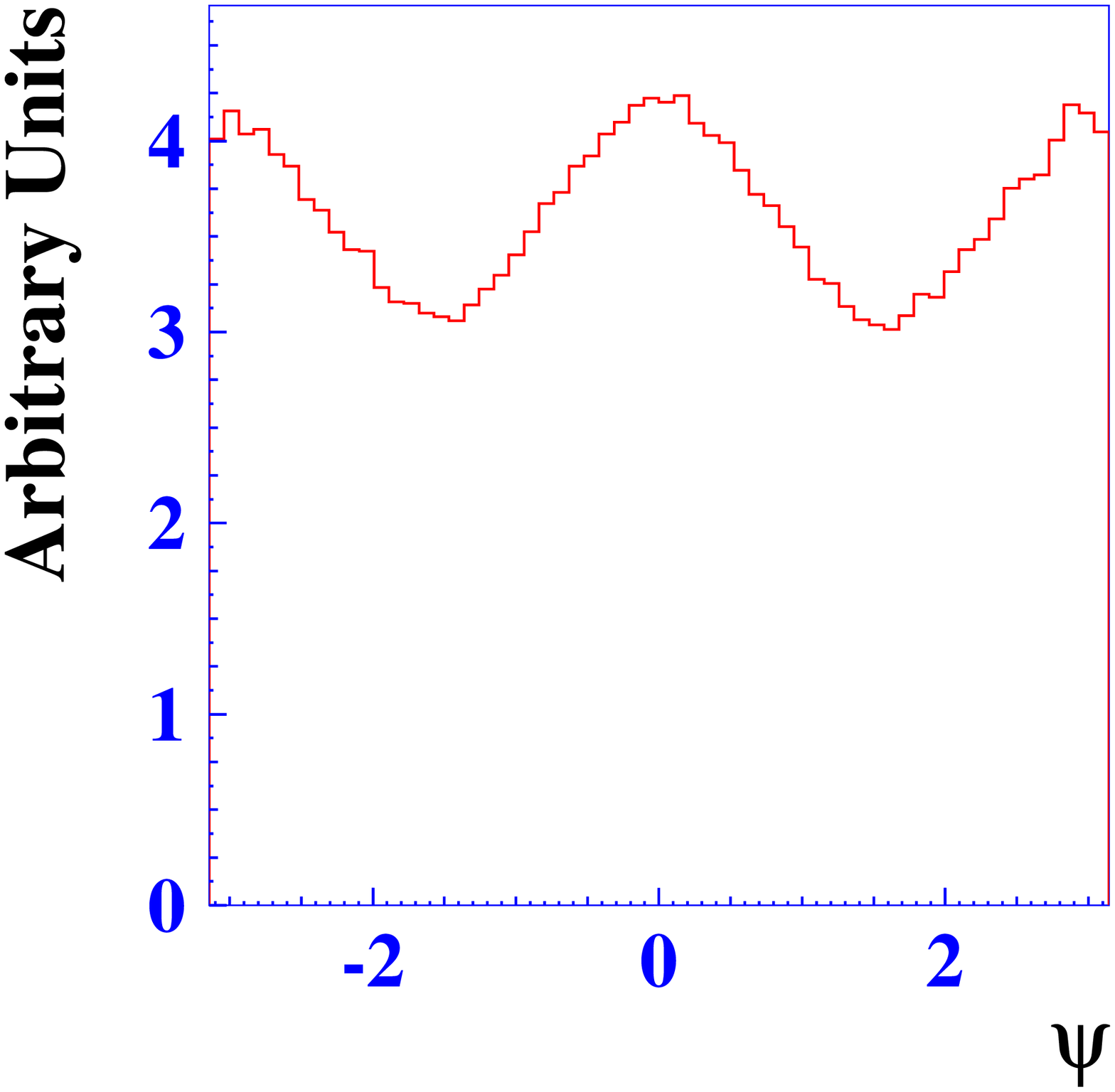} \\
e1) & e2) & e3)\\
\end{longtable}
\caption{Simulated angular distributions for
the $\omega \pi$-resonances. The figures a1), b1), c1), d1), e1)
correspond to the $J^P=2^+$ ($b_2^-$) intermediate state;
a2), b2), c2), d2), e2) --- $2^-$ ($\rho_2^-$)-state;
a3), b3), c3), d3), e3) --- $3^-$ ($\rho_3(1690)^-$)-state.}
\label{fig6}
\end{figure}

\newpage
\begin{longtable}{c c c}
\includegraphics[scale=0.23]{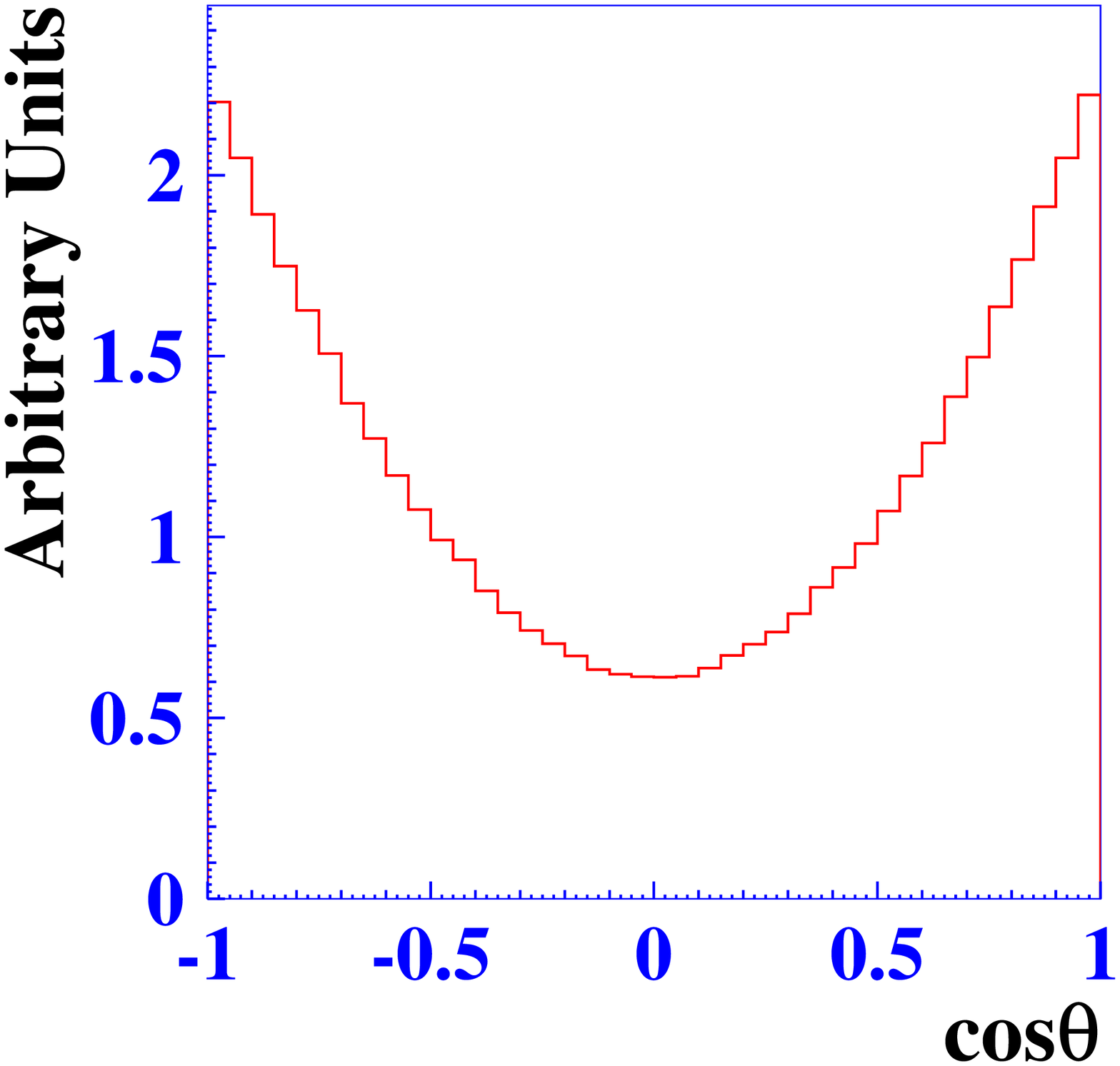} &
\includegraphics[scale=0.23]{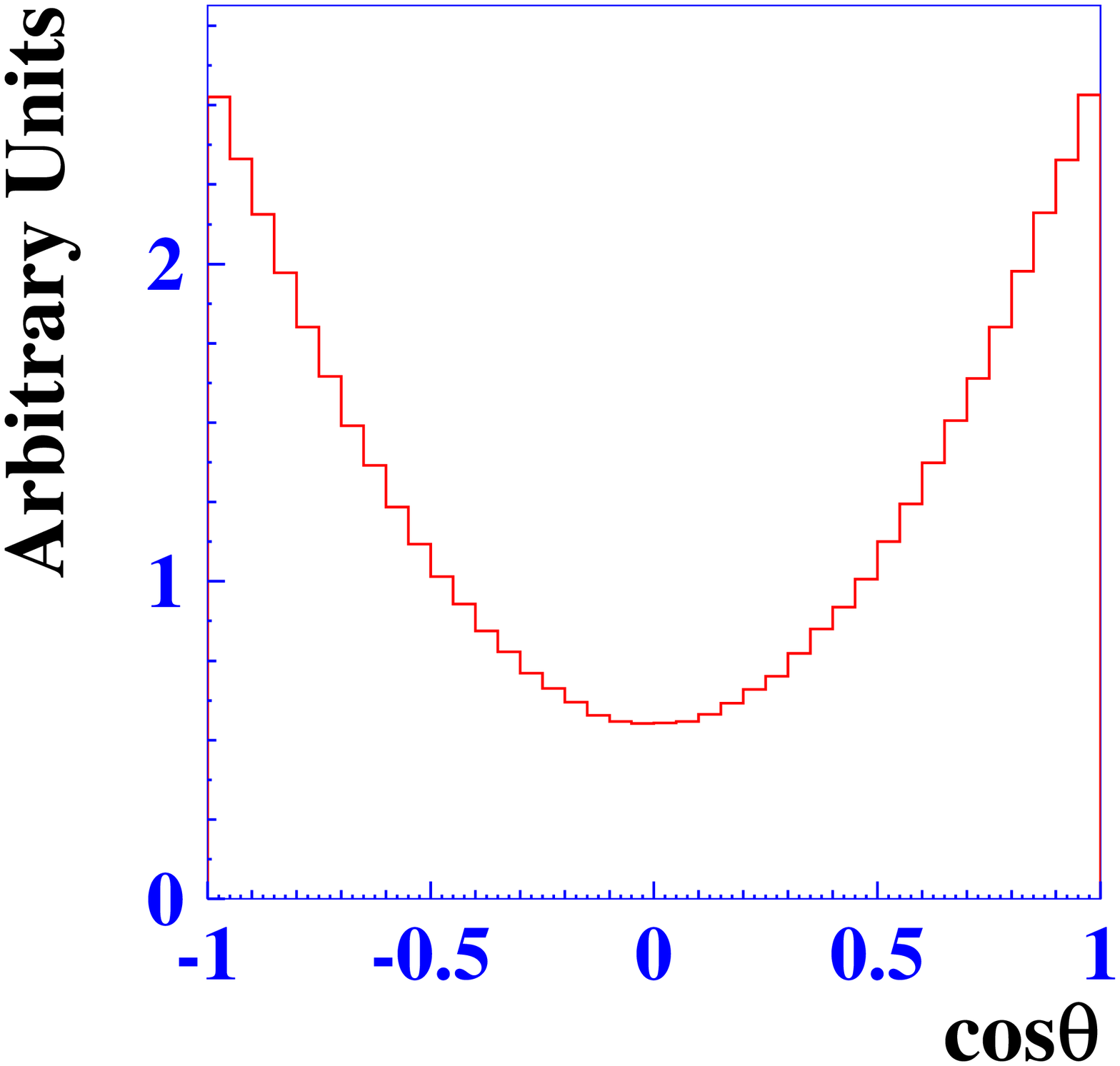}  &
\includegraphics[scale=0.23]{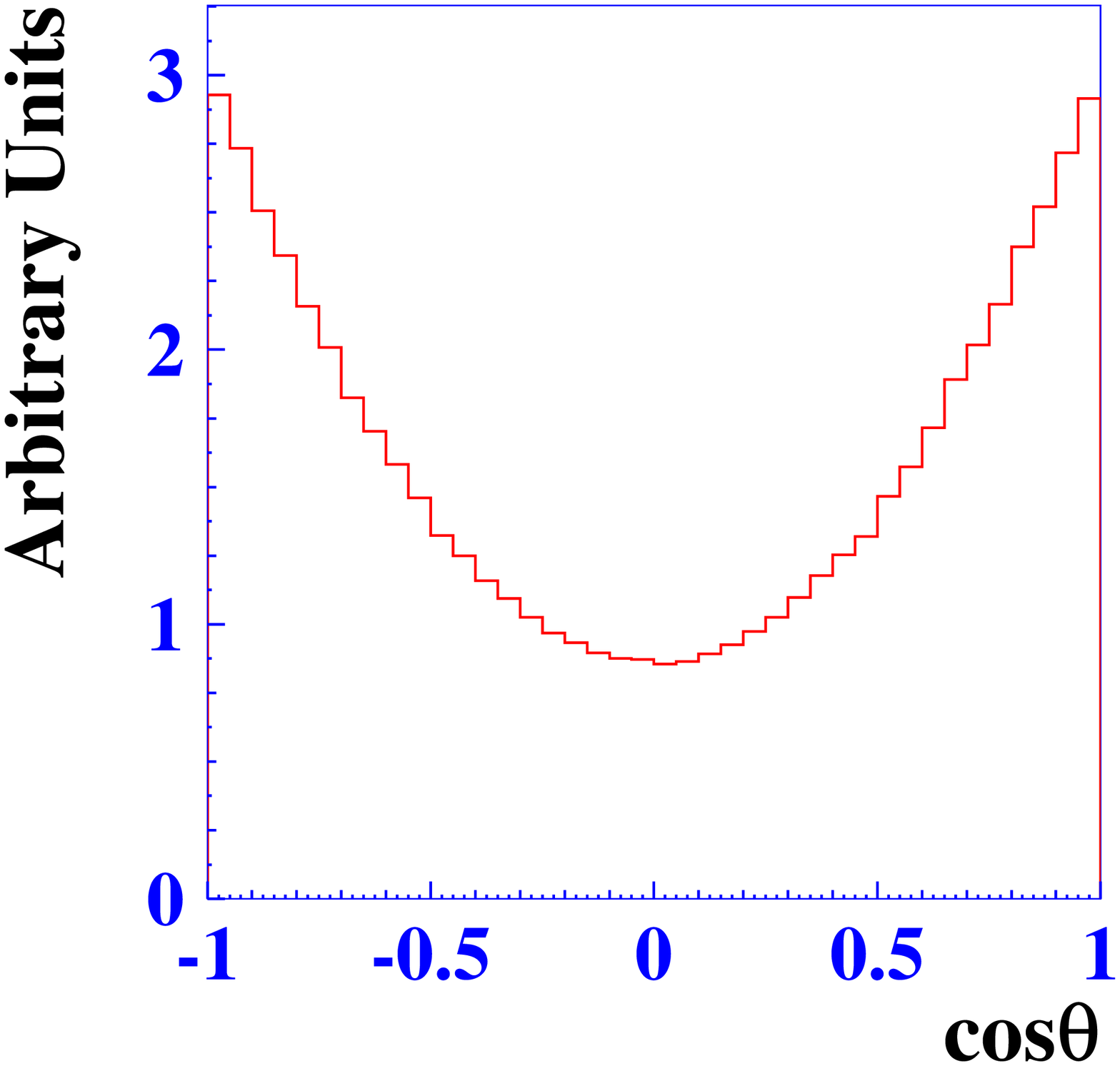}  \\
a1) & a2)  & a3) \\
\includegraphics[scale=0.23]{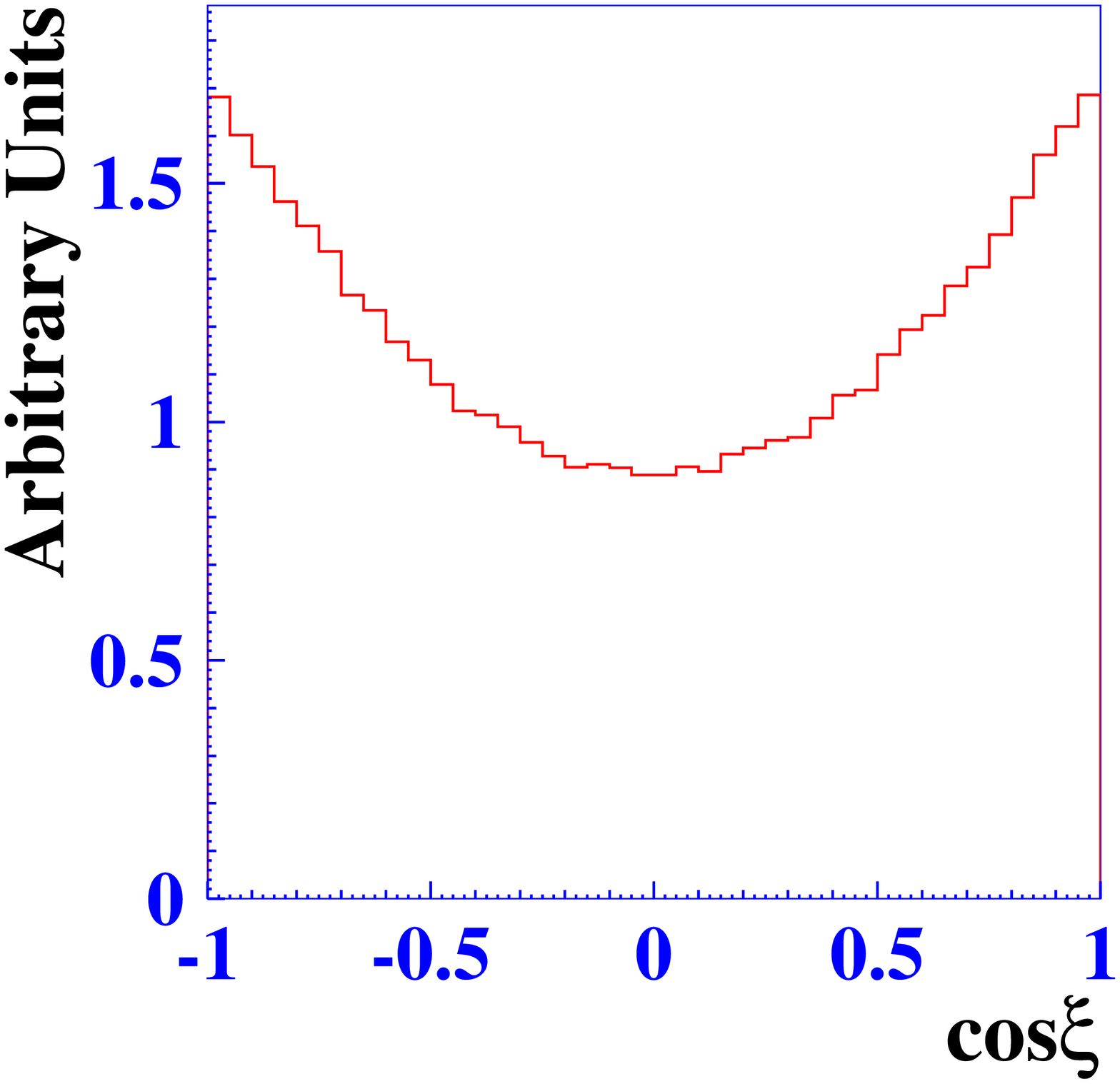} &
\includegraphics[scale=0.23]{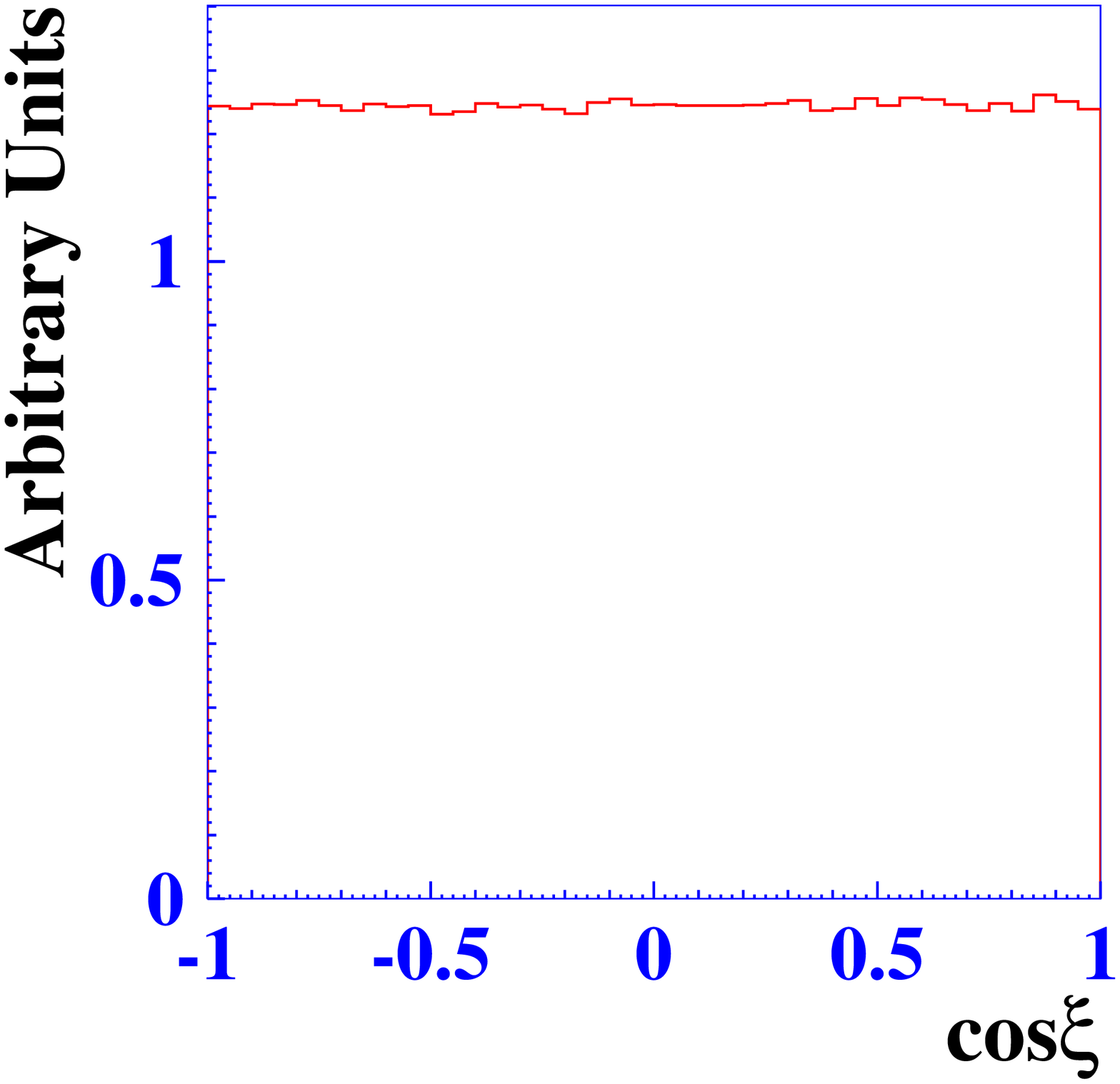}  &
\includegraphics[scale=0.23]{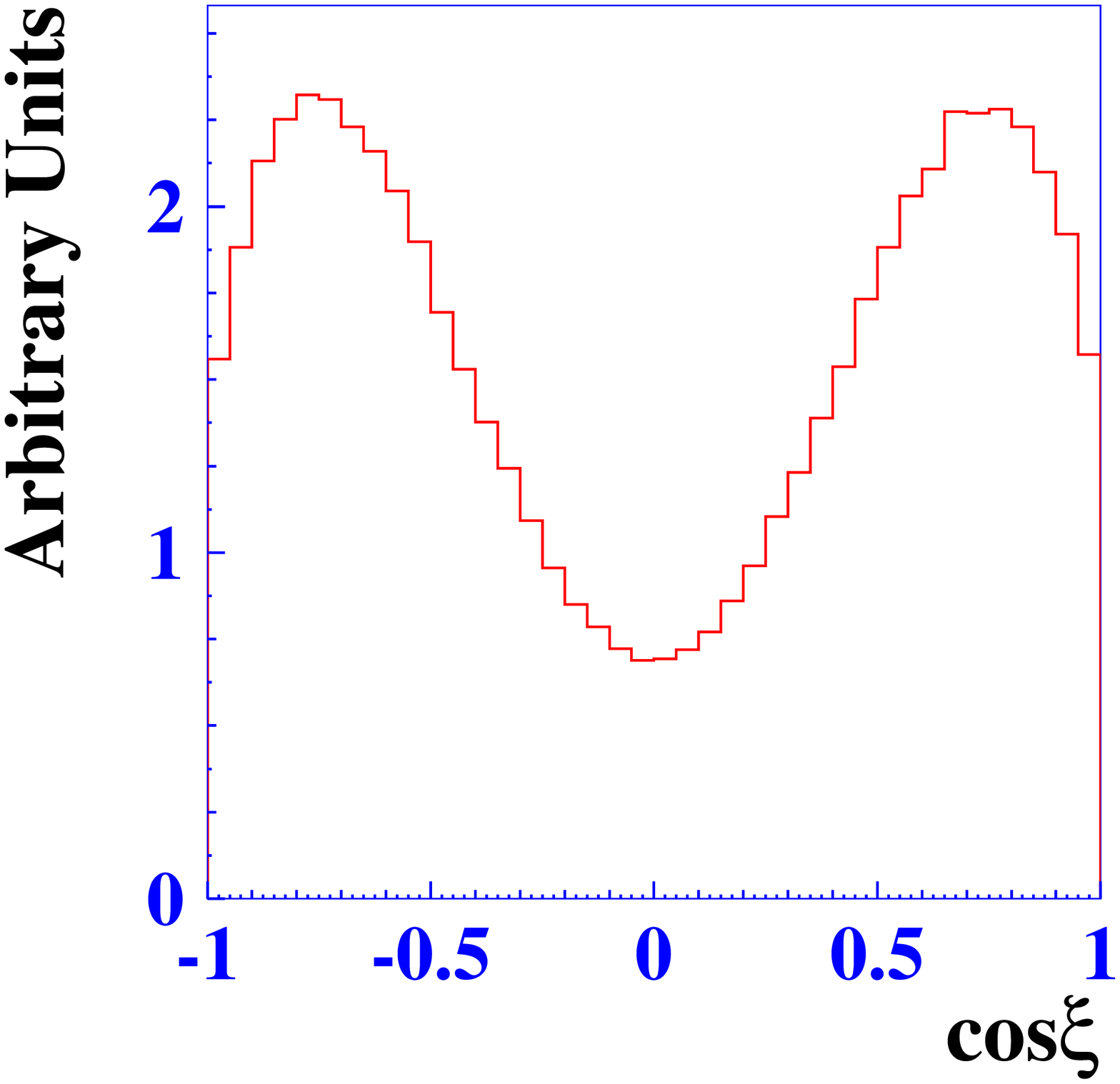}  \\
b1) & b2)  & b3) \\
\includegraphics[scale=0.23]{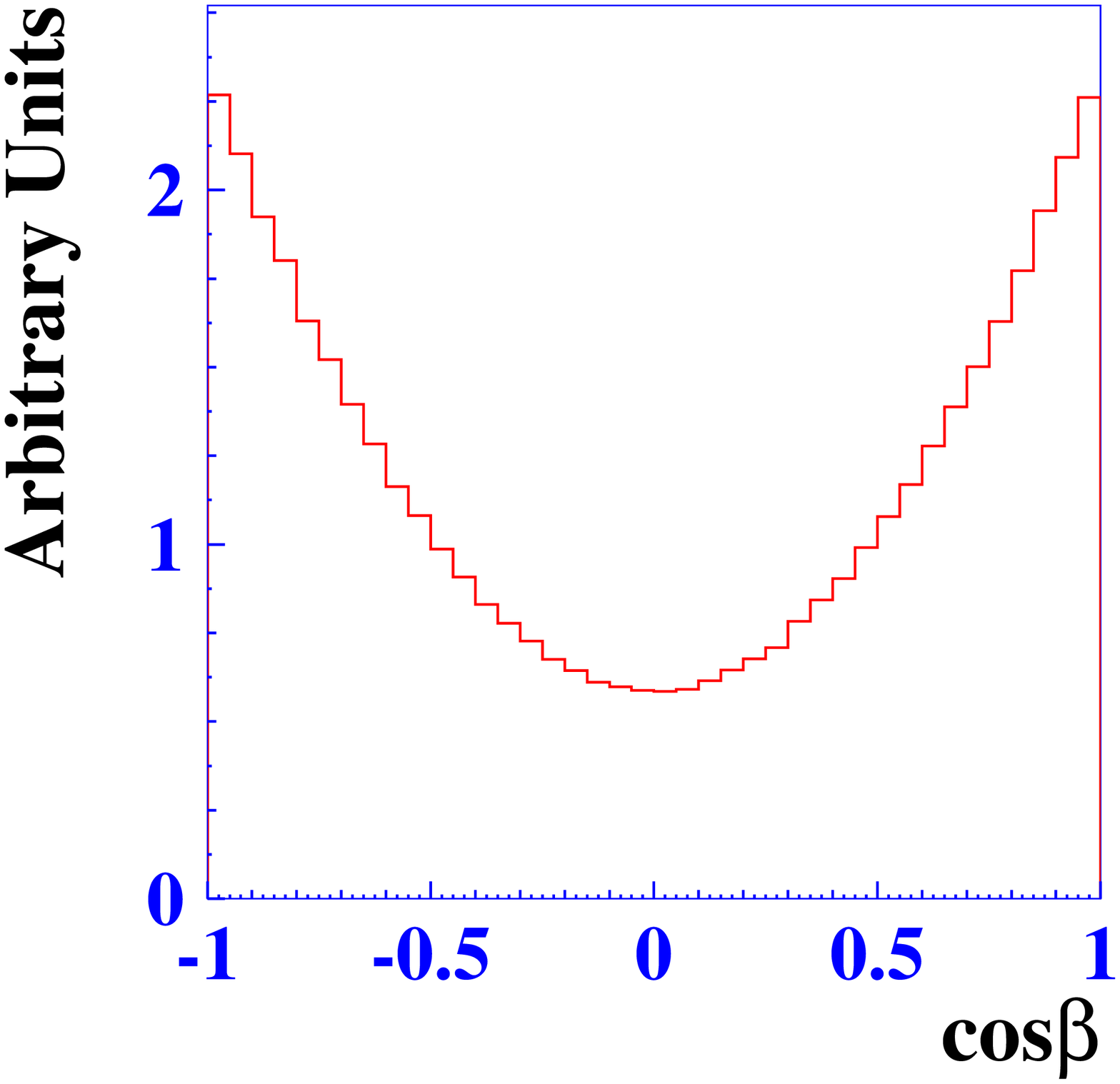} &
\includegraphics[scale=0.23]{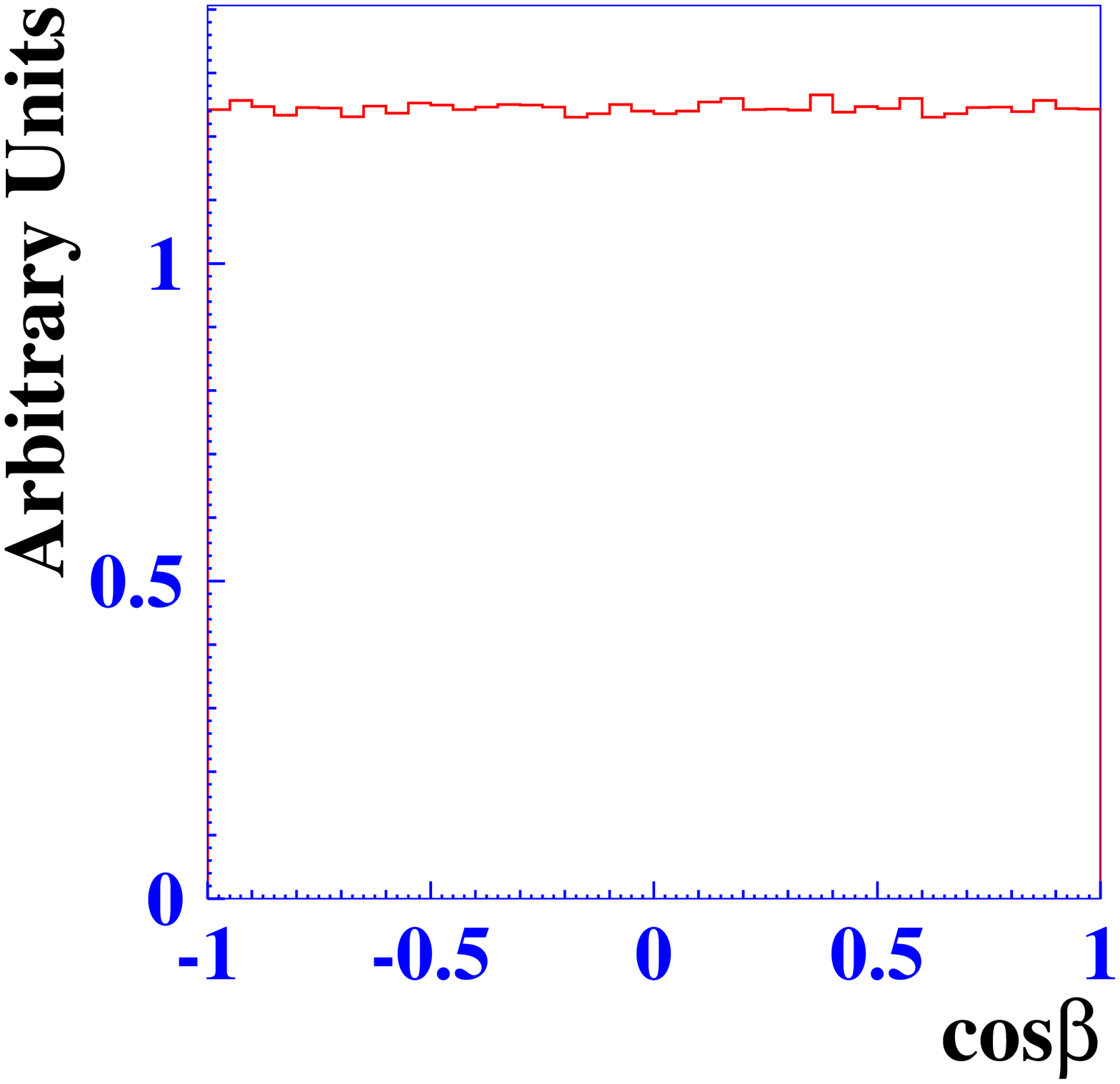}  &
\includegraphics[scale=0.23]{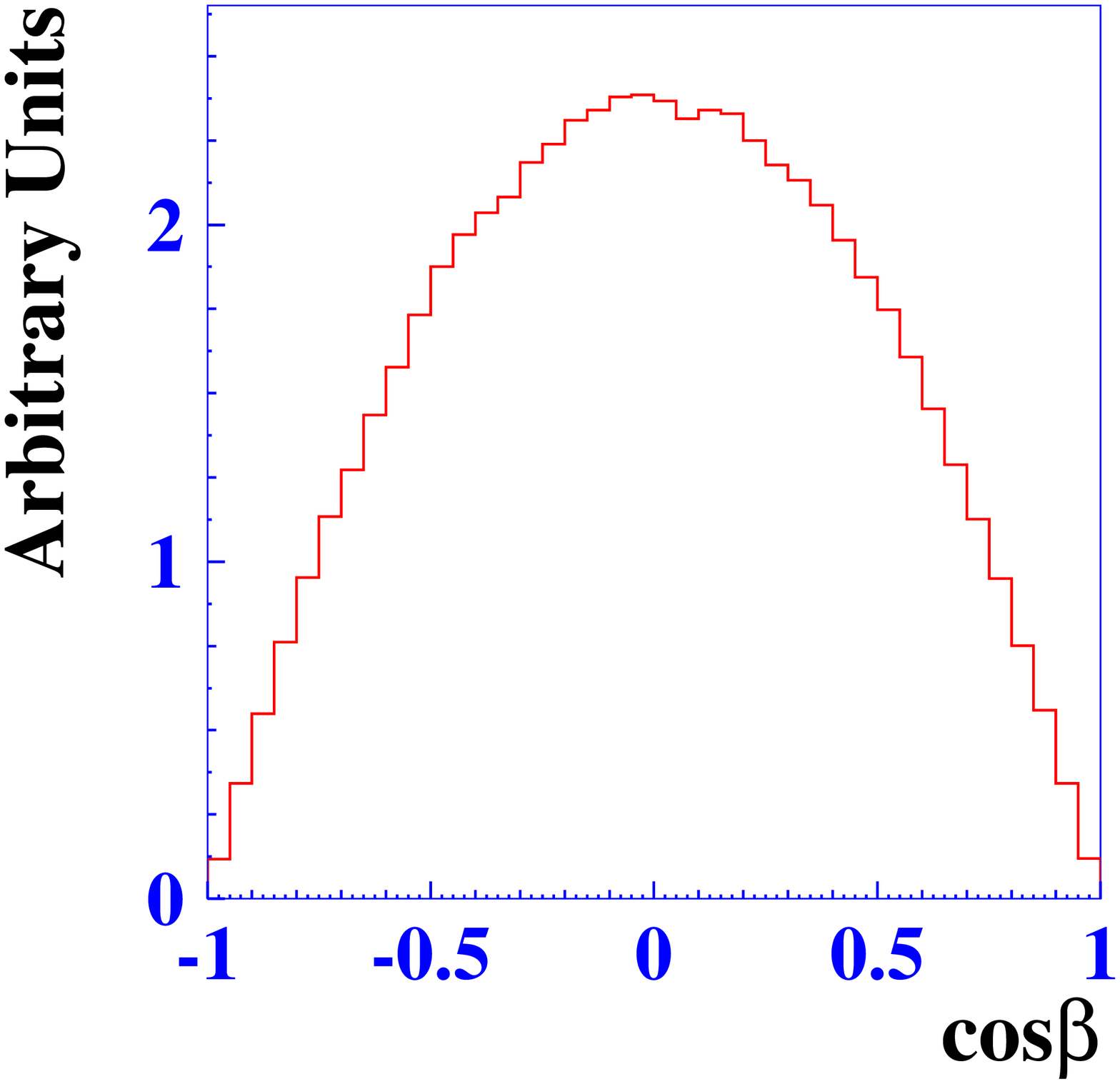}  \\
c1) & c2) & c3) \\
\includegraphics[scale=0.23]{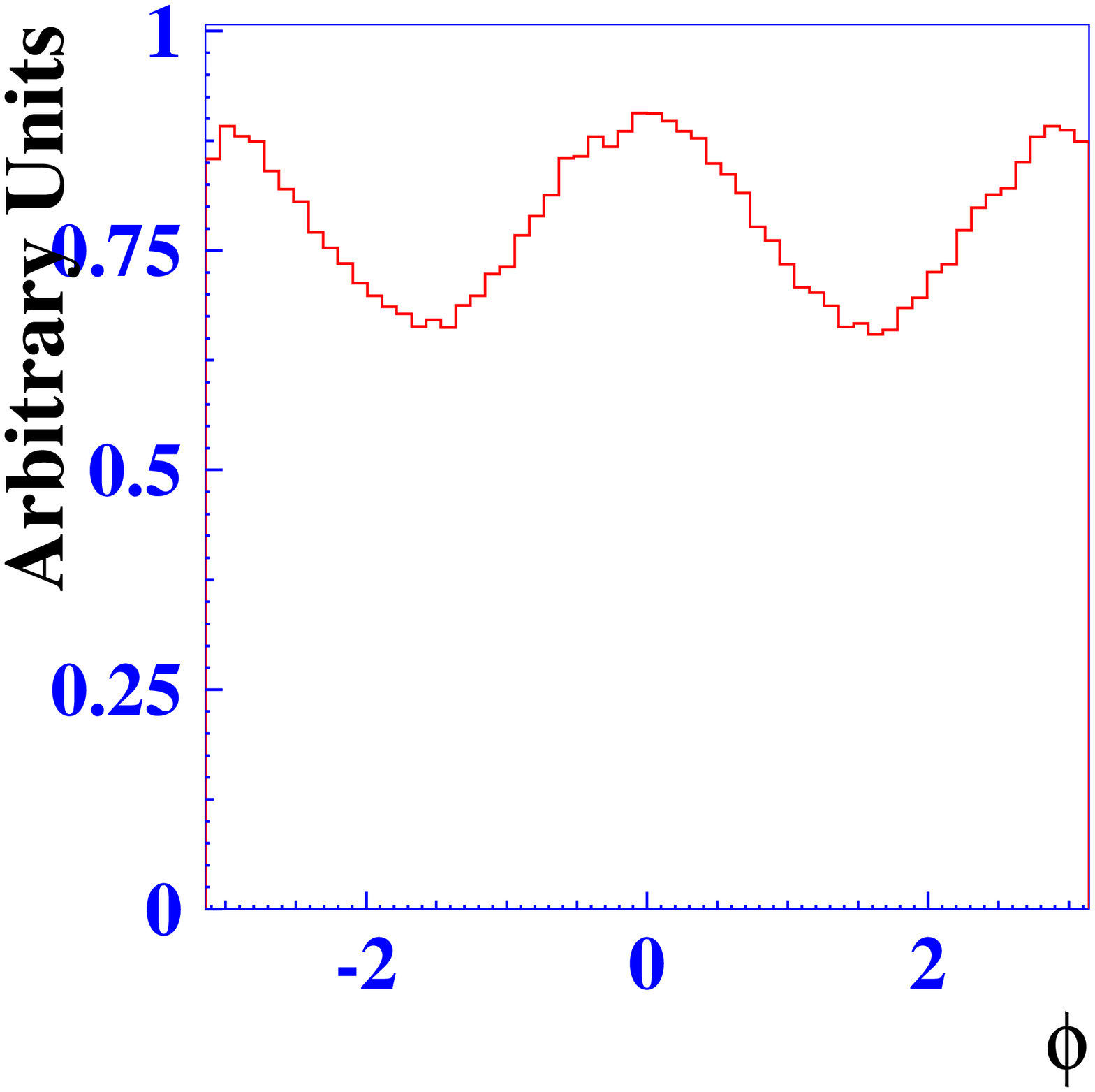} &
\includegraphics[scale=0.23]{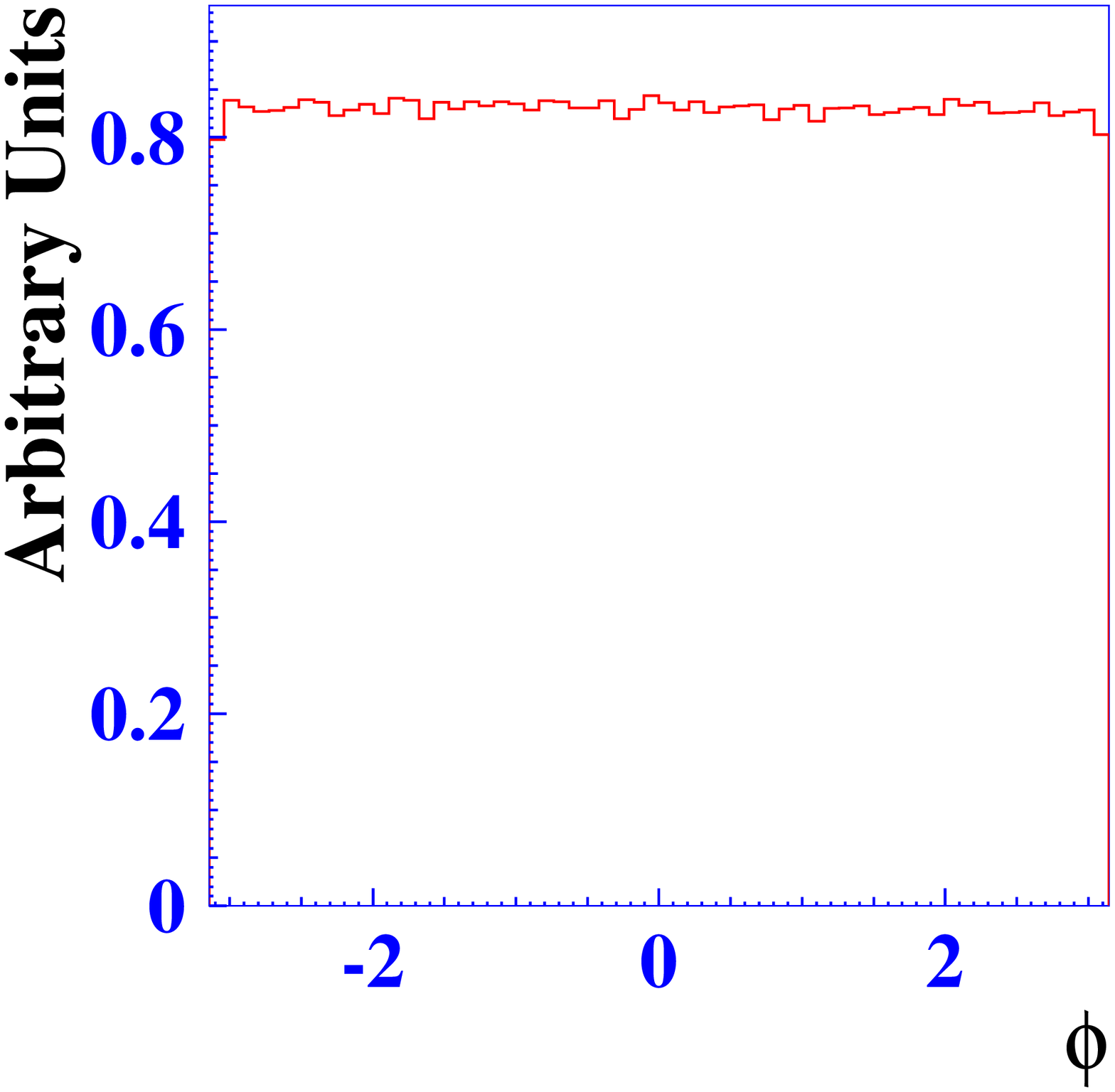}  &
\includegraphics[scale=0.23]{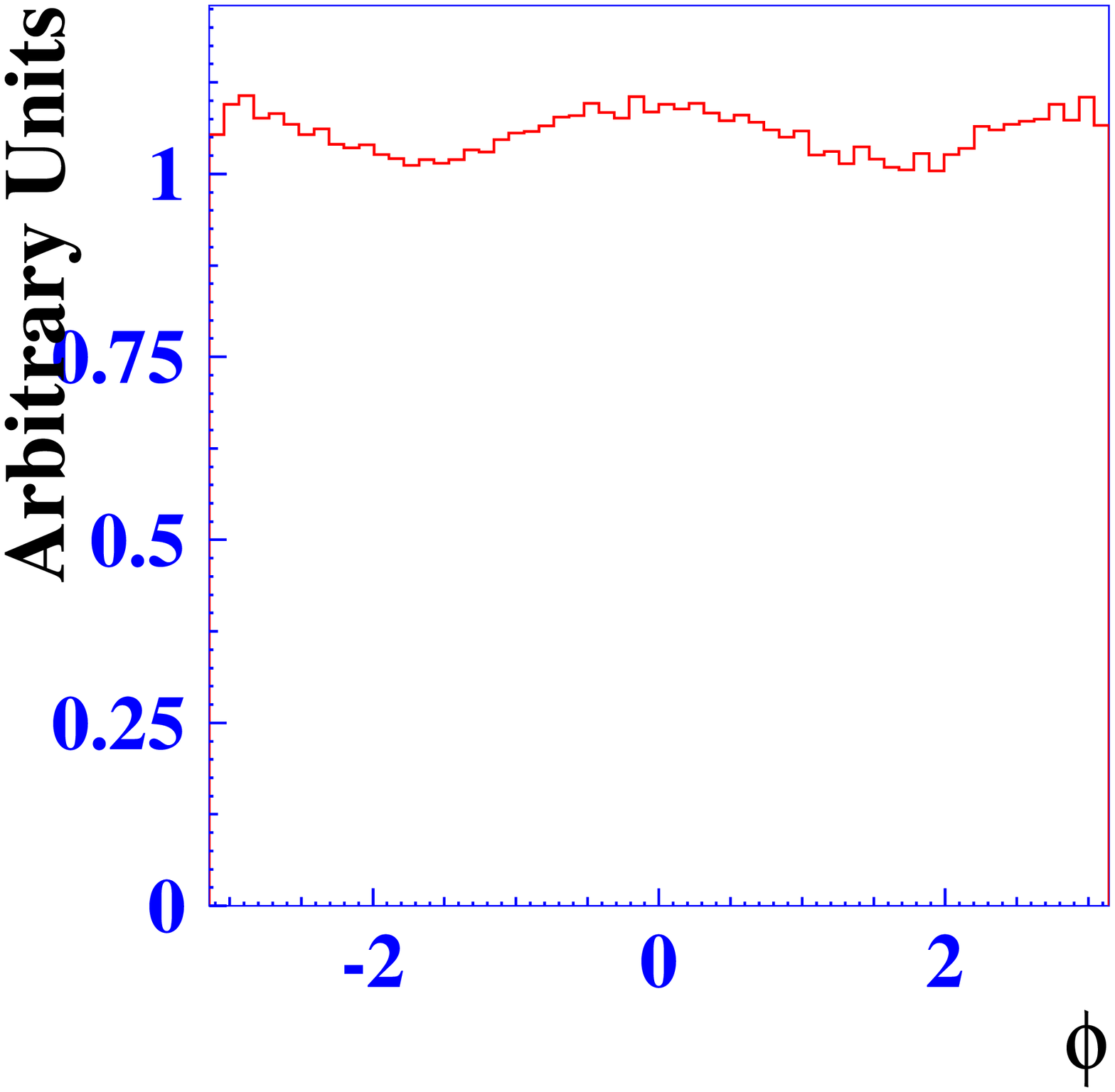} \\
d1) & d2) & d3) \\
\end{longtable}
\newpage
\begin{figure}[h]
\begin{longtable}{c c c}
\includegraphics[scale=0.23]{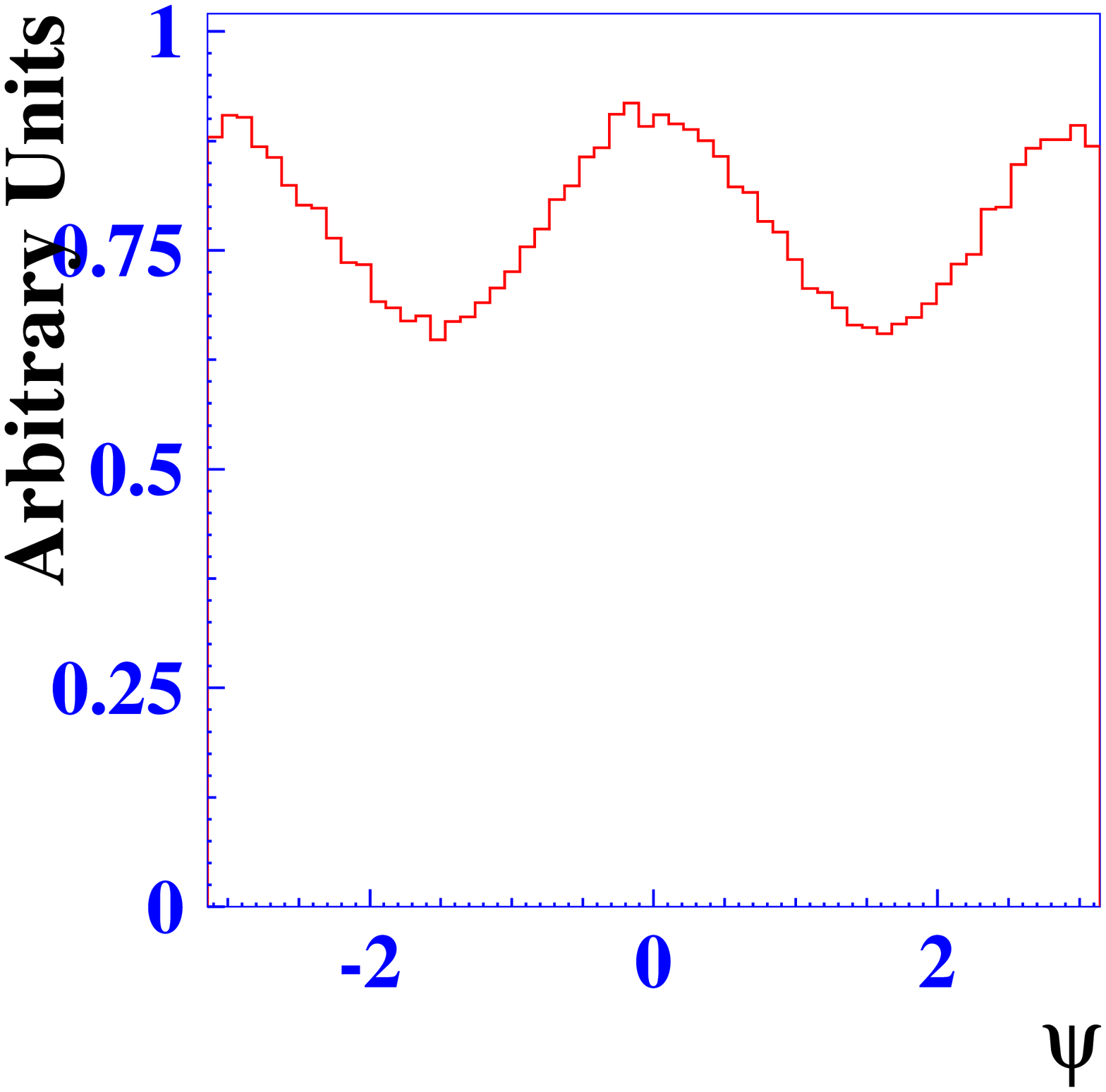} &
\includegraphics[scale=0.23]{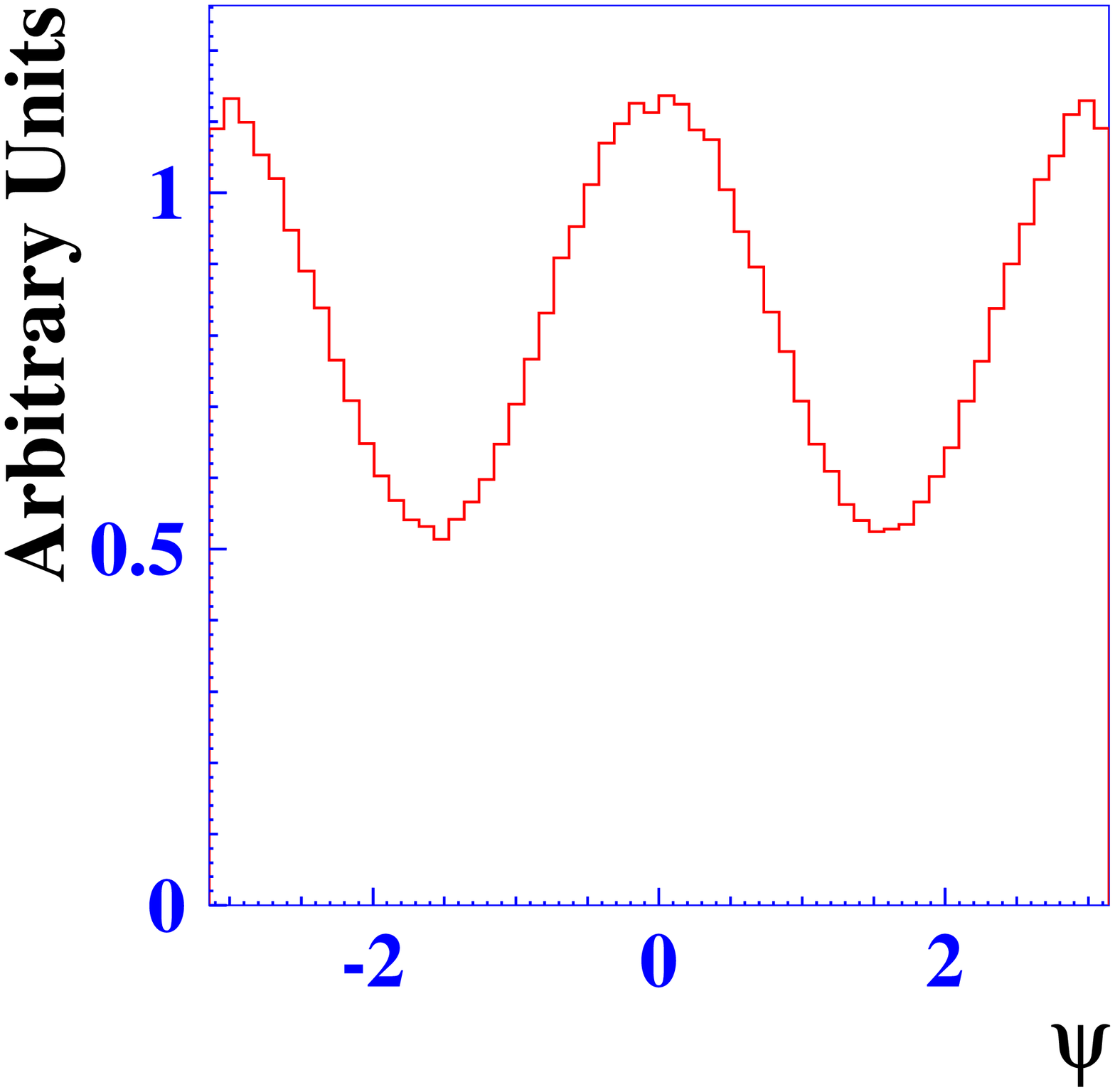}  &
\includegraphics[scale=0.23]{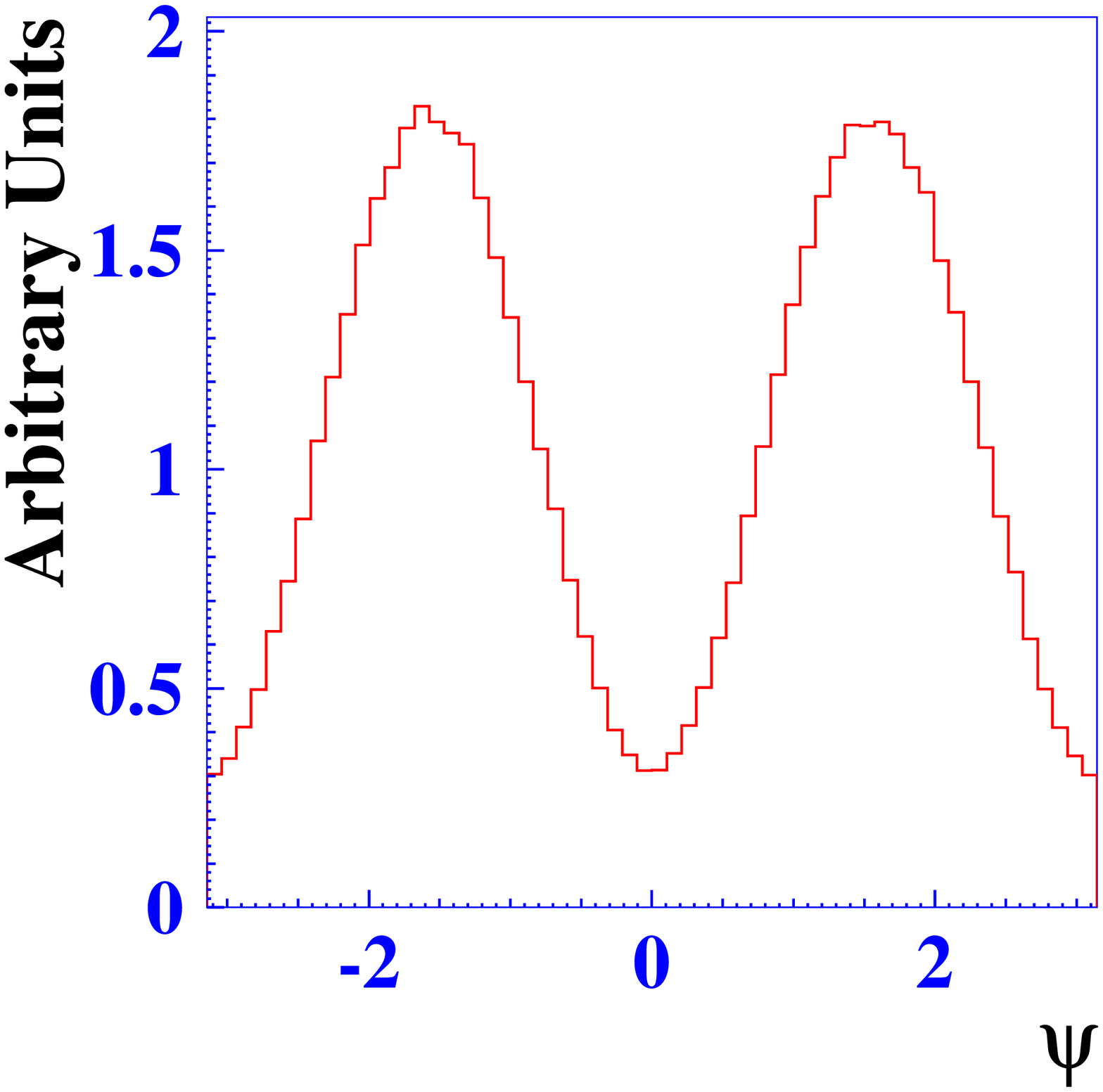} \\
e1) & e2) & e3)  \\
\end{longtable}
\caption{Simulated angular distributions for
the $P$-wave $D^{**}$-resonances. The figures a1), b1), c1), d1), e1)
correspond to the $J^P_{j_u}=1^+_{3/2}$ narrow state;
a2), b2), c2), d2), e2) ---  $J^P_{j_u}=1^+_{1/2}$ broad state;
a3), b3), c3), d3), e3) --- $J^P_{j_u}=2^+_{3/2}$ narrow state.}
\label{fig7}
\end{figure}

\begin{longtable}{c c c c}
\includegraphics[scale=0.18]{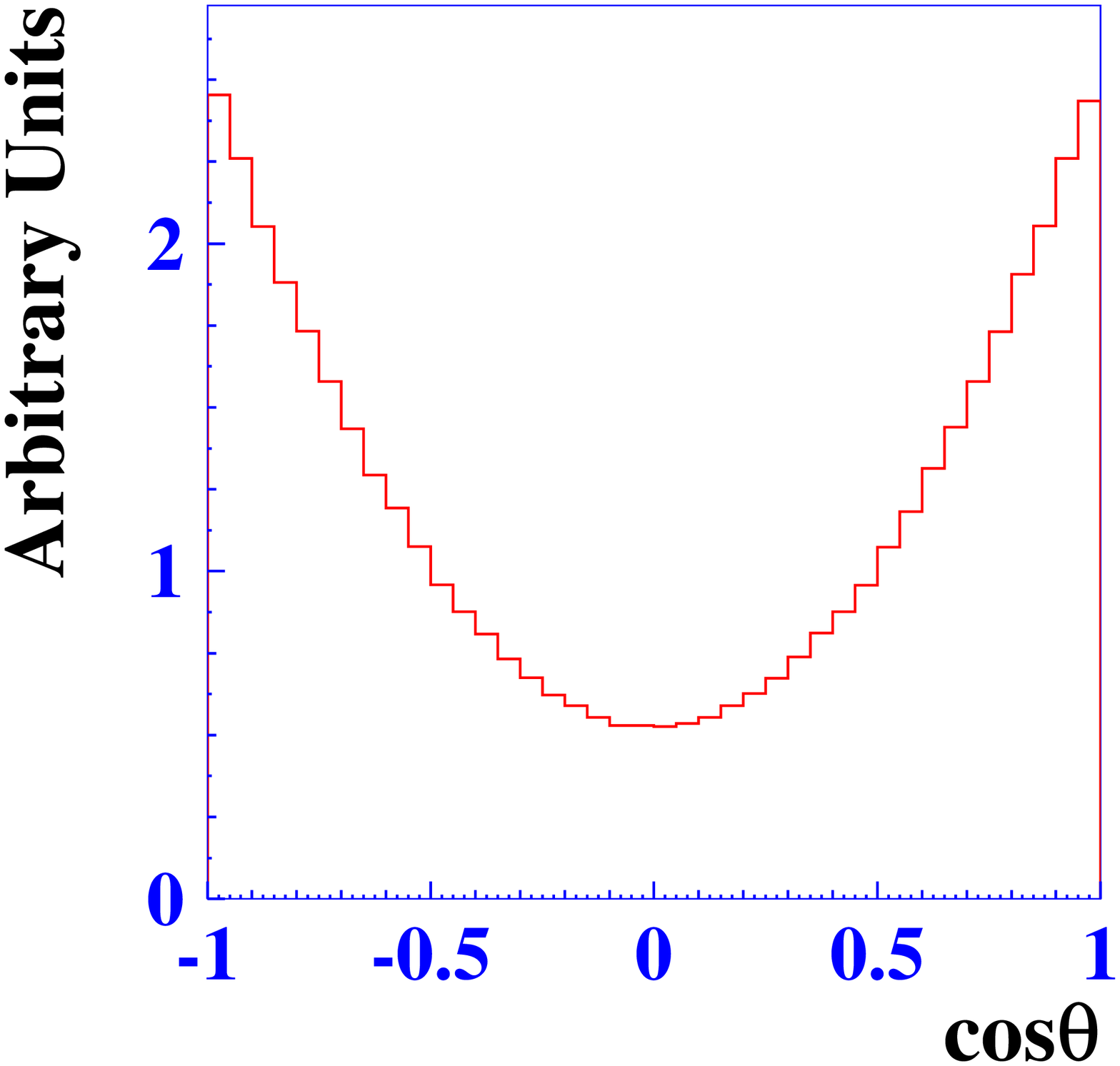}  &
\includegraphics[scale=0.18]{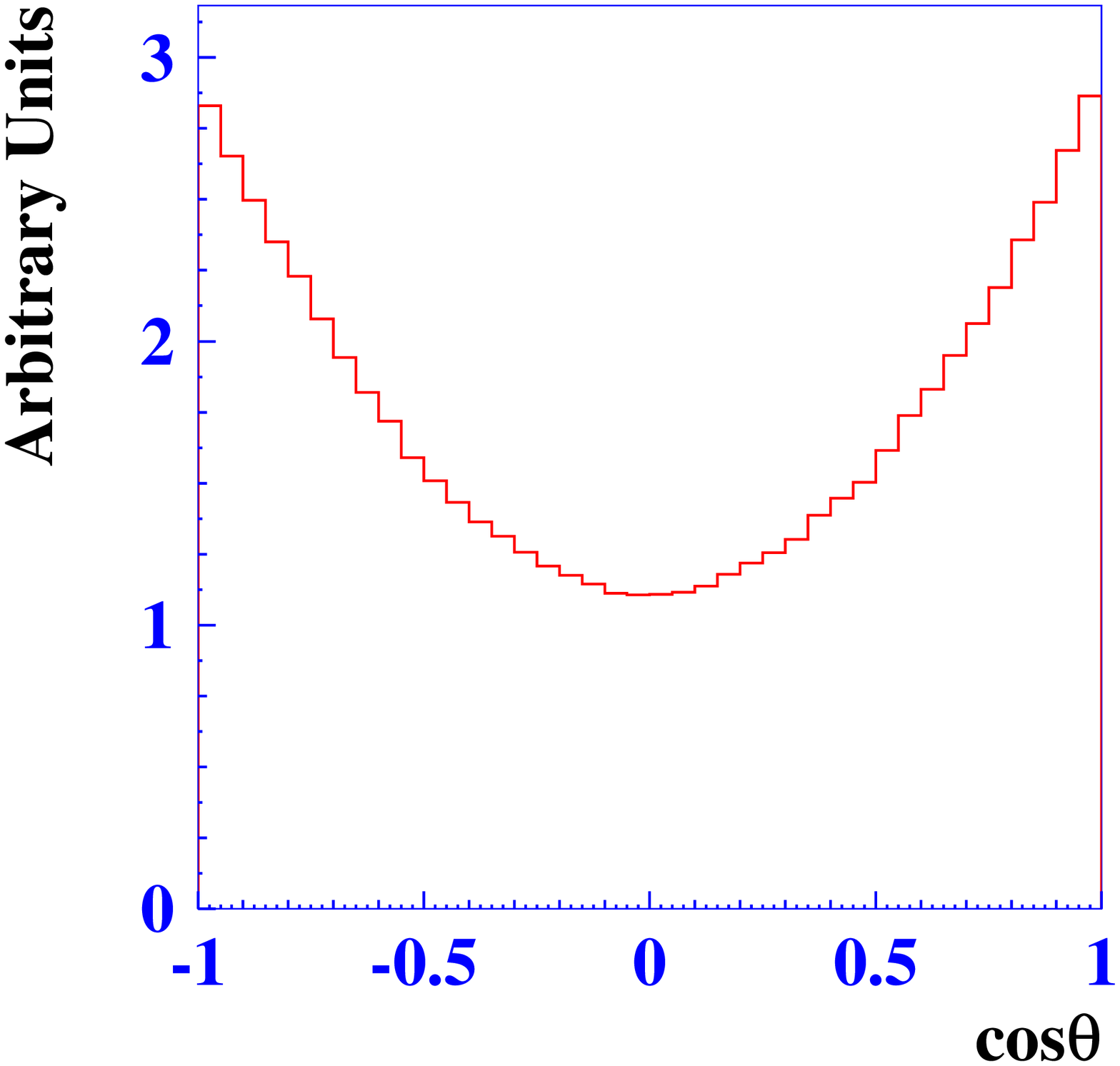}  &
\includegraphics[scale=0.18]{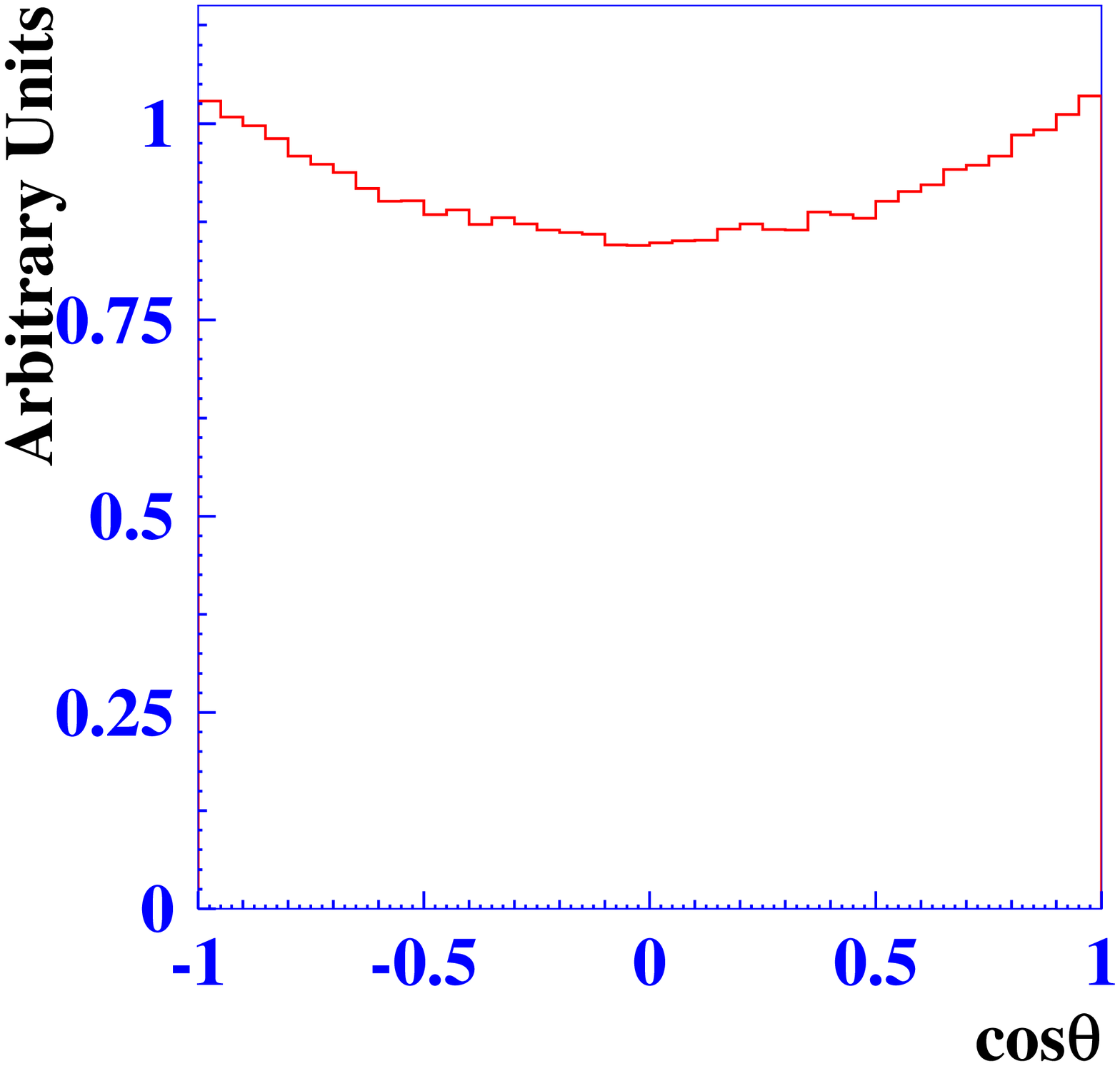}  &
\includegraphics[scale=0.18]{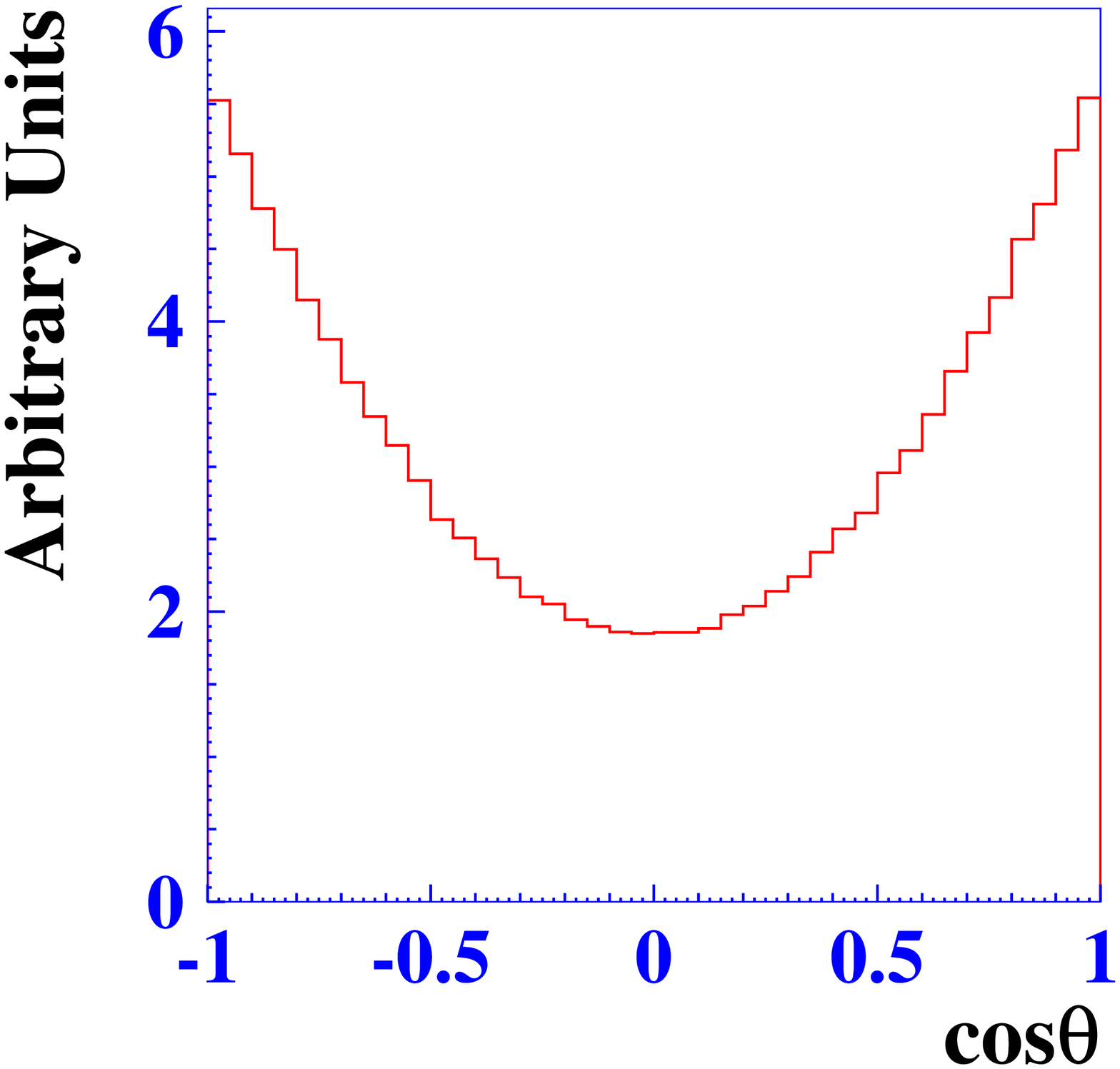}  \\
a1) & a2)  & a3) & a4) \\
\includegraphics[scale=0.18]{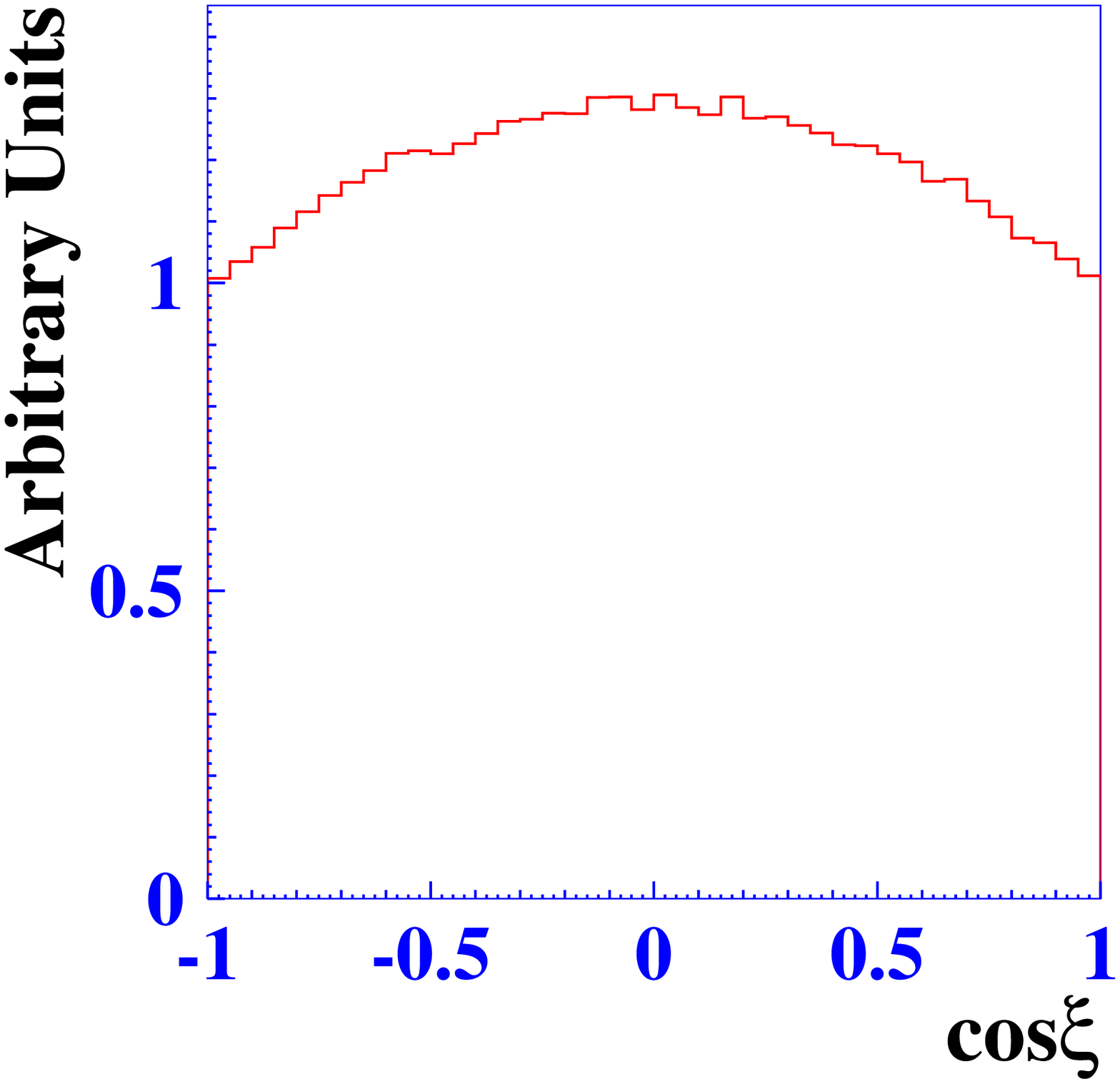}  &
\includegraphics[scale=0.18]{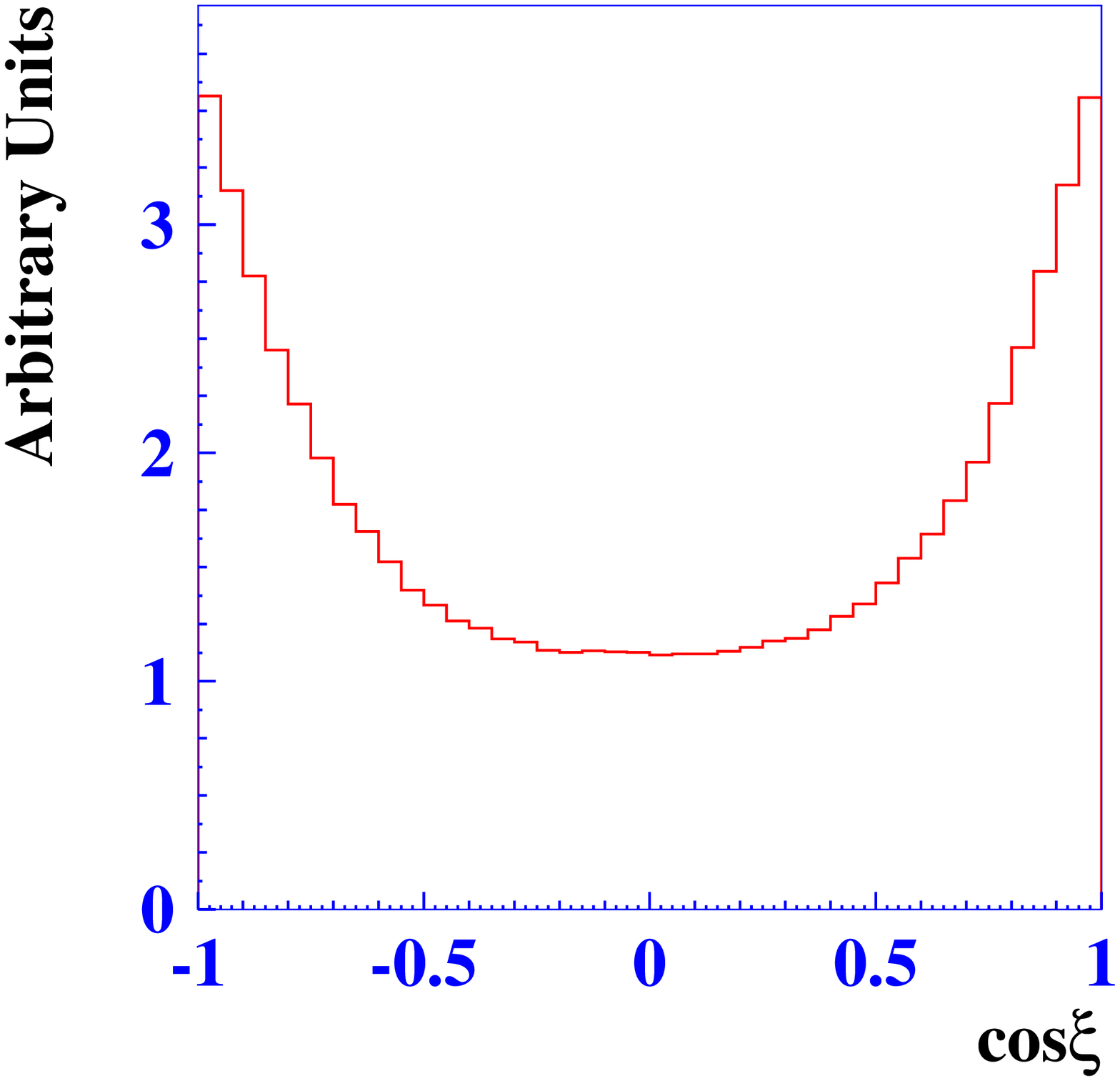}  &
\includegraphics[scale=0.18]{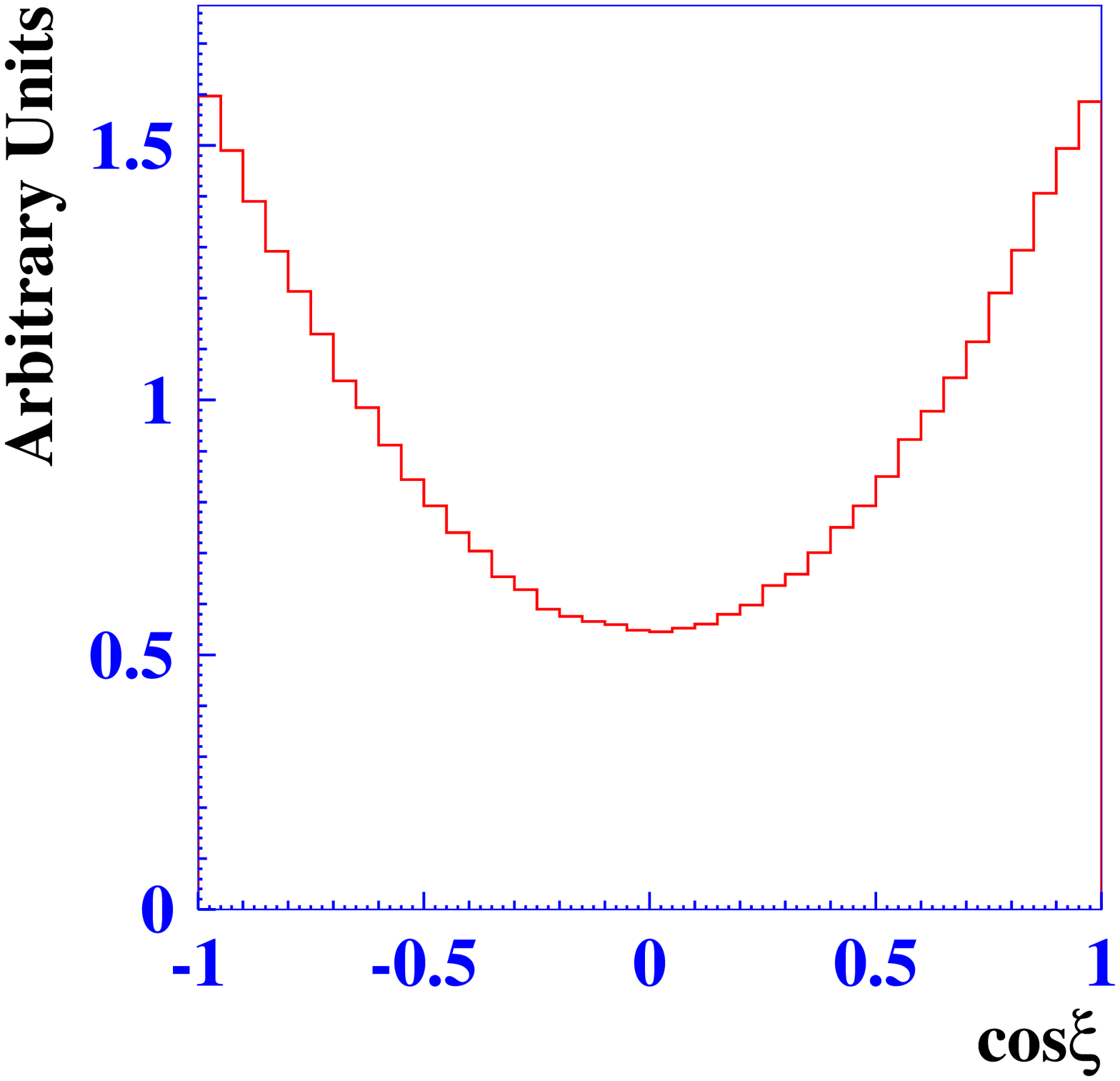}  &
\includegraphics[scale=0.18]{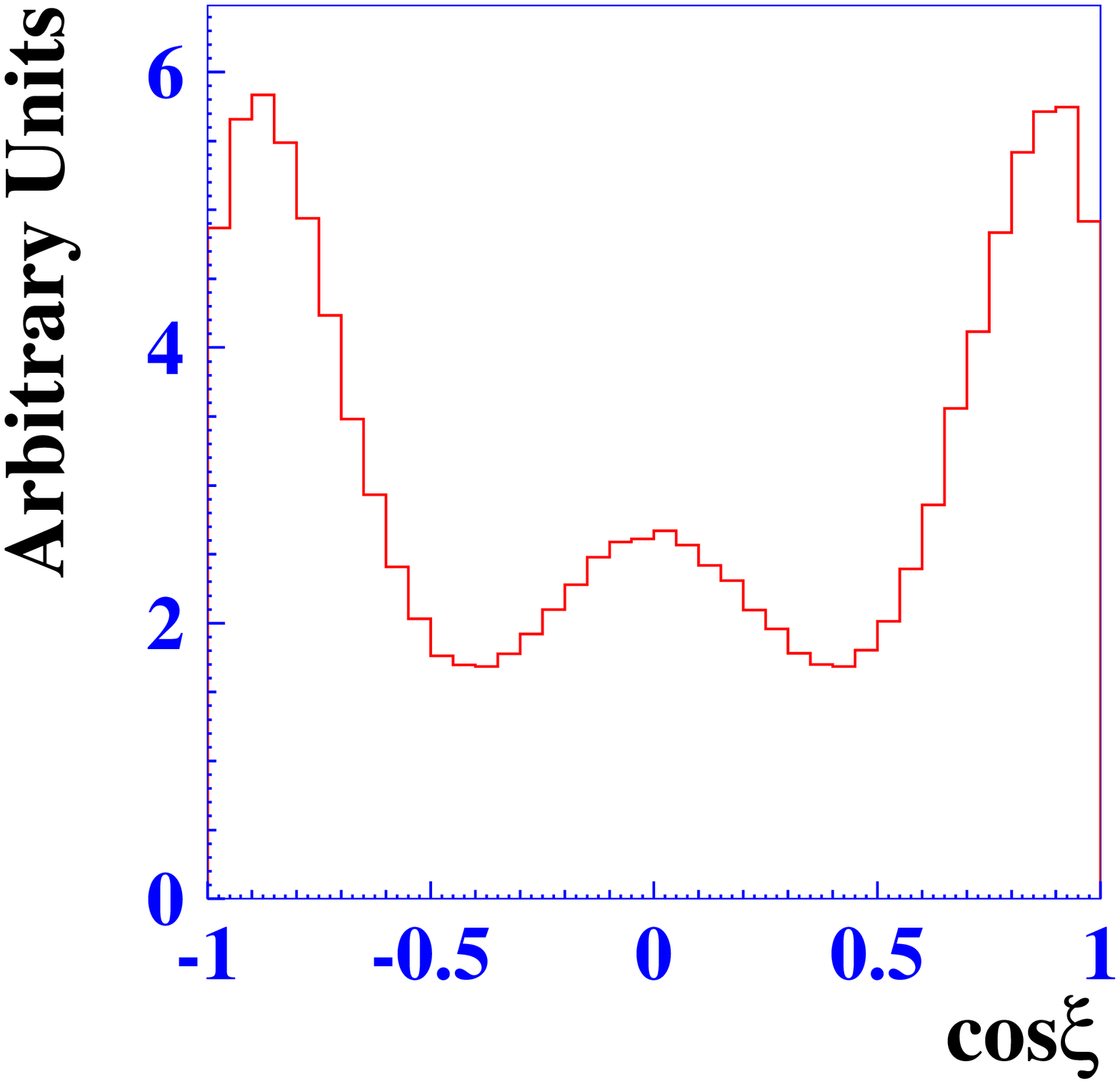} \\
b1) & b2) & b3) & b4) \\
\includegraphics[scale=0.18]{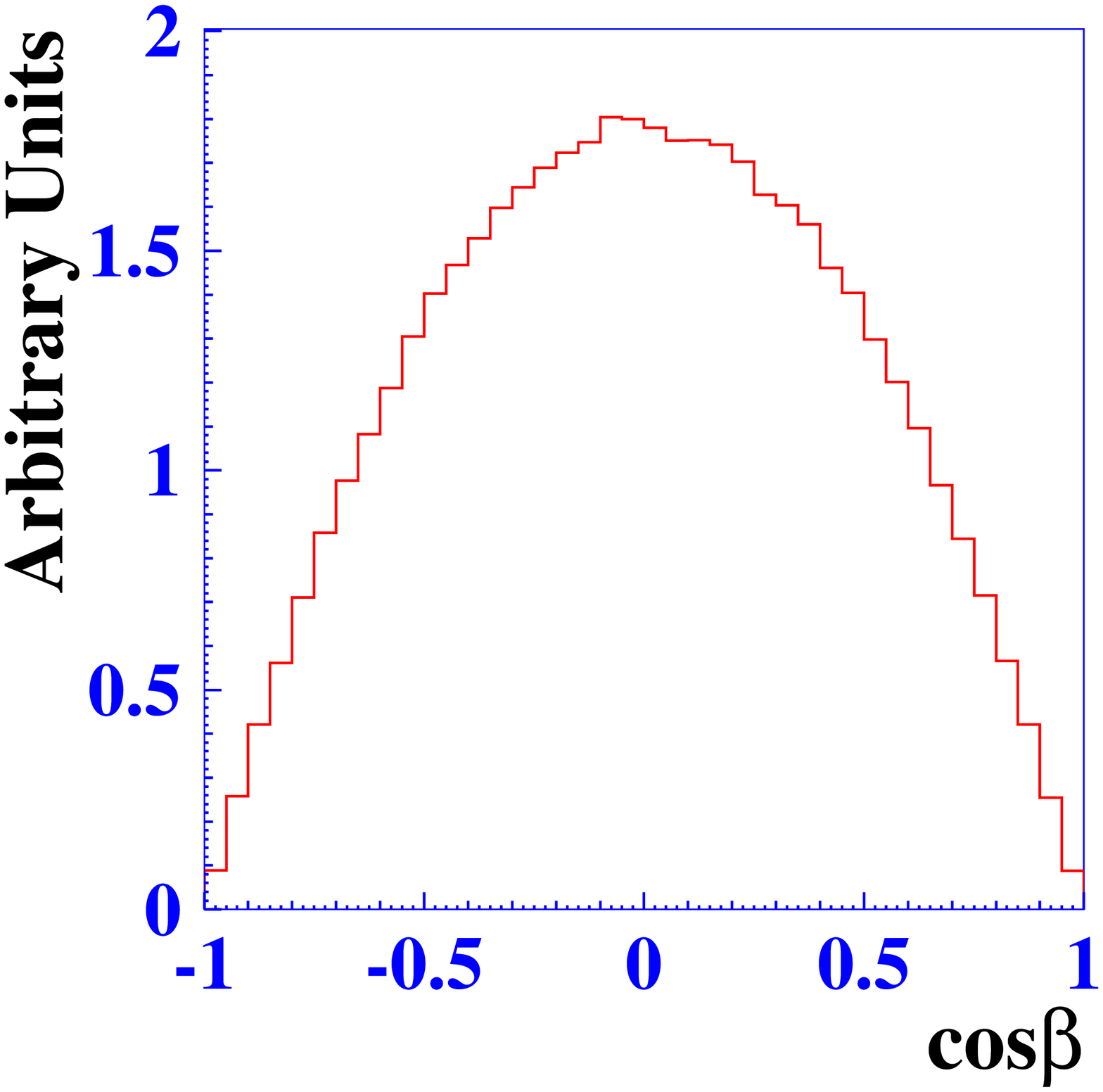}  &
\includegraphics[scale=0.18]{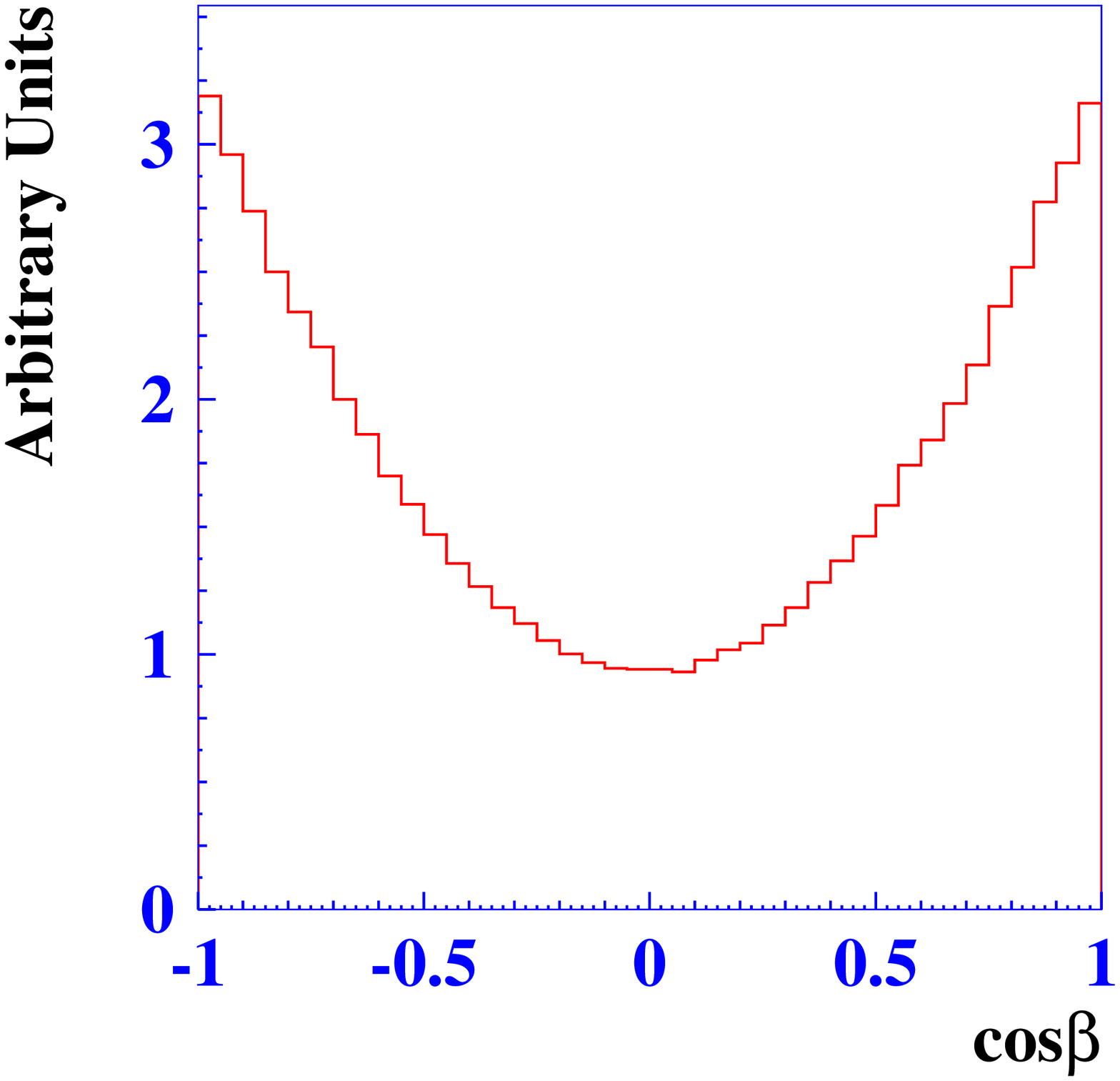} &
\includegraphics[scale=0.18]{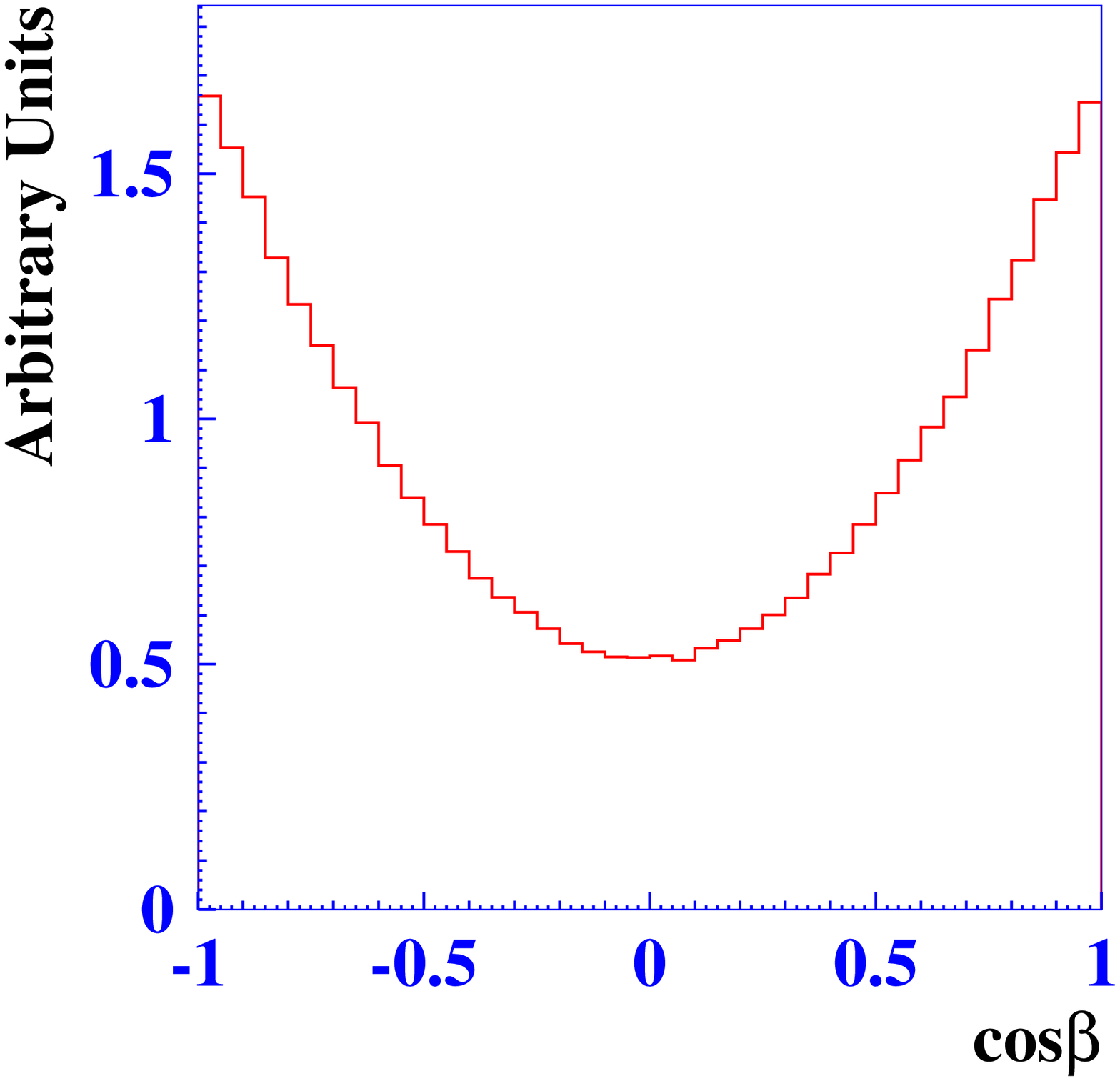}  &
\includegraphics[scale=0.18]{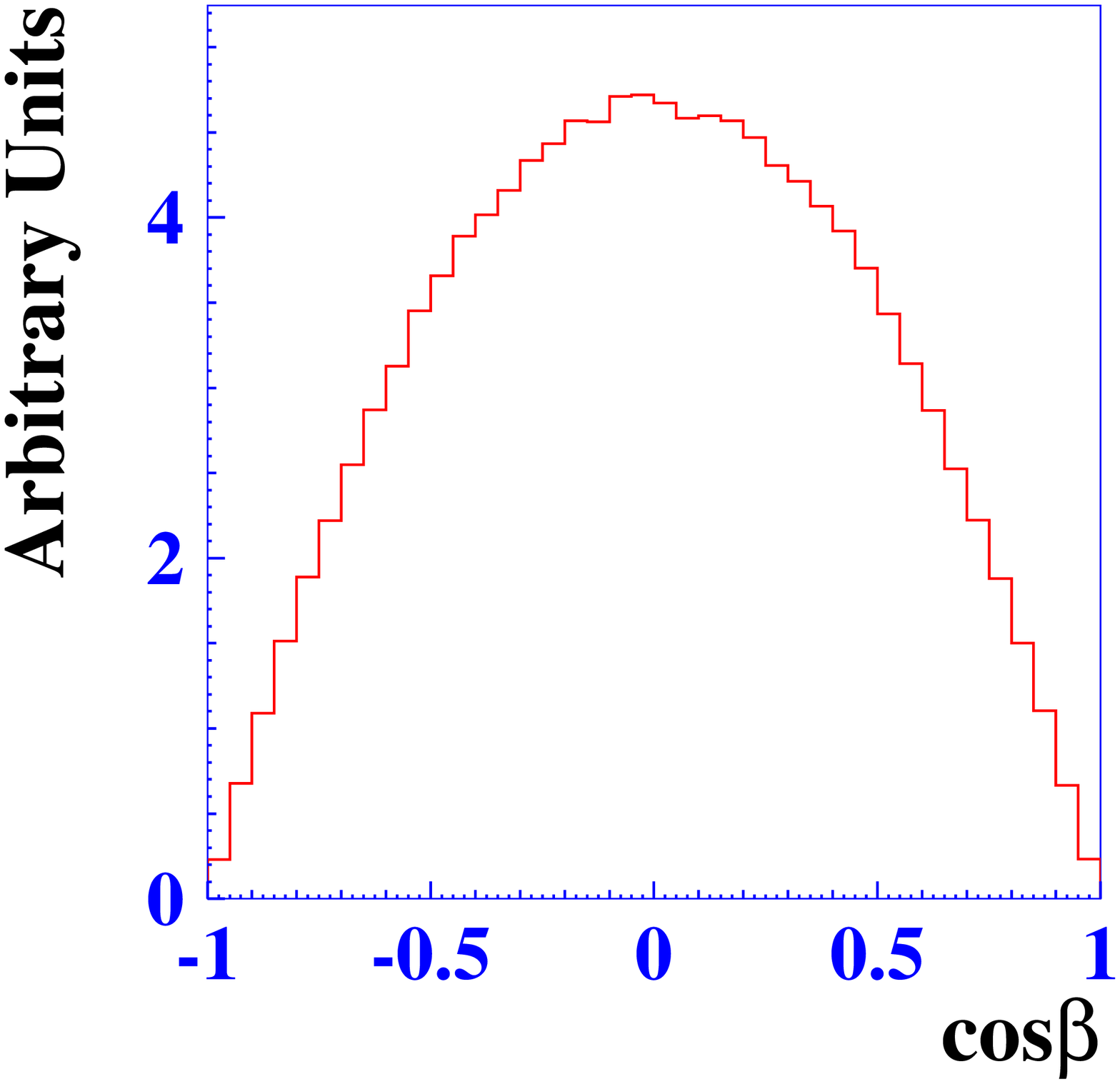} \\
c1) & c2) & c3) & c4) \\
\end{longtable}
\newpage
\begin{figure}[h]
\begin{longtable}{c c c c}
\includegraphics[scale=0.18]{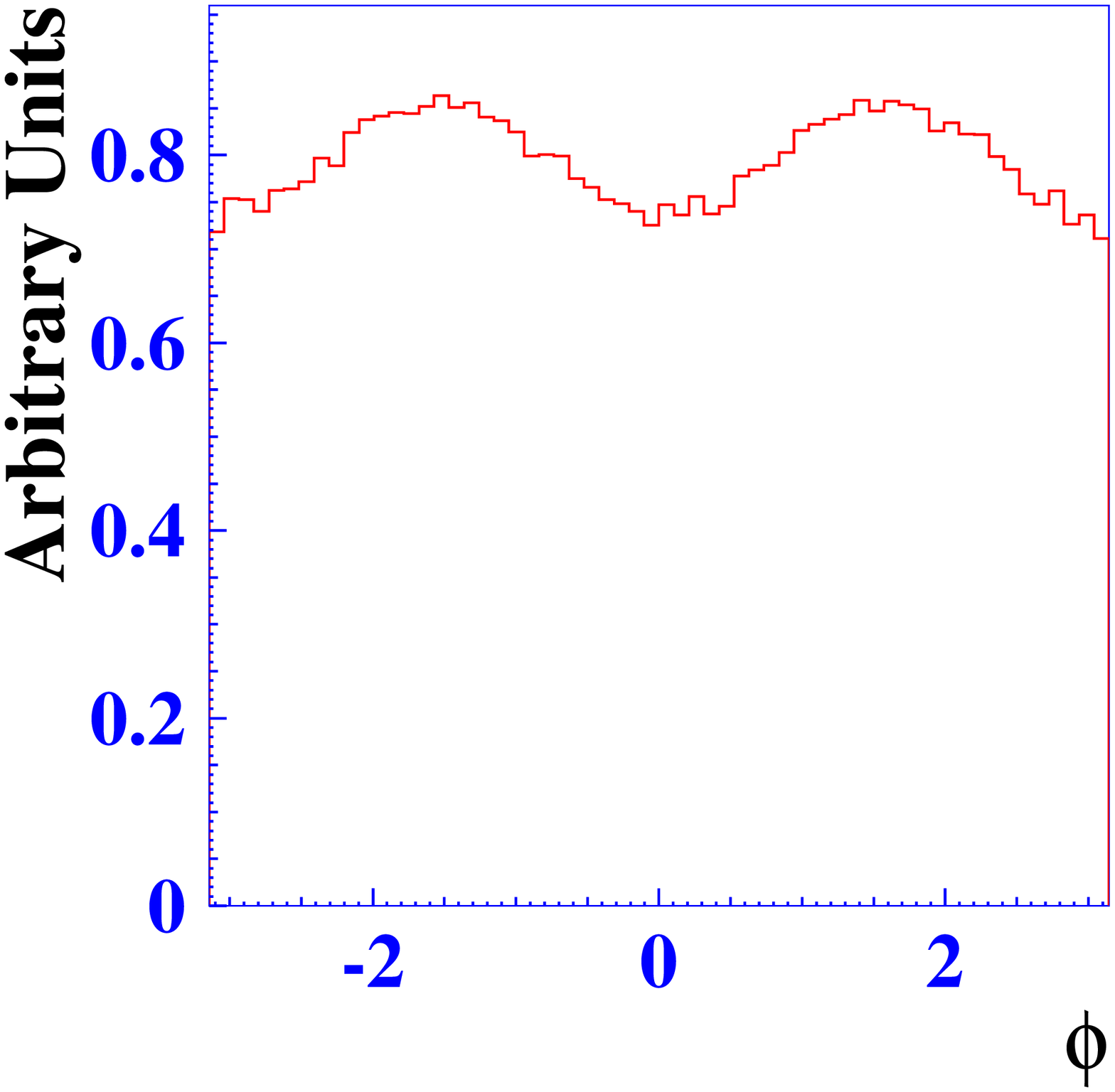}  &
\includegraphics[scale=0.18]{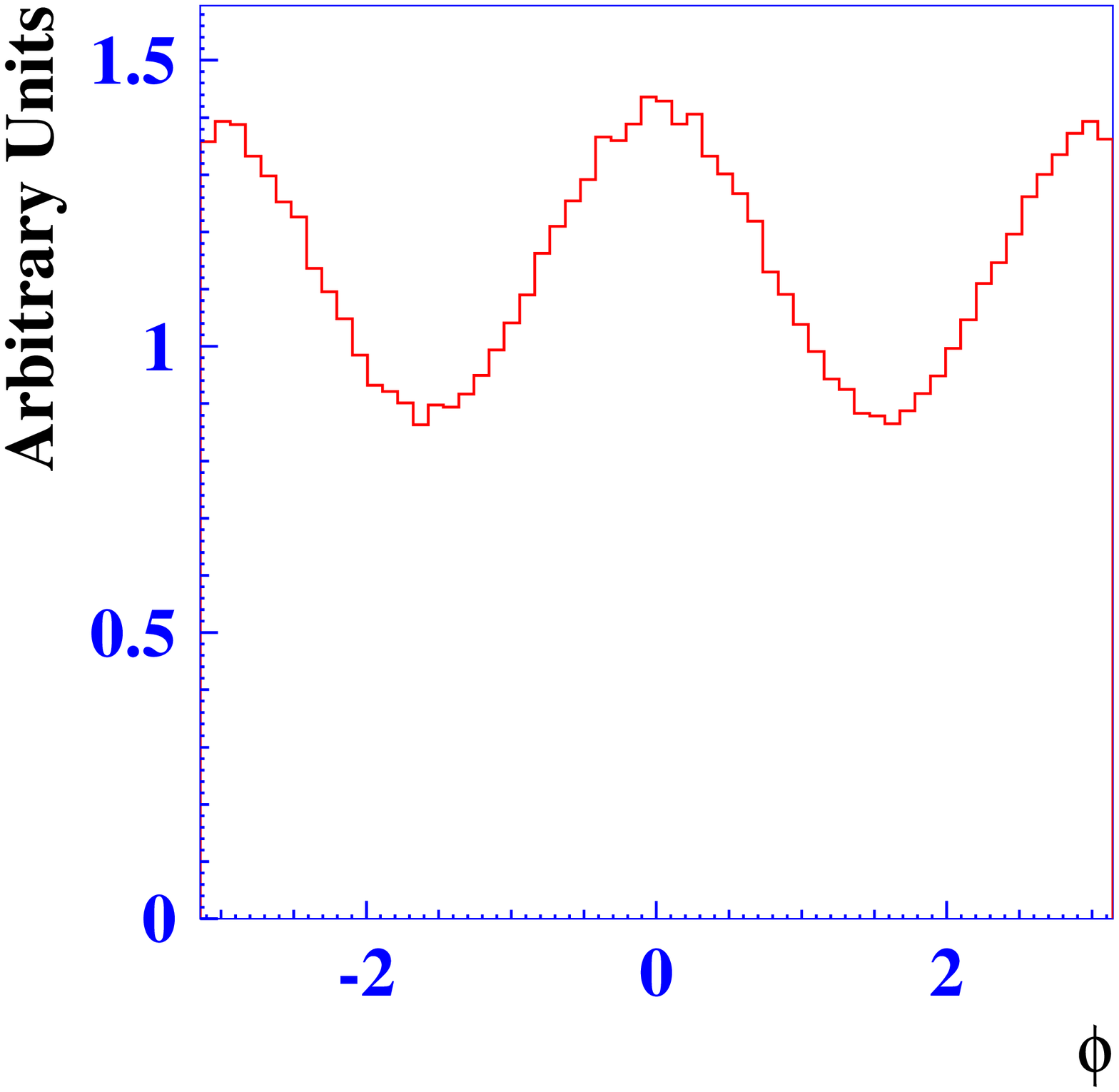} &
\includegraphics[scale=0.18]{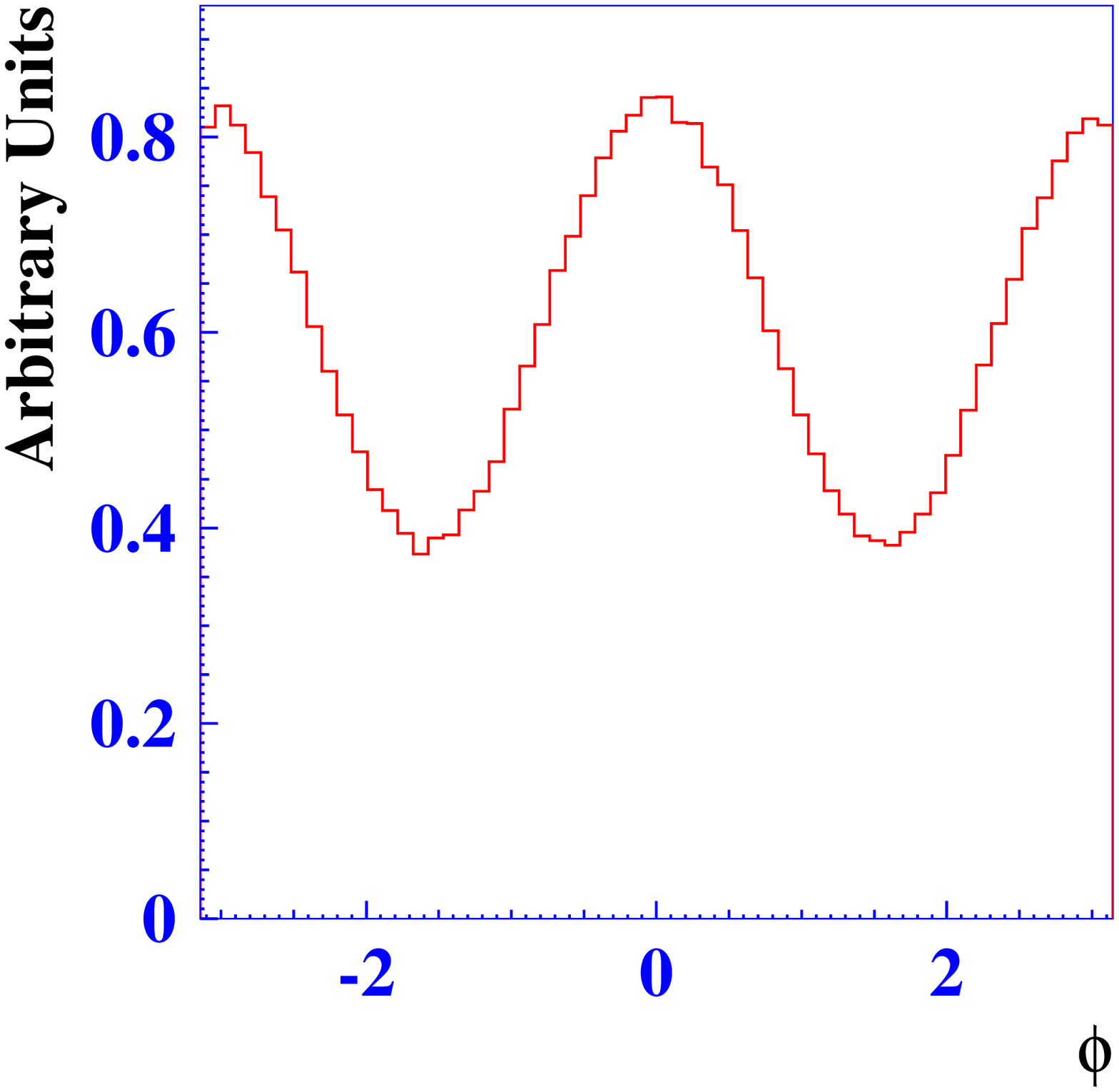}  &
\includegraphics[scale=0.18]{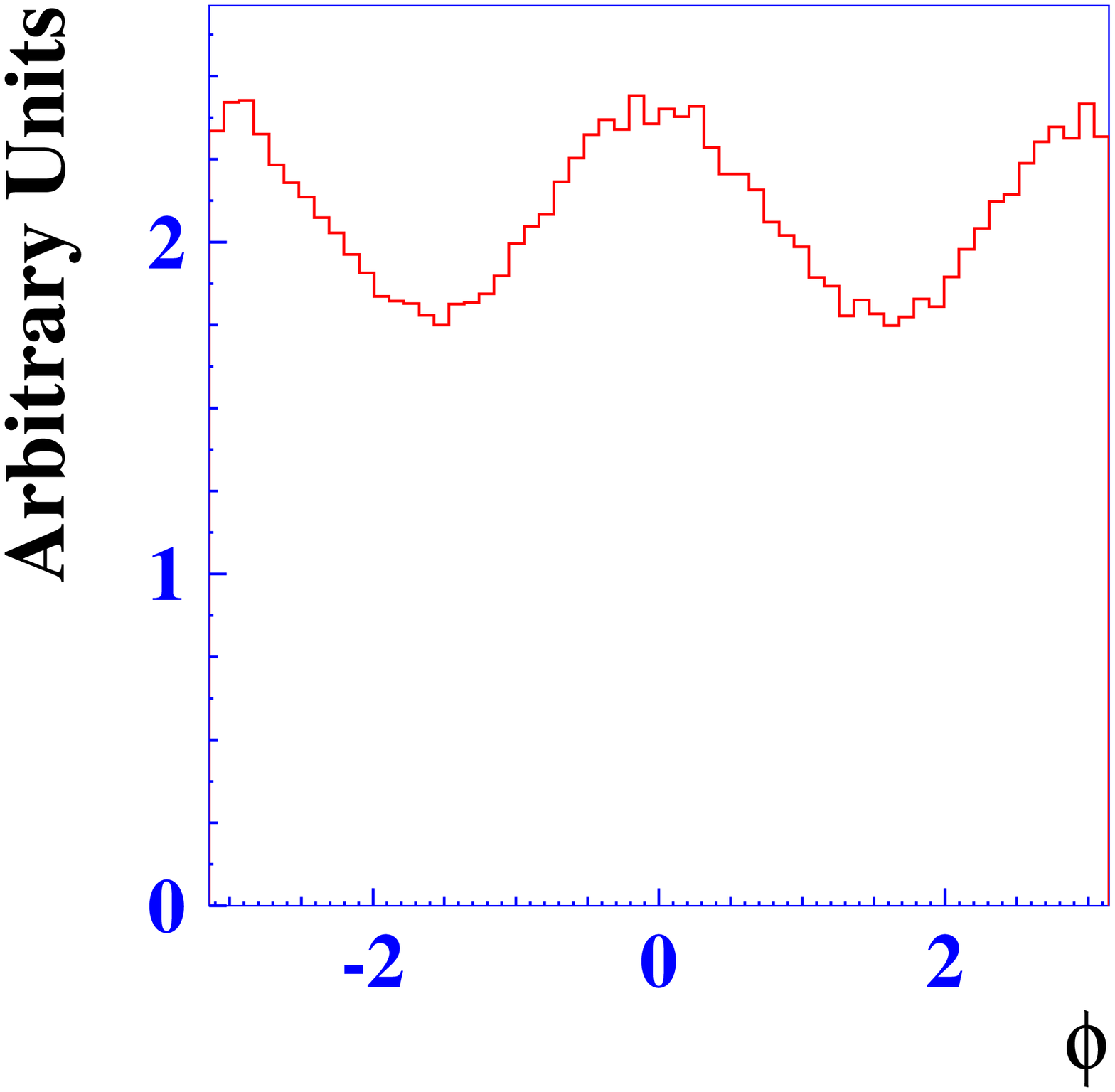} \\
d1) & d2) & d3) & d4) \\
\includegraphics[scale=0.18]{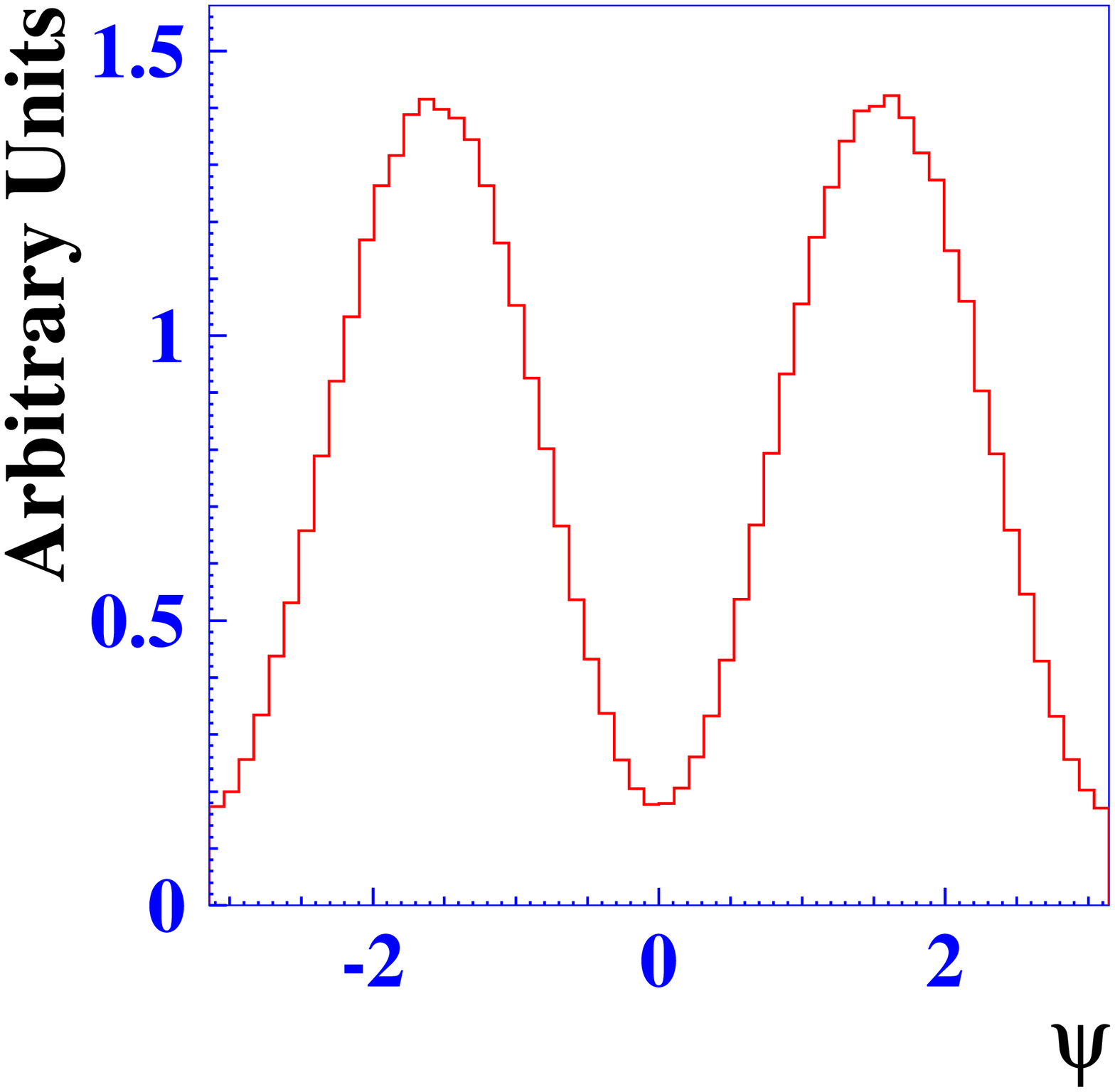}  &
\includegraphics[scale=0.18]{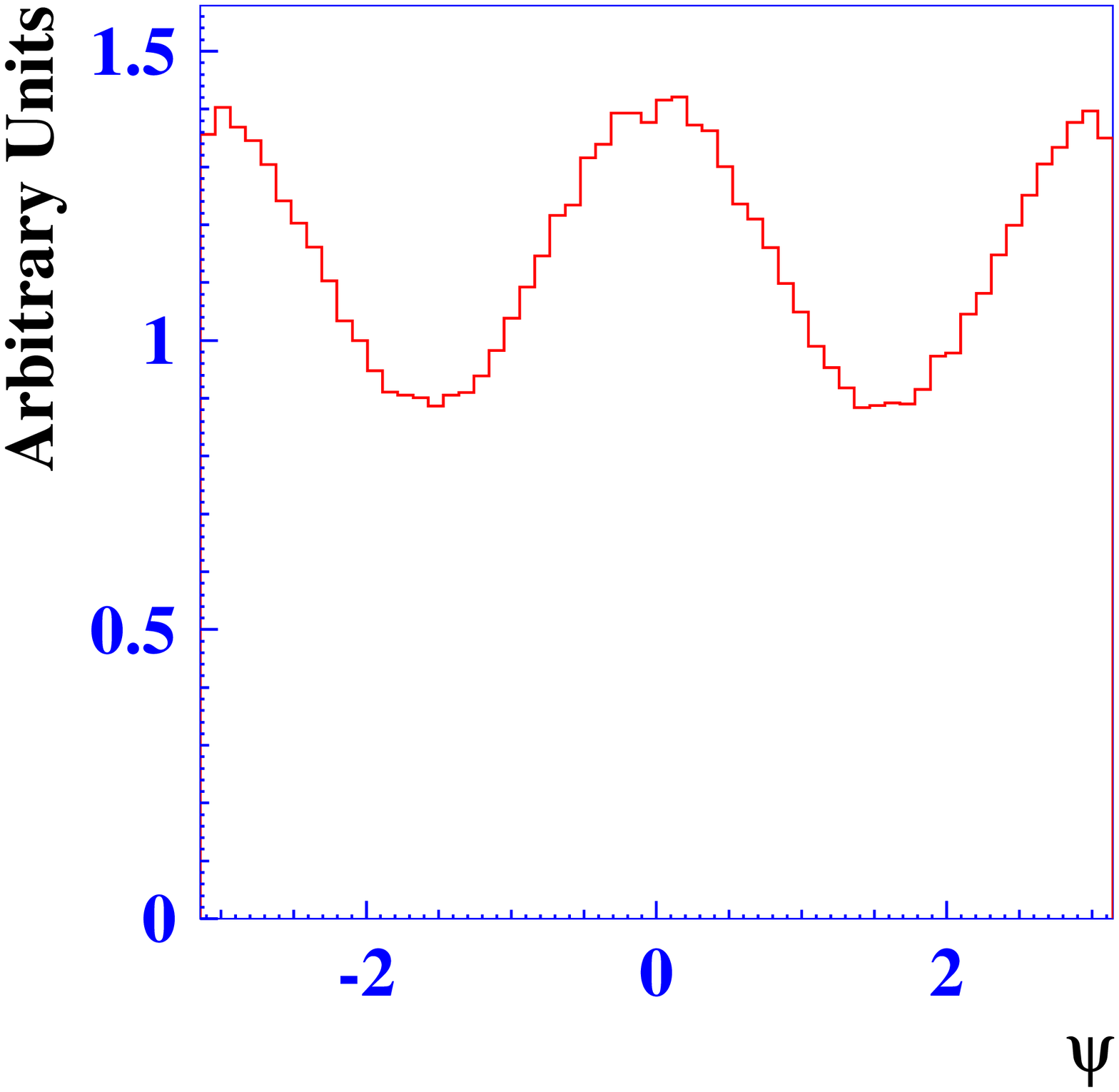} &
\includegraphics[scale=0.18]{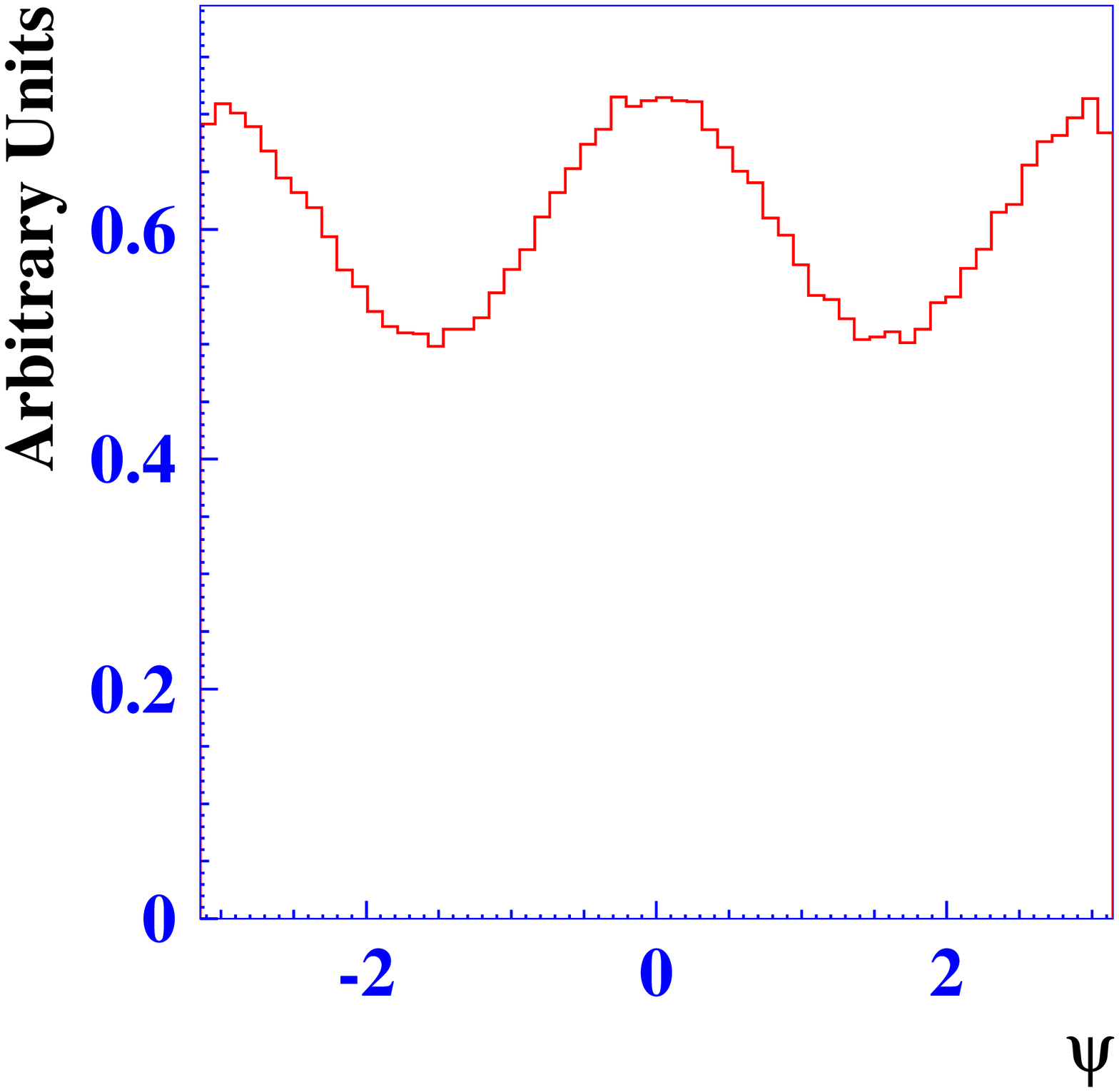}  &
\includegraphics[scale=0.18]{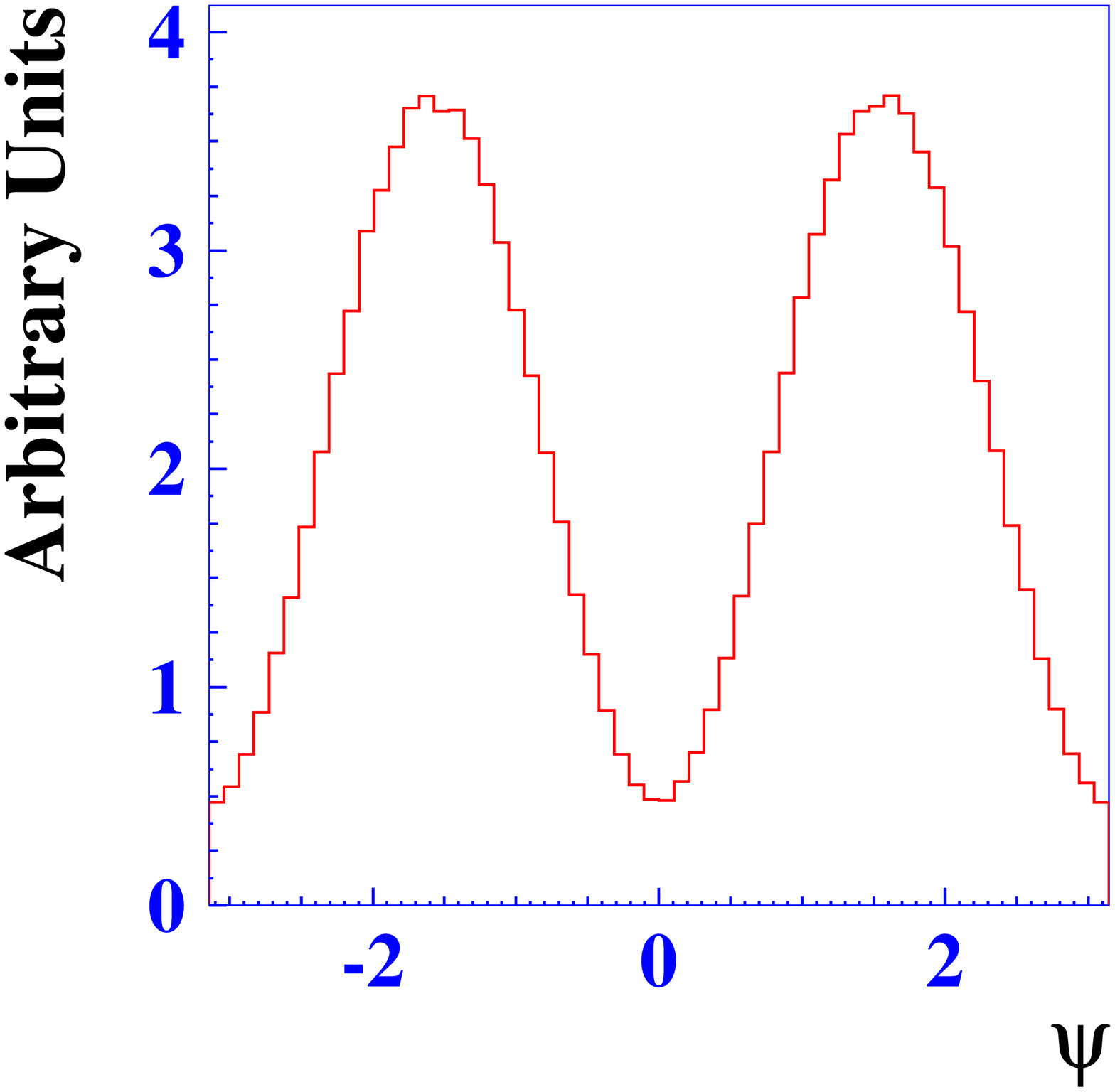} \\
e1) & e2) & e3) & e4)  \\
\end{longtable}
\caption{Simulated angular distributions for the
$D$-wave $D^{**}$-resonances. The figures a1), b1), c1), d1), e1)
correspond to the $J^P_{j_u}=1^-_{3/2}$ broad state;
a2), b2), c2), d2), e2) --- $J^P_{j_u}=2^-_{5/2}$ narrow state;
a3), b3), c3), d3), e3) --- $J^p_{j_u}=2^-_{3/2}$ broad state;
a4), b4), c4), d4), e4) --- $J^P_{j_u}=3^-_{5/2}$ narrow state.}
\label{fig8}
\end{figure}

If we consider one angular variable only, the distributions can be
the same for different resonant hypotheses.
Efficient separation between resonances is possible, when all angular
variables are taken into account.
This statement is demonstrated in Figs.~\ref{fig5} and~\ref{fig6} for the \(\omega\pi\)-
states and in Figs.~\ref{fig7} and~\ref{fig8} for the \(D^*\pi\)-states.
As mentioned above, $J^P=1^+$ $P$-wave and $J^P=2^-$ $D$-wave states
are a mixture of pure states.
However, for demonstration purposes we consider and show
angular distributions for pure states.
Moreover, we use a simple relativistic quark model of mesons
to estimate constant ratios $C_{J-1}/C_J$ and $C_{J+1}/C_J$ in (\ref{bdstr}),
which are responsible for relative contributions of amplitudes
with different orbital momenta in the total matrix element~\cite{wsb}.
The constant ratios for all discussed states are chosen roughly
as follows: \(C_{J-1}/C_{J}=3/2,\,C_{J+1}/C_{J}=2\).

For Dalitz plot analysis, interference between resonances should be taken
into account. For a one-dimensional distribution, an interference term
for resonances, which decay to the same final state, can cancel out
after integration over other variables. However, in a real experiment
such cancellation can disappear due to the nonuniform detection efficiency,
so that a finite interference term can be observed.
For resonances, which decay to the different final states \(\omega\pi\)
and \(D^*\pi\), the interference term cannot be neglected.
For demonstration purpose we show distributions between
\(b_1(1235)^-\) and pure \(D^0_1\) as well as \(\rho(1450)^-\) and pure \(D^{'0}_1\).
However, there is possible interference between the resonant and
non-resonant structures.

For simulation we use BW functions
for \(b_1(1235)^-\), \(\rho(1450)^-\), \(D^0_1\) and \(D^{'0}_1\). Thus,
the \(q^2\)-dependent widths have to be obtained.
For a $q^2$-dependent width of the $b_1$ we consider the dominant decay
to the $\omega\pi$~\cite{pdg}:
\begin{align}
\Gamma_{b_1}(q^2)\,&=\,\frac{m_{b_1}}{\sqrt{q^2}}\frac{m^4_{b_1} \tilde{F}^2_S(q^2)+b_D \tilde{f}^2_{1,2}(q^2)\tilde{F}^2_D(q^2)}{m^4_{b_1} + b_D \tilde{f}^2_{1,2}(m^2_{b_1})}\frac{\mathbf{p}}{\mathbf{p}_{0}}\Gamma_{b_1}{,}
\end{align}
where \(\mathbf{p}_{0}\) is the magnitude of the \(\omega\) momentum in the
resonance rest frame, when  $q^2=m^2_{b_1}$. Here, we use the fact that
the experimental ratio of the amplitudes with \(L=2\) and \(L=0\)
is about \(0.3\)~\cite{pdg} and thus the constant \(b_D \approx 53\).
For a $q^2$-dependent width of the pure $D_1$ and pure $D'_1$ we consider
the decay to the $D^*\pi$:
\begin{align}
\Gamma_{D_1,D'_1}(q^2)\,&=\,\frac{m_{D_1,D'_1}}{\sqrt{q^2}}\tilde{F}^2_{D,S}(q^2)\frac{\mathbf{Q}}{\mathbf{Q}_{0}}\Gamma_{D_1,D'_1}{,}
\end{align}
where \(\mathbf{Q}_{0}\) is the magnitude of the \(D^*\) momentum in
the resonance rest frame, when  $q^2=m^2_{D_1,D'_1}$.
For a $q^2$-dependent width of the $\rho(1450)$ we consider its decays
into the $a_1(1260)\pi$ and $\omega\pi$ modes:
\begin{align}
\Gamma_{\rho(1450)}\,&=\,(1-a)\frac{m_{\rho(1450)}}{\sqrt{q^2}}\frac{m^4_{\rho(1450)}\tilde{F}^2_S(q^2)+b_D \tilde{f}^2_{1,2}(q^2)\tilde{F}^2_D(q^2)}{m^4_{\rho(1450)} + b_D \tilde{f}^2_{1,2}(m^2_{\rho(1450)})}\frac{\mathbf{k}_{(a_1)}}{\mathbf{k}_{0(a_1)}}\Gamma_{\rho(1450)} + \nonumber \\& +
a\frac{\sqrt{q^2}}{m_{\rho(1450)}}\tilde{F}^2_P(q^2)\frac{\mathbf{p}^3}{\mathbf{p}^3_{0}}\Gamma_{\rho(1450)}{,}
\end{align}
where the parameter $a=2/5$,
when $\sqrt{q^2} > m_{a_1}+m_{\pi^-}$~\cite{akhmet}, and $a=1$,
when $\sqrt{q^2} \leq m_{a_1}+m_{\pi^-}$, $\mathbf{k}_{(a_1)}$ is the momentum
of the $a_1$ in the $\rho(1450)$ rest frame,
$\mathbf{k}_{0 (a_1)}$ is the same momentum, when $\sqrt{q^2}=m_{\rho(1450)}$.
Here, we use the fact that the experimental ratio of the amplitudes
with \(L=2\) and \(L=0\) in the \(a_1(1260) \to \rho \pi\) decay
is about \(-0.06\)~\cite{pdg} and accept the same value for the
\(\rho(1450) \to a_1(1260) \pi\) decay. Thus, we can estimate the constant
\(b_D \approx 182\).

\begin{center}
\begin{figure}[h]
\begin{tabular}{c c}
\includegraphics[scale=0.35]{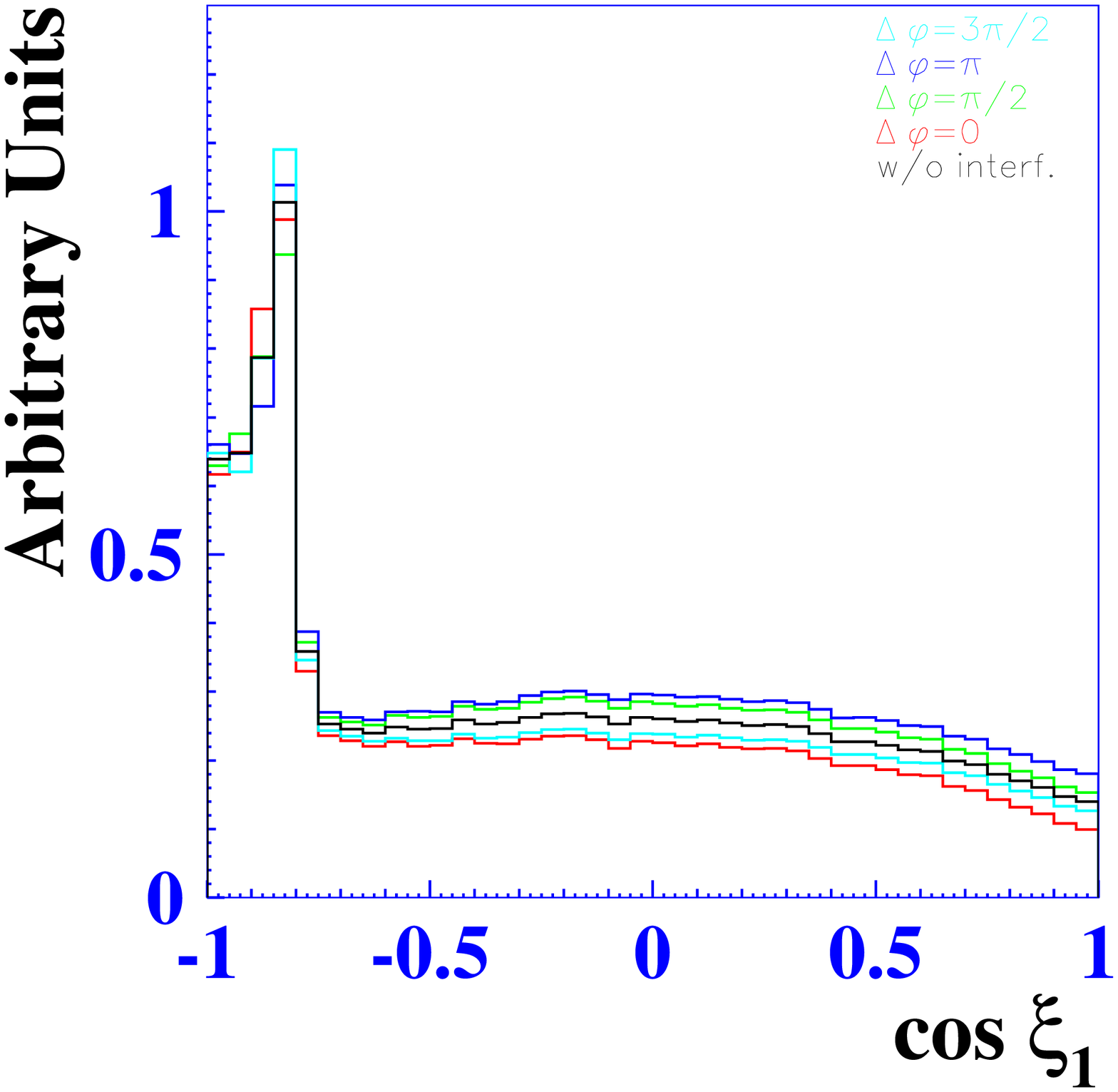} &
\includegraphics[scale=0.35]{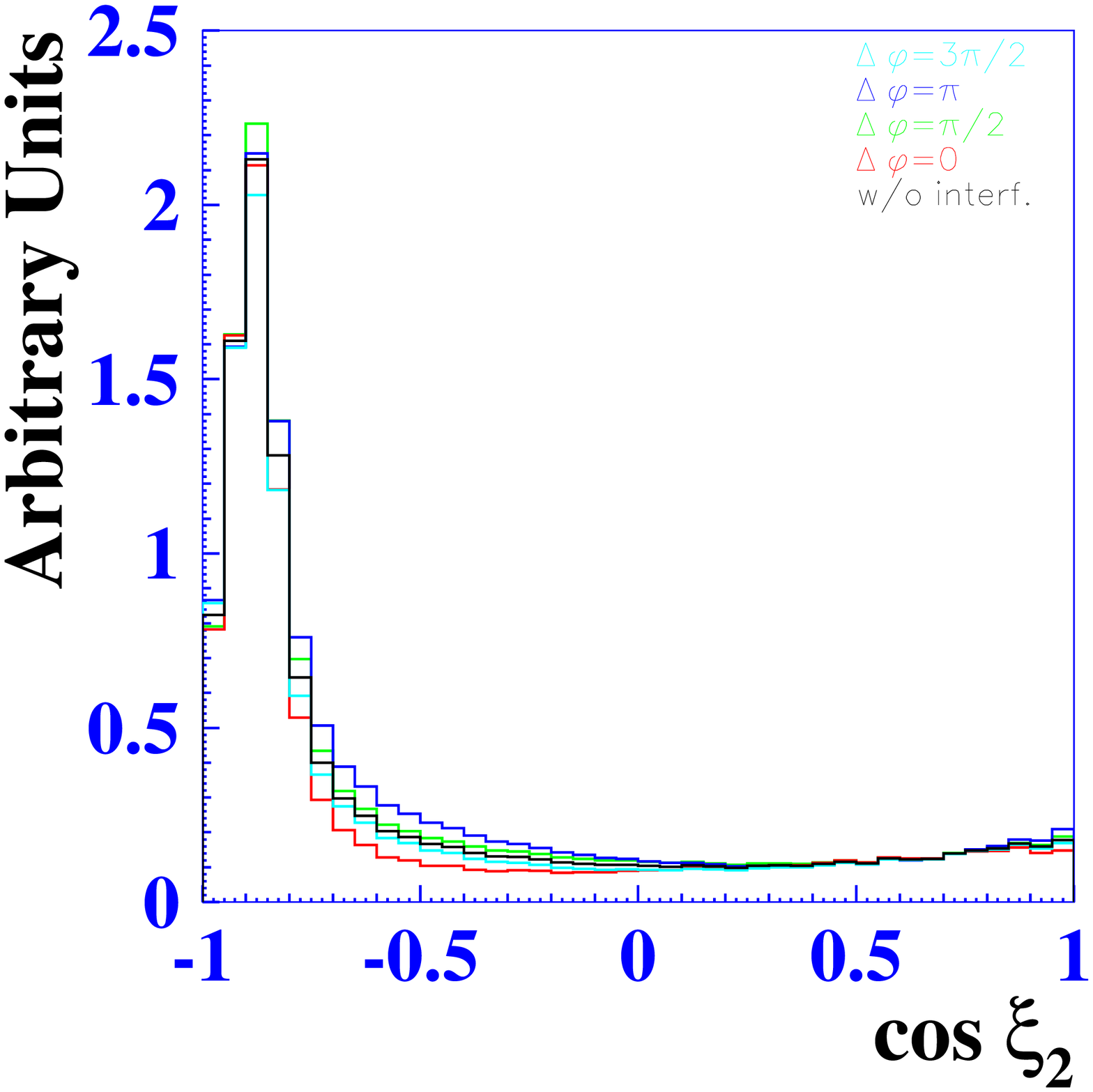} \\
a1) \(b_1\,D_1\) &  a2) \(b_1\,D_1\) \\
\includegraphics[scale=0.35]{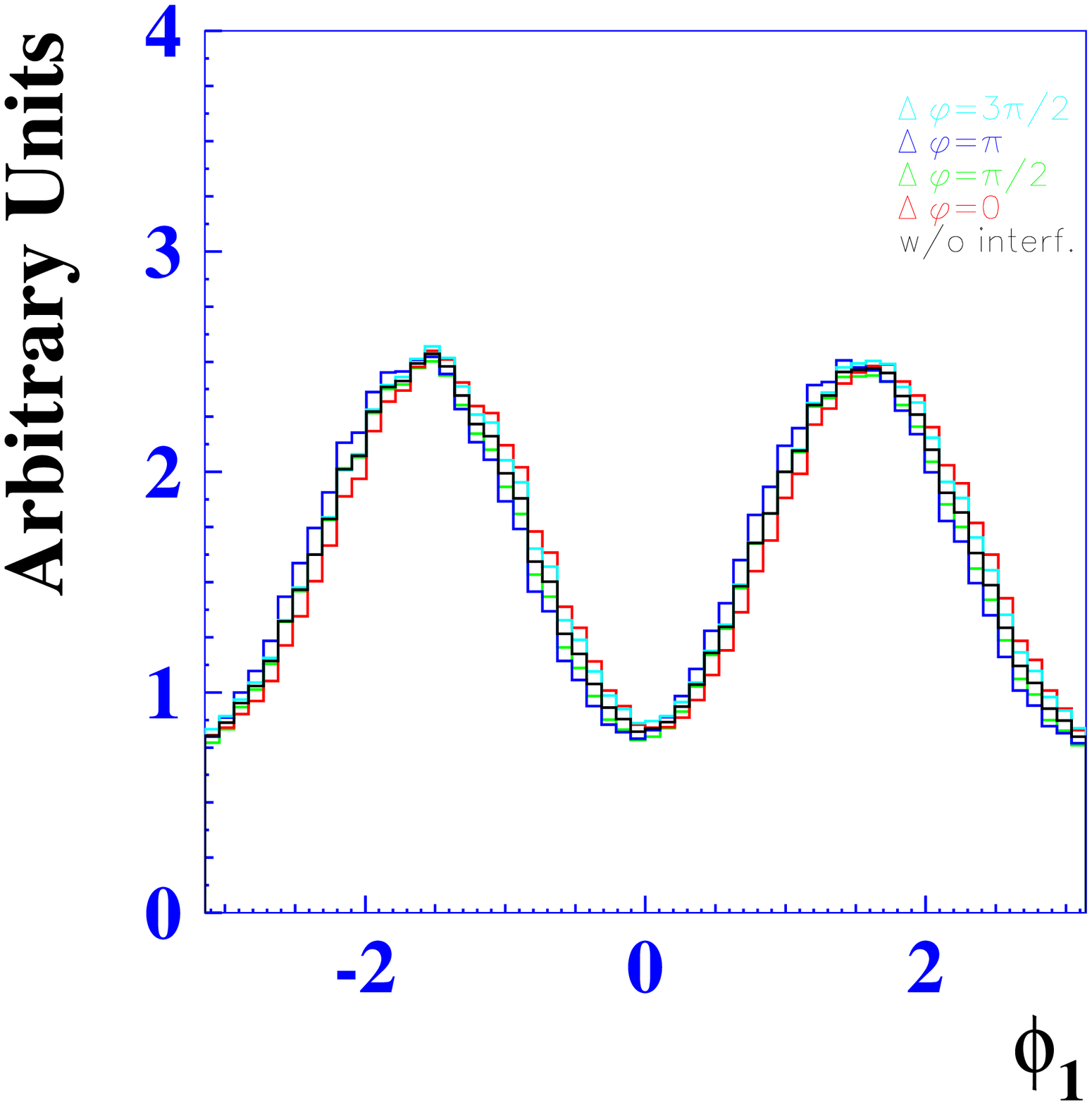} &
\includegraphics[scale=0.35]{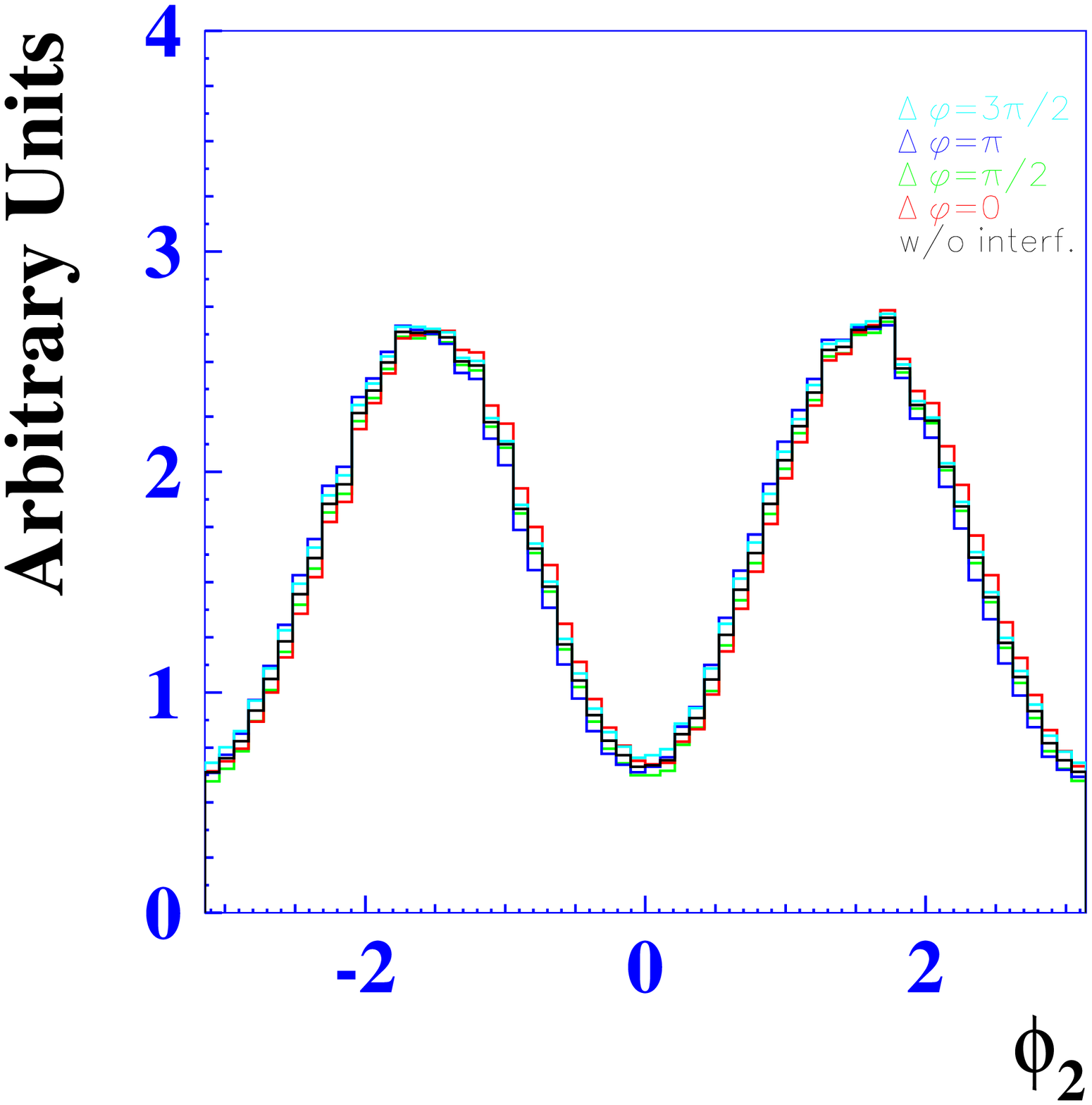} \\
b1) \(\rho' \,D'_1\) & b2) \(\rho' \,D'_1\) \\
\end{tabular}
\caption{Demonstrative interference distributions.
The figures a1), a2) correspond to the distributions over
angles \(\cos\xi_1\) and \(\cos\xi_2\) for interference between
the \(b_1(1235)^-\) and pure \(D^{0}_1\); b1), b2) correspond to the
distributions over angles \(\phi_1\) and \(\phi_2\) for
interference between the \(\rho(1450)^-\) and pure \(D^{'0}_1\).
The subscripts \(1\) and \(2\) correspond to the \(\omega\pi\)- and
\(D^{**}\)-resonances, respectively.}
\label{fig9}
\end{figure}
\end{center}

Obviously, it is impossible to analyse spectra without knowledge of
the relative phases in the amplitude.
Thus, in Fig.~\ref{fig9} we show some typical distributions for different
relative phases \(\Delta \varphi\), such as \(0\), \(\pi/2\), \(\pi\),
\(3\pi/2\) and the distribution without interference. The relative
constant amplitudes between resonant matrix elements squared are chosen
of one order of magnitude for the \(b_1(1235)^-\) and \(D^{0}_1\)
for simplicity and
one order of magnitude smaller for the \(D^{'0}_1\) than for
the \(\rho(1450)^-\) according to experiment~\cite{babar}.
Although small, the interference effects are not negligible.
\newpage


\section{Conclusion}

We have described a model of the
\(\bar{B}^0 \to D^{*+}\omega\pi^-\) decay, 
in which a total amplitude is a sum of contributions of different
intermediate states.  
In our study we consider different resonant contributions to the matrix element,
such as light \(\omega\pi\)-hadrons with the spin-parities of
\(J^P=0^-,1^{\pm},2^{\pm},3^-\),
and heavy-light hadrons, which are excitations of the $c\bar{u}$-states
in $P$- and $D$-waves.
All resonances are described by the
relativistic Breit-Wigner factors. The resonant matrix elements are
parameterized in the angular basis, which is convenient for the
experimental Dalitz plot analysis and is natural from the physical
point of view.

Monte-Carlo simulation based on the obtained expressions has been performed.
The angular distributions obtained for the listed above intermediate states
and their interference effects are demonstrated.

\section*{Acknowledgments}
This work was supported in part by the RFBR grants 11-02-112-a, 
11-02-90458-a, and grant DFG GZ: HA1457/7-1. 

\appendix
\section{Appendix}

In this section we present the phase integral \(W(p^2)\)
at the decay rate defined by (\ref{decompi}).
\setcounter{equation}{0}
\renewcommand{\theequation}{A\arabic{equation}}
The integral
\begin{align}
W(p^2)\,&=\,\pi\int_{(2m_+)^2}^{(\sqrt{p^2}-m_0)^2}dM^2_0\int_{M^2_{+\,min}}^{M^2_{+\,max}}dM^2_+\frac{\varDelta(p,P_+,P_0)}{p^2}g^2_{\omega\rho\pi}(p^2)\times \nonumber \\ & \times \left|a_{3\pi}+\sum_{i=\pm,0}\frac{g_{\rho\pi\pi}}{D_{\rho^i}(M^{2}_i)Z(M^{2}_i)}\right|^2
\end{align}
is a standard phase space factor for $\omega$-decay~\cite{achasov}.
Here, the Kibble determinant $\varDelta(p,P_+,P_0)$, which zeros
determine the phase-space boundary, is presented as follows:
\[\varDelta(p,P_+,P_0)\,=\,
\begin{vmatrix}
p^2 & pP_+ & pP_0 \\
pP_+ & m^2 & P_+P_0 \\
pP_0 & P_+P_0 & m^2_0
\end{vmatrix}{,}\]
where $m$ and $m_0$ are the charged and neutral pions masses, respectively;
the range limits of $M^2_+$ are
\begin{eqnarray}
M^2_{+\,min}\,&=&\,(E^{(+ -)}_++E^{(+ -)}_0)^2-(\sqrt{E^{(+ -)\,2}_+-m^2}+\sqrt{E^{(+ -)\,2}_0-m^2_0})^2{,}\nonumber\\
M^2_{+\,max}\,&=&\,(E^{(+ -)}_++E^{(+ -)}_0)^2-(\sqrt{E^{(+ -)\,2}_+-m^2}-\sqrt{E^{(+ -)\,2}_0-m^2_0})^2{,}
\end{eqnarray}
where
\begin{eqnarray}
E^{(+ -)}_+\,&=&\,\frac{M^2_0 - 2m^2}{2M_0}{,}\nonumber\\
E^{(+ -)}_0\,&=&\,\frac{p^2-M^2_0 - m^2_0}{2M_0}
\end{eqnarray}
are the energies of $\pi^+$ and $\pi^0$ from $\omega$ decay in
the $\pi^+\pi^-$ rest frame;
\begin{equation}
\label{gomrhopi}
g_{\omega\rho\pi}(p^2)\,=\,g_{\omega\rho\pi}(m^2_{\omega})\frac{1+(m_{\omega}r)^2}{1+(\sqrt{p^2}r)^2}
\end{equation}
is the form factor which restricts too fast growth of the width
$\Gamma_{\omega}(p^2)$ with $p^2$, so that
$\Gamma_{\omega}(p^2)\to \mathrm{const}$ as $p^2 \to \infty$ (here $r$ is a hadron scale)~\cite{nachasov};
the quantity $g_{\omega\rho\pi}(m^2_{\omega})\simeq 16 \,\mathrm{GeV}^{-1}$~\cite{cgomrhopi,lubl,lnsv};
the $a_{3\pi}$ and $g_{\rho\pi\pi}$ amplitudes are assumed to be
real constants and, thus,
$a_{3\pi}=(0.01\pm0.23\pm0.25)\times10^{-5}\,\mathrm{MeV}^{-2}$~\cite{achasov} and $g_{\rho\pi\pi}\simeq 6$ \cite{achasov,cgomrhopi,lnsv};
\begin{equation}
D_{\rho^{\pm,0}}(M^2_{\pm,0})\,=\,M^2_{\pm,0}-m^2_{\rho^{\pm,0}}+i m_{\rho{\pm,0}}\Gamma_{\rho^{\pm,0}}(M^2_{\pm,0}){,}
\end{equation}
where
\begin{eqnarray}
M^2_{-}\,&=&\,p^2-M^2_+-M^2_0 + m^2_0 + 2m^2{,}\nonumber\\
\Gamma_{\rho^{\pm, 0}}(M^2_{\pm, 0})\,&=&\,\frac{m_{\rho^{\pm, 0}}}{M_{\pm, 0}}\left(\frac{k_{\pm, 0}(M^2_{\pm,0})}{k_{\pm, 0}(m^2_{\rho^{\pm,0}})}\right)^3\Gamma_{\rho^{\pm,0}}{,}
\end{eqnarray}	
$k_i$ is an absolute value of pion momentum in the $\pi^+\pi^-$ rest
frame for $i=0$, in the $\pi^+\pi^0$ rest frame for $i=+$ and
in the $\pi^-\pi^0$ rest frame for $i=-$;
\begin{equation}
Z(M_{\pm,0})\,=\,1-i s_1 \Phi(M_{\pm,0},\sqrt{p^2})
\end{equation}
is the factor taking into account the interaction of the
$\rho$ and $\pi$ mesons in the final $\omega$ decay state,
where the parameter $s_1=1 \pm 0.2$ corresponds to the
prediction of~\cite{nachasov2}, where the specific form of the
$\Phi(M_{\pm, 0},\sqrt{p^2})$ function can be found. The couplings
$g_{\omega\rho\pi}$ and $g_{\rho\pi\pi}$ are the same for $\rho^{\pm,0}$
because of isotopic invariance. As emphasized in the text,
the angle $\xi$ is related to the intermediate resonance mass
by a simple expression.

Finally, let us give here the $p^2$-dependent width of the $\omega$-meson
taking into account the $3\pi$ and $\pi^0\gamma$ modes~\cite{pdg}:
\begin{equation}
\label{gamom}
\Gamma_{\omega}(p^2)\,=\,\frac{W(p^2)}{W(m^2_{\omega})}\mathcal{B}_{\omega\to 3 \pi}\Gamma_{\omega}+
\frac{m^2_{\omega}}{p^2}\frac{(p^2-m_0^2)^3}{(m^2_{\omega}-m^2_0)^3}\frac{g^2_{\omega\pi\gamma}(p^2)}{g^2_{\omega\pi\gamma}(m^2_{\omega})}\mathcal{B}_{\omega\to\pi\gamma}\Gamma_{\omega}{.}
\end{equation}
A form factor $g_{\omega\pi\gamma}(p^2)$ has a similar form (\ref{gomrhopi})~\cite{achasov} and \(g_{\omega\pi\gamma}(m^2_{\omega}) \simeq 0.7 \, \mathrm{GeV}^{-1}\)~\cite{lnsv}.


\end{document}